\documentclass[english,12pt]{article}

\usepackage[T1]{fontenc}
\usepackage[utf8]{inputenc}
\usepackage{array}
\usepackage{multirow}
\usepackage{amsmath}
\usepackage{amsthm}
\usepackage{amssymb}
\usepackage{graphicx}
\usepackage[authoryear]{natbib}

\providecommand{\tabularnewline}{\\}

\theoremstyle{plain}
\newtheorem{theorem}{Theorem}
\theoremstyle{plain}

\theoremstyle{plain}
\newtheorem{corollary}{Corollary}

\usepackage{cmap}
\usepackage[T1]{fontenc}

\usepackage[letterpaper,margin=1.0in]{geometry}
\usepackage{threeparttable}
\usepackage{hyperref}
\usepackage{bookmark}
\usepackage{bbm}
\usepackage{enumitem}

\usepackage{chngcntr}

\usepackage{setspace}
\usepackage{footmisc}
\newlength{\myfootnotesep}
\allowdisplaybreaks

\newtheorem{assumptionx}{Assumption}

\newcommand{\ind}{\mathbbm{1}}

\usepackage{etoolbox}
\makeatletter
\patchcmd{\NAT@citex}
  {\@citea\NAT@hyper@{%
     \NAT@nmfmt{\NAT@nm}%
     \hyper@natlinkbreak{\NAT@aysep\NAT@spacechar}{\@citeb\@extra@b@citeb}%
     \NAT@date}}
  {\@citea\NAT@nmfmt{\NAT@nm}%
   \NAT@aysep\NAT@spacechar\NAT@hyper@{\NAT@date}}{}{}

\patchcmd{\NAT@citex}
  {\@citea\NAT@hyper@{%
     \NAT@nmfmt{\NAT@nm}%
     \hyper@natlinkbreak{\NAT@spacechar\NAT@@open\if*#1*\else#1\NAT@spacechar\fi}%
       {\@citeb\@extra@b@citeb}%
     \NAT@date}}
  {\@citea\NAT@nmfmt{\NAT@nm}%
   \NAT@spacechar\NAT@@open\if*#1*\else#1\NAT@spacechar\fi\NAT@hyper@{\NAT@date}}
  {}{}
\makeatother

\usepackage{babel}

\begin{document}
\title{What Do We Get from Two-Way Fixed Effects Regressions? Implications
from Numerical Equivalence}
\author{Shoya Ishimaru\footnote{Hitotsubashi University, Department of Economics (email: shoya.ishimaru@r.hit-u.ac.jp). \newline I thank Mark Colas, Jonathan Meer, Nobuhiko Nakazawa, Masayuki Sawada, Christopher Taber, Kensuke Teshima, Takahide Yanagi, and many seminar participants for helpful comments and suggestions. Support from the Japan Society for the Promotion of Science (Grant No. 21K13310) is gratefully acknowledged. All errors are mine.}}
\maketitle

    \begin{abstract}

This paper develops numerical and causal interpretations of two-way
fixed effects (TWFE) regressions in settings with nonbinary, nonstaggered
treatments and time-varying covariates. Using the equivalence between
TWFE and pooled first-difference (FD) regressions, I express the TWFE
coefficient as a weighted average of FD coefficients across all horizons,
clarifying how short- and long-run changes contribute to the estimate.
Causal interpretation of the TWFE coefficient relies on common trends
assumptions at all horizons simultaneously, whereas each FD coefficient
relies on the assumption only at its own horizon. This structure opens
the identifying assumptions to empirical scrutiny: I propose diagnostic
procedures that assess common trends horizon by horizon, and illustrate
them by reexamining TWFE estimates of minimum-wage effects on employment.\vfill

    \end{abstract}


\section{Introduction}

Linear regression methods are widely used in empirical economics for
their simplicity, but their ability to deliver clear causal insights
is debated when treatment effects are heterogeneous \citep{angrist2010credibility,heckman2010comparing}.
Two-way fixed effects (TWFE) regressions, a workhorse method in panel
data settings, illustrate this tension. They build on the linear regression
framework with unit and time fixed effects, drawing conceptual motivation
from the canonical two-period difference-in-differences (DID) design.
Yet in settings with multiple periods and nonbinary treatments, the
link between the data and the TWFE coefficients becomes less transparent.
Recent methodological work has proposed alternative estimators, but
few simultaneously accommodate nonbinary treatments, time-varying
covariates, and complex treatment paths. TWFE regressions remain widely used in practice, highlighting the need to understand their behavior in general settings.

This paper examines TWFE regressions from both numerical and causal
perspectives, without assuming that their motivating linear equation
fully reflects the true causal relationships. My analysis builds on
a simple yet powerful insight: TWFE regressions can be understood
through their equivalence to pooled first-difference (FD) regressions.
This equivalence holds in panel data with units $i=1,\ldots,N$ and
periods $t=1,\ldots,T$, where TWFE regressions are based on the
equation
\begin{equation}
Y_{it}=\alpha_{i}+X_{it}'\beta+C_{i}'\gamma_{t}+\mu_{t}+\varepsilon_{it},\label{eq:FE_model}
\end{equation}
where $Y_{it}$ is a scalar outcome, $\alpha_{i}$ is a unit-specific
effect, $X_{it}$ is a vector of explanatory variables, $C_{i}$ is
a vector of time-invariant covariates, $\mu_{t}$ is a period-specific
effect, and $\varepsilon_{it}$ is a residual. Consider the corresponding
time-series differences of the TWFE equation:
\begin{equation}
\Delta_{k}Y_{it}=\Delta_{k}X_{it}'\beta+C_{i}'\Delta_{k}\gamma_{t}+\Delta_{k}\mu_{t}+\Delta_{k}\varepsilon_{it}\thinspace\thinspace\thinspace\text{for }k=1,\ldots,T-1,\label{eq:FD_model}
\end{equation}
where $\Delta_{k}a_{t}\equiv a_{t+k}-a_{t}$ denotes a $k$-period
difference of any time series $\{a_{t}\}_{t=1}^{T}$. Remarkably,
a least-squares estimate of the TWFE equation \eqref{eq:FE_model}
and a pooled least-squares estimate of the FD equation \eqref{eq:FD_model}
across all $k=1,\ldots,T-1$ yield algebraically identical estimates
of the coefficient $\beta$.

This equivalence generalizes the well-known TWFE--FD identity from
two-period panels ($T=2$) to multiperiod settings ($T\ge 2$). It
follows from a basic U-statistics identity and has been applied to
bias correction of the within estimator under correctly specified
models \citep{HanLee2017,han2022bias}. My contribution lies not in
the algebra but in its use: interpreting TWFE regressions without
assuming the linear model's causal validity. The interpretation
applies across binary, discrete, and continuous regressors, regardless
of treatment timing or the inclusion of covariates, including settings
where related decomposition results exist and settings where they do not.

Building on the equivalence, Section \ref{sec:Numerical-Equivalence}
decomposes the TWFE coefficient into a weighted average of the
$k$-period FD coefficients. While existing decompositions express
TWFE estimators as averages of comparisons between treatment and
comparison groups, revealing \emph{what units} are compared
(e.g., \citealp{goodman2018difference}), this
decomposition reveals \emph{over what time spans} comparisons are
made. Because it presumes no well-defined treatment and comparison
groups, it applies to arbitrary treatment paths. The FD coefficients
serve as building blocks of a different kind from the group-based DID
comparisons: their usefulness rests not primarily on being parameters
that empirical work ultimately targets, but on their diagnostic value.

This diagnostic value originates in the identifying assumptions the
decomposition calls for. Because TWFE regressions aggregate $k$-period changes, their causal
content rests on a common trends assumption stated at the level of
$k$-period changes:
\[
E\left[\Delta_{k}Y_{it}(d)|\Delta_{k}D_{it},\Delta_{k}W_{it},C_{i}\right]=E\left[\Delta_{k}Y_{it}(d)|\Delta_{k}W_{it},C_{i}\right],
\]
imposing conditional mean independence of the potential outcome change
$\Delta_{k}Y_{it}(d)$, evaluated at treatment level $d$, from the
concurrent treatment change $\Delta_{k}D_{it}$, given time-invariant covariates ($C_{i}$) and changes in time-varying covariates ($\Delta_{k}W_{it}$).
Crucially, this assumption conditions only on concurrent changes.
Strict exogeneity in panel settings, of which group exogeneity in
staggered designs is a special case, instead conditions on full
histories, so its $k=1$ statement automatically implies
the corresponding statements at all horizons. By contrast, this
assumption consists of separate restrictions, one per horizon $k$,
and its plausibility can be assessed one horizon at a time. 
Section \ref{sec:Common-Trends-Diagnostics} formally states
this assumption and develops diagnostic regressions that exploit
nonconcurrent treatment and outcome changes to flag its violations
horizon by horizon. Unlike the familiar pre-trend tests developed
for staggered adoption designs, these diagnostics apply under arbitrary
treatment paths and highlight the often-overlooked question of how
far across horizons one is willing to assume common trends.

Section \ref{sec:Causal-Interpretation} then formally establishes
the causal interpretations that these assumptions support, revealing
a basic asymmetry. The TWFE coefficient, because it aggregates all horizons, admits a weighted-average interpretation of treatment effects
only when common trends hold at all horizons $k=1,\ldots,T-1$. Each $k$-period
FD coefficient, by contrast, admits the same interpretation under
these conditions imposed at its own horizon alone. This asymmetry
is what gives the FD coefficients their diagnostic value: each is
causally interpretable under assumptions targeted to a single horizon,
so systematic variation in FD estimates
across $k$ signals that the conditions justifying TWFE's aggregation fail
somewhere. The diagnostics locate the failure at specific horizons.
When they support common trends across all horizons, researchers may
proceed with TWFE or, where applicable, heterogeneity-robust alternatives
that likewise rely on long-horizon common trends. When they reveal
violations at longer horizons, a more defensible approach is provided
by FD regressions at shorter horizons, or by
heterogeneity-robust alternatives that do not rely on common trends
across the entire panel. Section
\ref{sec:Empirical-Illustration} demonstrates this strategy in an
empirical application.

This paper contributes to the recent literature on TWFE regressions,
which has primarily focused on settings with binary or staggered treatments.
A growing number of studies investigate numerical and causal properties
of TWFE estimators in such settings, diagnose their issues, and propose
alternative estimators. The literature covers binary and staggered
treatment cases (e.g., \citealp{athey2018design,callaway2020difference,goodman2018difference,wooldridge2021two}),
binary and nonstaggered cases (e.g., \citealp{Chaisemartin2020}),
event-study settings (e.g., \citealp{Borusyak2018,lin2022interpreting,schmidheiny2020event,sun2020estimating}),
cases with multiple binary treatment variables (e.g., \citealp{de2022several}),
and continuous and staggered treatment settings (e.g., \citealp{callaway2021difference}).\footnote{\citet{wooldridge2021two} applies to general settings numerically
but focuses on binary, staggered cases for causal interpretation.
\citet{Chaisemartin2020} cover nonbinary treatments in an appendix;
the relationship to the present paper is discussed in Section \ref{subsec:TWFE-Interpretation}.} While these existing studies offer valuable insights into specific
well-defined settings, this paper provides general results for TWFE
regressions that hold across a broader range of specifications, including
binary, discrete, or continuous regressors, with or without covariates,
and under arbitrary treatment paths.

\section{Numerical Properties}\label{sec:Numerical-Equivalence}

This section develops a numerical interpretation of the TWFE estimator,
revealing how it aggregates information across time horizons. Unlike
existing approaches that decompose the TWFE estimator into between-group
comparisons, the approach here isolates the contribution of each time
horizon without presuming well-defined treatment and comparison groups.
Applicable to arbitrary treatment paths, this perspective enables 
horizon-specific assumptions for causal interpretation and diagnostics 
for their violations (Section \ref{sec:Common-Trends-Diagnostics}).

I focus on a balanced panel for simplicity; Appendix \ref{Asec:Unbalanced}
discusses how the results in the main paper extend to, or differ in,
an unbalanced panel. For any time series $\{a_{t}\}_{t=1}^{T}$, I
use $\Delta_{k}a_{t}\equiv a_{t+k}-a_{t}$ to denote a $k$-period
difference and $\overline{a}\equiv\frac{1}{T}\sum_{t=1}^{T}a_{t}$
to denote the time-series average.

\subsection{Equivalence of Least-Squares Objectives}

In a two-period panel, it is well known that TWFE and FD regressions
yield the same coefficient estimates. This equivalence extends to
multiperiod panels. The following theorem shows that the TWFE estimator
can be obtained from a least-squares problem that pools across all
$k$-period first differences. This result holds for both univariate
and multivariate regressors, whether they are binary, discrete, or
continuous.
\begin{theorem}
\label{thm:FE_FD_DID}Let $\widehat{\beta}_{\text{FE}}$ be the coefficient
on $X_{it}$ from the least-squares problem:
\begin{equation}
\underset{\beta,\{\alpha_{i}\}_{i=1}^{N},\{\gamma_{t},\mu_{t}\}_{t=1}^{T}}{\min}\sum_{i=1}^{N}\sum_{t=1}^{T}\left(Y_{it}-\alpha_{i}-X_{it}'\beta-C_{i}'\gamma_{t}-\mu_{t}\right)^{2}.\label{eq:FE_reg}
\end{equation}
Then $\beta=\widehat{\beta}_{\text{FE}}$ also solves:

\begin{align}
\underset{\beta,\{\gamma_{t},\mu_{t}\}_{t=1}^{T}}{\min} & \sum_{i=1}^{N}\sum_{k=1}^{T-1}\sum_{t=1}^{T-k}\left(\Delta_{k}Y_{it}-\Delta_{k}X_{it}'\beta-C_{i}'\Delta_{k}\gamma_{t}-\Delta_{k}\mu_{t}\right)^{2}.\label{eq:PFD_reg}
\end{align}
\end{theorem}
The equivalence follows from a well-known U-statistics identity (see
Appendix \ref{sec:Proofs}). A version of this result without period-specific
parameters $\{\gamma_{t},\mu_{t}\}_{t=1}^{T}$ appears in
\citet{HanLee2017,han2022bias}, who use it for bias correction under a
correctly specified linear model; Theorem \ref{thm:FE_FD_DID} extends
this representation to period-specific parameters through the same
identity. The same algebraic fact, however,
enables a fundamentally different use in this paper: reinterpreting TWFE regressions
without assuming a linear causal model, and doing so under arbitrary
treatment paths and covariate structures. For binary
treatment settings, \citet{goodman2018difference} and \citet{strezhnev2018semiparametric}
show that TWFE estimators can be expressed as averages of two-unit,
two-period DID comparisons; their results stem from the same numerical structure, as detailed in Appendix \ref{sec:Comparison-GB}.

\subsection{Weighted-Average Relationship}\label{subsec:Weighted-Average-Relationship}

A least-squares estimate of equation \eqref{eq:FD_model} using only
$k$-period differences is obtained from:
\begin{equation}
\underset{\beta,\{\gamma_{t},\mu_{t}\}_{t=1}^{T}}{\min}\sum_{i=1}^{N}\sum_{t=1}^{T-k}\left(\Delta_{k}Y_{it}-\Delta_{k}X_{it}'\beta-C_{i}'\Delta_{k}\gamma_{t}-\Delta_{k}\mu_{t}\right)^{2},\label{eq:FD_LS}
\end{equation}
which yields $\widehat{\beta}_{\text{FD},k}$ as the coefficient on
$\Delta_{k}X_{it}$ for each $k=1,\ldots,T-1$. The least-squares problem \eqref{eq:PFD_reg}, which produces $\widehat{\beta}_{\text{FE}}$
according to Theorem \ref{thm:FE_FD_DID}, pools the objective \eqref{eq:FD_LS}
across all $k=1,\ldots,T-1$. The similarity of the two objectives
indicates a tight numerical connection between the TWFE coefficient
$\widehat{\beta}_{\text{FE}}$ and the FD coefficients $\left\{ \widehat{\beta}_{\text{FD},k}\right\} _{k=1}^{T-1}$.

In fact, this connection admits an exact matrix-weighted-average interpretation: the TWFE coefficient can be written as a matrix-weighted average of the FD coefficients, with weight matrices that sum to the identity. This result is derived in the proof of Theorem~\ref{theorem:FE_not_equal_FD} in Appendix \ref{sec:Proofs} and is stated as equation \eqref{eq:FE_FD_multi}.

The interpretation of this matrix-weighted representation depends
on whether $X_{it}$ is univariate or multivariate. When $X_{it}$
is univariate, the matrix weights reduce to scalars and the TWFE coefficient
is simply a convex combination of the $k$-period FD coefficients.
When $X_{it}$ is multivariate, the aggregation generally involves
matrix-valued weights, which are harder to interpret directly. In
practice, however, empirical applications typically focus on the coefficient
on a single treatment variable, motivating a decomposition that isolates
that coefficient even in multivariate specifications.

I therefore consider a common empirical setting in which $X_{it}$
consists of a scalar treatment variable $D_{it}$ and a vector of
time-varying covariates $W_{it}$, and focus on the interpretation
of the coefficient on $D_{it}$. Using $D_{it}$ as the first element
of $X_{it}$, I write:
\[
X_{it}=\left[\begin{array}{c}
D_{it}\\
W_{it}
\end{array}\right],\thinspace\thinspace\thinspace\widehat{\beta}_{\text{FE}}=\left[\begin{array}{c}
\widehat{\beta}_{\text{FE}}^{D}\\
\widehat{\beta}_{\text{FE}}^{W}
\end{array}\right],\thinspace\thinspace\thinspace\text{and}\,\thinspace\thinspace\widehat{\beta}_{\text{FD},k}=\left[\begin{array}{c}
\widehat{\beta}_{\text{FD},k}^{D}\\
\widehat{\beta}_{\text{FD},k}^{W}
\end{array}\right].
\]
I also define 
\begin{equation}
\widetilde{Y}_{it}\equiv Y_{it}-\left(\begin{array}{c}
1\\
C_{i}
\end{array}\right)'\left(\sum_{i=1}^{N}\left(\begin{array}{c}
1\\
C_{i}
\end{array}\right)\left(\begin{array}{c}
1\\
C_{i}
\end{array}\right)'\right)^{-1}\sum_{i=1}^{N}\left(\begin{array}{c}
1\\
C_{i}
\end{array}\right)Y_{it}\label{eq:Resid}
\end{equation}
to be a residual from a regression of $Y_{it}$ on $C_{i}$ performed separately for each $t$, and analogously for other variables.

The following theorem establishes that $\widehat{\beta}_{\text{FE}}^{D}$
can be expressed as a weighted average of $\widehat{\beta}_{\text{FD},k}^{D}$
after adjusting for the influence of time-varying covariates.
\begin{theorem}
\label{theorem:FE_not_equal_FD}The TWFE coefficient on $D_{it}$
is given by:
\begin{equation*}
\widehat{\beta}_{\text{FE}}^{D}=\sum_{k=1}^{T-1}\widehat{w}_{k}\widehat{\beta}_{\text{FD},k}^{D}+\frac{\sum_{k=1}^{T-1}\sum_{i=1}^{N}\sum_{t=1}^{T-k}\left(\widehat{\delta}_{\text{FD},k}^{W}-\widehat{\delta}_{\text{FE}}^{W}\right)'\left(\Delta_{k}\widetilde{W}_{it}\Delta_{k}\widetilde{W}_{it}'\right)\left(\widehat{\beta}_{\text{FD},k}^{W}-\widehat{\beta}_{\text{FE}}^{W}\right)}{\sum_{k=1}^{T-1}\sum_{i=1}^{N}\sum_{t=1}^{T-k}\Delta_{k}\widetilde{D}_{it}\left(\Delta_{k}\widetilde{D}_{it}-\Delta_{k}\widetilde{W}_{it}'\widehat{\delta}_{\text{FE}}^{W}\right)},
\end{equation*}
where $\widehat{\delta}_{\text{FE}}^{W}$ and $\widehat{\delta}_{\text{FD},k}^{W}$
are the coefficients from the TWFE and $k$-period FD regressions
of $D_{it}$ on $W_{it}$, and the weights are defined as:
\begin{equation}
\widehat{w}_{k}\equiv\frac{\sum_{i=1}^{N}\sum_{t=1}^{T-k}\Delta_{k}\widetilde{D}_{it}\left(\Delta_{k}\widetilde{D}_{it}-\Delta_{k}\widetilde{W}_{it}'\widehat{\delta}_{\text{FE}}^{W}\right)}{\sum_{\ell=1}^{T-1}\sum_{i=1}^{N}\sum_{t=1}^{T-\ell}\Delta_{\ell}\widetilde{D}_{it}\left(\Delta_{\ell}\widetilde{D}_{it}-\Delta_{\ell}\widetilde{W}_{it}'\widehat{\delta}_{\text{FE}}^{W}\right)}.\label{eq:weight_cov}
\end{equation}
\end{theorem}
This theorem shows that $\widehat{\beta}_{\text{FE}}^{D}$ can be
decomposed into two components. The first is a weighted average of
the FD coefficients. The weights sum to one and reflect variation in treatment
changes not explained by covariates or time effects. The second is an adjustment term, which arises from discrepancies
between the TWFE and FD regressions in how they account for covariate
effects. While a TWFE regression applies the same coefficients ($\widehat{\beta}_{\text{FE}}^{W}$
and $\widehat{\delta}_{\text{FE}}^{W}$) to all $k$-period covariate
changes, FD regressions allow the coefficients on covariate changes
($\widehat{\beta}_{\text{FD},k}^{W}$ and $\widehat{\delta}_{\text{FD},k}^{W}$)
to vary across $k=1,\ldots,T-1$, leading to a deviation from a clean
weighted-average interpretation. Although the adjustment term complicates the decomposition, 
it provides diagnostic value in its own right by signaling a violation of time homogeneity, 
one of the assumptions underlying TWFE's causal interpretation (Section \ref{sec:Causal-Interpretation}).

This decomposition isolates how different time horizons contribute
to the TWFE estimator. Unlike decompositions based on between-group comparisons
(e.g., \citealp{goodman2018difference}), which reveal \emph{what units}
are being compared, this decomposition reveals \emph{over what time spans}
comparisons are made. Group-based decompositions presume well-defined
treatment and comparison groups and are most naturally interpreted
when common trends assumptions hold across all time horizons. By contrast, the present decomposition applies to arbitrary treatment paths and organizes the TWFE estimator by time horizon, so that systematic variation in $\widehat{\beta}_{\text{FD},k}^{D}$ across $k$ may signal that these assumptions do not hold uniformly across horizons. Section \ref{sec:Common-Trends-Diagnostics} develops diagnostic tools to evaluate these signals more formally.

\section{Common Trends and Diagnostics}\label{sec:Common-Trends-Diagnostics}

Section \ref{sec:Numerical-Equivalence} shows that the TWFE estimator
aggregates $k$-period changes across all horizons $k=1,\ldots,T-1$.
Causal interpretation of the TWFE coefficient therefore naturally
rests on assumptions about $k$-period changes: whether treatment
changes over a $k$-period horizon are unrelated to how a potential outcome would
have evolved over the same horizon. This
section states this common trends assumption for a given horizon $k$
and develops diagnostics that flag its violations. Section
\ref{sec:Causal-Interpretation} then formalizes the causal
interpretations that build on it: each $k$-period FD coefficient
relies on common trends only at its own horizon, whereas the TWFE
coefficient, aggregating all horizons, invokes it at every horizon
simultaneously.

\subsection{Setup and the Common Trends Assumption}\label{subsec:CT-Assumption}

In the remainder of the paper, I focus on the case in which $X_{it}$
consists of a scalar treatment ($D_{it}$) and a vector of time-varying
covariates ($W_{it}$). I consider a large-$N$, fixed-$T$ setting,
with $\left(Y_{it},D_{it},W_{it}\right)_{t=1}^{T}$ and $C_{i}$ being
independent and identically distributed across $i=1,\ldots,N$. The
cross-sectional mean is represented by $E\left[Y_{it}\right]$, while
the time-series mean is denoted by $\overline{Y}_{i}$. I adopt a
potential outcome framework.

\renewcommand{\theassumptionx}{PO}
\begin{assumptionx}\label{assu:PO}\textup{(Potential Outcome, Static)} 
	For each $t=1,\ldots,T$,
	$\{Y_{it}(d):d\in(\underline{d},\overline{d})\}$ is a stochastic
	process that defines a potential outcome associated with each possible
	treatment level $d\in(\underline{d},\overline{d})$, where $-\infty\le\underline{d}<\overline{d}\le\infty$.
	The observed outcome is given by $Y_{it}=Y_{it}(D_{it})$.
\end{assumptionx}

Assumption \ref{assu:PO} restricts potential outcomes to depend only
on current treatment status, ruling out dynamic treatment effects.
In staggered adoption settings with binary treatments, TWFE regressions
can be analyzed while allowing dynamics (e.g., \citealp{callaway2020difference,goodman2018difference}).
With continuous treatments
and arbitrary treatment paths, allowing both heterogeneity and dynamics
would make potential outcomes intractably high-dimensional.\footnote{Appendix \ref{subsec:Dynamic-Effects}
considers dynamics under a homogeneous
treatment effect restriction, demonstrating that the TWFE coefficient
remains difficult to interpret even under this strong assumption.}

The key identifying condition is a common trends assumption, stated
for a given difference length $k\in\{1,\ldots,T-1\}$:
\renewcommand{\theassumptionx}{CT--k}
\begin{assumptionx}\label{assu:PT-k}\textup{(Conditional Common Trends, $k$-period)}
	There exists $d^{0}\in(\underline{d},\overline{d})$ such that the
	following holds for any $t\in\{1,\ldots,T-k\}$:
	\[
	E\left[\Delta_{k}Y_{it}(d^{0})|\Delta_{k}D_{it},\Delta_{k}W_{it},C_{i}\right]=E\left[\Delta_{k}Y_{it}(d^{0})|\Delta_{k}W_{it},C_{i}\right].
	\]
\end{assumptionx}

Assumption \ref{assu:PT-k} requires that trends in potential outcomes
$\Delta_{k}Y_{it}(d^{0})$ at baseline treatment level $d^{0}$
be mean independent from treatment changes $\Delta_{k}D_{it}$, conditional
on time-invariant covariates $C_{i}$ and the concurrent changes $\Delta_{k}W_{it}$
in time-varying covariates. It requires mean independence from the
concurrent treatment changes $\Delta_{k}D_{it}$, rather than from
the entire treatment path $(D_{i1},\ldots,D_{iT})$. Therefore, when
considered for any individual $k$, the condition is weaker than the
strict exogeneity assumption standard in panel data and commonly invoked
in DID settings.\footnote{In staggered adoption designs, the standard assumption imposes exogeneity
	with respect to treatment group (defined by adoption timing), which
	is by construction equivalent to exogeneity with respect to the entire
	treatment path.} However, the causal interpretation of the TWFE coefficient
rests on this condition at all difference lengths
$k=1,\ldots,T-1$ simultaneously, as formalized in Section
\ref{subsec:TWFE-Interpretation}.

Assumption \ref{assu:PT-k} presumes the existence of a natural baseline
treatment level with economic meaning, such as ``no treatment''
($d^{0}=0$) in a binary treatment setting. The baseline level $d^{0}$
serves as the counterfactual reference point for causal interpretation.
In particular, under Assumption \ref{assu:PO}, the baseline level $d^{0}$ also defines
\begin{equation}
\tau_{it}\equiv\begin{cases}
	\frac{Y_{it}(D_{it})-Y_{it}(d^{0})}{D_{it}-d^{0}} & D_{it}\ne d^{0}\\
	0 & D_{it}=d^{0}
\end{cases}\label{eq:tau_it}
\end{equation}
as a per-unit effect of a deviation of the treatment $D_{it}$ from
the baseline level $d^{0}$. The diagnostics below use $\tau_{it}$ to
link observed outcome changes to potential outcome changes.

The baseline level $d^{0}$ need not be experienced in the data as
long as it represents an economically meaningful reference point and
the common trends assumption at $d^{0}$ can be economically justified.\footnote{For instance, a tariff rate of zero in trade policy analysis may provide
	a natural baseline even if no country implements complete free trade,
	and common trends in the absence of trade barriers may be plausible
	on economic grounds.} When no natural baseline exists, such as with minimum wage policy
where all levels represent active intervention, it can be problematic
to assume common trends hold at one arbitrarily chosen level while
potentially failing at all others. Appendix \ref{subsec:Causal-Interpretation-without} addresses this issue by invoking the common trends assumption for
every treatment level.

\subsection{Diagnosing Violations of the Common Trends Assumption}\label{subsec:Diagnosing-Common-Trends}

The common trends assumption cannot be directly tested, since potential
outcome changes $\Delta_{k}Y_{it}(d^{0})$ are unobserved. In staggered
adoption settings, pre-trend tests provide suggestive evidence about
the plausibility of common trends. By contrast, under general treatment
paths there is no obvious way to scrutinize this assumption. Yet the
plausibility of this assumption can be assessed through testable implications
involving the relationship among nonconcurrent treatment and outcome
changes. This section develops diagnostic procedures that exploit
the temporal structure of panel data to flag violations of common
trends across different time horizons.

\subsubsection*{Sufficient Conditions for Common Trends}

The key insight underlying these diagnostics is that the common trends
assumption becomes implausible when treatment changes are systematically
related to nonconcurrent outcome changes. Suppose that for given $k\in\{2,\ldots,T-1\}$,
potential outcome changes in the first $\ell$ periods are correlated
with treatment changes in the subsequent $k-\ell$ periods, or treatment
changes in the first $\ell$ periods are correlated with potential
outcome changes in the subsequent $k-\ell$ periods. This would suggest
intertemporal outcome-to-treatment feedback or unobserved factors
that drive both treatment and outcome dynamics over the horizon of
$k$ periods, indicating a potential violation of the common trends
assumption.

More formally, for each $\ell\in\{1,\ldots,k-1\}$, the following
two conditions jointly form a sufficient condition for Assumption
\ref{assu:PT-k}:
\begin{description}
	\item [{(Condition\,I)}] $E\left[\Delta_{\ell}Y_{it}(d^{0})|\Delta_{\ell}D_{it},\Delta_{k-\ell}D_{i,t+\ell},\Delta_{k}W_{it},C_{i}\right]=E\left[\Delta_{\ell}Y_{it}(d^{0})|\Delta_{k}W_{it},C_{i}\right]$
	for $t\in\{1,\ldots,T-k\}$.
	\item [{(Condition\,II)}] $E\left[\Delta_{k-\ell}Y_{i,t+\ell}(d^{0})|\Delta_{\ell}D_{it},\Delta_{k-\ell}D_{i,t+\ell},\Delta_{k}W_{it},C_{i}\right]=E\left[\Delta_{k-\ell}Y_{i,t+\ell}(d^{0})|\Delta_{k}W_{it},C_{i}\right]$
	for $t\in\{1,\ldots,T-k\}$.
\end{description}
Condition I requires that potential outcome changes over periods $t$
to $t+\ell$ be mean independent of future treatment changes over
periods $t+\ell$ to $t+k$ as well as concurrent treatment changes,
conditional on covariates. Condition II requires that potential outcome
changes over periods $t+\ell$ to $t+k$ be mean independent of past
treatment changes over periods $t$ to $t+\ell$ as well as concurrent
treatment changes, conditional on covariates.

\subsubsection*{Testable Implications and Diagnostic Regressions}

To overcome the challenge that potential outcome changes are unobserved,
I consider the following decomposition of observed outcome changes,
which follows from Assumption \ref{assu:PO} and the definition of
per-unit treatment effects $\tau_{it}$ in \eqref{eq:tau_it}.
\[
\Delta_{k}Y_{it}=\Delta_{k}Y_{it}(d^{0})+\left(\frac{\tau_{i,t+k}+\tau_{it}}{2}\right)\Delta_{k}D_{it}+\Delta_{k}\tau_{it}\left(\frac{D_{i,t+k}+D_{it}}{2}-d^{0}\right).
\]
While heterogeneity of per-unit treatment effects $\tau_{it}$ complicates
the analysis in general, a constant effects assumption delivers clean
testable implications.
\renewcommand{\theassumptionx}{CE}
\begin{assumptionx}
	\label{assu:CE}\textup{(Constant Effects)} The per-unit treatment
	effect satisfies $\tau_{it}=\tau$ for all $(i,t)$.
\end{assumptionx}
Combining Assumption \ref{assu:CE} and Condition I yields:
\[
E\left[\Delta_{\ell}Y_{it}|\Delta_{\ell}D_{it},\Delta_{k-\ell}D_{i,t+\ell},\Delta_{k}W_{it},C_{i}\right]=E\left[\Delta_{\ell}Y_{it}(d^{0})|\Delta_{k}W_{it},C_{i}\right]+\tau\Delta_{\ell}D_{it}.
\]
Similarly, under Assumption \ref{assu:CE} and Condition II:
\[
E\left[\Delta_{k-\ell}Y_{i,t+\ell}|\Delta_{\ell}D_{it},\Delta_{k-\ell}D_{i,t+\ell},\Delta_{k}W_{it},C_{i}\right]=E\left[\Delta_{k-\ell}Y_{i,t+\ell}(d^{0})|\Delta_{k}W_{it},C_{i}\right]+\tau\Delta_{k-\ell}D_{i,t+\ell}.
\]
These expressions provide the foundation for empirical diagnostics,
implemented by the following regression specifications:
\begin{eqnarray}
	\Delta_{\ell}Y_{it} & = & \beta_{0}+\beta_{1}\Delta_{k-\ell}D_{i,t+\ell}+\beta_{2}\Delta_{\ell}D_{it}+C_{i}'\gamma_{t}+\Delta_{k}W_{it}'\delta+\epsilon_{it},\label{eq:lead_reg}\\
	\Delta_{k-\ell}Y_{i,t+\ell} & = & \beta_{0}+\beta_{1}\Delta_{\ell}D_{it}+\beta_{2}\Delta_{k-\ell}D_{i,t+\ell}+C_{i}'\gamma_{t}+\Delta_{k}W_{it}'\delta+\epsilon_{it}.\label{eq:lag_reg}
\end{eqnarray}

Each specification regresses outcome changes on lead/lag treatment
changes, controlling for concurrent treatment changes and covariates.
Under the joint hypothesis of Conditions I and II (sufficient for
Assumption \ref{assu:PT-k}) and constant effects (Assumption \ref{assu:CE}),
we should observe $\beta_{1}=0$ for each regression. These diagnostics
can be implemented for increasing values of $k$, starting with $k=2$.
Since Conditions I and II also demand exogeneity of concurrent treatment
changes, violation of common trends for some $k^{*}$ implies violation
of Conditions I and II for larger $k>k^{*}$, which in turn signals
(though does not imply) violation of Assumption \ref{assu:PT-k}.

These diagnostics resemble distributed-lag regressions but differ
in a key respect: they exploit treatment changes over
nonoverlapping time periods, whereas distributed-lag regressions,
being a form of multivariate TWFE regression, pool all possible FDs (by Theorem \ref{thm:FE_FD_DID})
and therefore incorporate numerous overlapping treatment changes,
obscuring the connection to specific common trends assumptions.

\subsubsection*{Limitations and Interpretation}

These diagnostic procedures should not be viewed as direct tests
of the common trends assumption---a limitation inherent to all such
diagnostics, including standard pre-trend tests. The link between
diagnostic results and the common trends assumption involves multiple
inferential steps, each with important caveats.

First, Conditions I and II are sufficient but not necessary for the
common trends assumption (Assumption \ref{assu:PT-k}). Their violation
therefore does not in itself imply that the assumption fails. Nevertheless,
for the common trends assumption to hold despite violations of Conditions
I and II, correlations in different subperiods must coincidentally
cancel out when aggregated over the full $k$-period horizon. While
such cancellation is theoretically possible, it would demand a precise
offsetting of intertemporal feedback, which is difficult to justify
on economic grounds.

Second, $\beta_{1}=0$ in the lead and lag diagnostics is not sufficient
for Conditions I and II to hold, since the diagnostics are silent
about the exogeneity of concurrent treatment changes---a key part
of these conditions. This mirrors a well-known limitation of pre-trend
tests in staggered adoption designs, where absence of differential
pre-trends does not guarantee common trends in post-treatment periods.

Third, the purest interpretation of the diagnostics requires the constant
effects assumption (Assumption \ref{assu:CE}). When treatment effects
$\tau_{it}$ vary and depend on treatment paths, the coefficients
$\beta_{1}$ can be contaminated by this heterogeneity rather than
purely reflecting violations of the sufficient conditions. Yet, this
limitation does not undermine diagnostic value: if $\beta_{1}\ne0$
is detected, it indicates either a violation of the sufficient conditions
for common trends or the presence of treatment effect heterogeneity,
which complicates the causal interpretation of TWFE coefficients
formalized in Section \ref{subsec:TWFE-Interpretation}. Both scenarios suggest that
standard TWFE estimates may be problematic, making the diagnostic
informative regardless of which interpretation applies.

These caveats notwithstanding, the diagnostics remain informative
for assessing the plausibility of common trends at different horizons,
especially since intertemporal outcome--treatment correlation captures
economically meaningful threats to identification, such as anticipation,
policy responses to past outcomes, or persistent unobserved shocks.
They also have the ancillary benefit of detecting dynamic treatment
effects, since $\beta_{1}\neq0$ can arise when treatment effects
depend on treatment history (see Appendix \ref{subsec:Dynamic-Effects}).
Accordingly, the diagnostics help evaluate whether TWFE estimates
are likely to be interpretable as causal effects, and whether estimators
that rely more heavily on short-run (or long-run) variation may be
preferable, as discussed in Section \ref{subsec:Discussion}.

\section{Causal Interpretation}\label{sec:Causal-Interpretation}

\subsection{TWFE Interpretation}\label{subsec:TWFE-Interpretation}

This section clarifies the conditions under which the population TWFE
coefficient can be interpreted causally, maintaining the setting
introduced in Section \ref{subsec:CT-Assumption}. In addition to the
notation introduced there, I define
\[
\widetilde{Y}_{it}\equiv Y_{it}-\left(\begin{array}{c}
	1\\
	C_{i}
\end{array}\right)'E\left[\left(\begin{array}{c}
	1\\
	C_{i}
\end{array}\right)\left(\begin{array}{c}
	1\\
	C_{i}
\end{array}\right)'\right]^{-1}E\left[\left(\begin{array}{c}
	1\\
	C_{i}
\end{array}\right)Y_{it}\right]
\]
to be a residual from the population projection of $Y_{it}$ on $(1,C_{i})$.
This parallels the sample definition in \eqref{eq:Resid}; for simplicity
of notation I reuse the same symbol.

The equivalence result in Section \ref{sec:Numerical-Equivalence}
reduces a TWFE regression to a pooled regression of $\Delta_{k}\widetilde{Y}_{it}$
on $\Delta_{k}\widetilde{D}_{it}$ controlling for $\Delta_{k}\widetilde{W}_{it}$.
The population TWFE coefficient on $D_{it}$ is:
\begin{equation}
	\beta_{\text{FE}}^{D}=\frac{\sum_{k=1}^{T-1}\sum_{t=1}^{T-k}E\left[\Delta_{k}Y_{it}\left(\Delta_{k}\widetilde{D}_{it}-\Delta_{k}\widetilde{W}_{it}'\delta_{\text{FE}}^{W}\right)\right]}{\sum_{k=1}^{T-1}\sum_{t=1}^{T-k}E\left[\Delta_{k}D_{it}\left(\Delta_{k}\widetilde{D}_{it}-\Delta_{k}\widetilde{W}_{it}'\delta_{\text{FE}}^{W}\right)\right]},\label{eq:pop_FE}
\end{equation}
where
\begin{equation*}
	\delta_{\text{FE}}^{W}=\left(\sum_{k=1}^{T-1}\sum_{t=1}^{T-k}E\left[\Delta_{k}\widetilde{W}_{it}\Delta_{k}\widetilde{W}_{it}'\right]\right)^{-1}\left(\sum_{k=1}^{T-1}\sum_{t=1}^{T-k}E\left[\Delta_{k}\widetilde{W}_{it}\Delta_{k}\widetilde{D}_{it}\right]\right)
\end{equation*}
represents the population version of $\widehat{\delta}_{\text{FE}}^{W}$.
With incidental unit-specific effect parameters eliminated,
this structure becomes analogous to a standard cross-sectional regression,
making the problem tractable with standard causal inference tools.

Together with Assumption \ref{assu:PO}, the following assumptions
characterize the conditions under which $\beta_{\text{FE}}^{D}$
admits causal interpretation. The goal is
not to justify these assumptions but to make explicit what the TWFE
approach relies on, enabling researchers to evaluate when it is appropriate.

\renewcommand{\theassumptionx}{CT}
\begin{assumptionx}\label{assu:PT}\textup{(Conditional Common Trends at All Horizons)}
	Assumption \ref{assu:PT-k} holds for every $k=1,\ldots,T-1$, with 
	common baseline level $d^{0}\in(\underline{d},\overline{d})$ across
	all $k$.
\end{assumptionx}

Because the TWFE estimator pools all difference lengths, its causal
interpretation invokes the common trends assumption at every horizon
simultaneously. This structure imposes collective restrictions on treatment
dynamics, as detailed in the remarks at the end of this section.

\renewcommand{\theassumptionx}{V}
\begin{assumptionx}\label{assu:Variation}\textup{(Variation)}
	$\sum_{k=1}^{T-1}\sum_{t=1}^{T-k}E\left[Var\left(\Delta_{k}D_{it}|\Delta_{k}W_{it},C_{i}\right)\right]>0$.
\end{assumptionx}

Assumption \ref{assu:Variation} requires that $\Delta_{k}D_{it}$
have some variation conditional on $\left(\Delta_{k}W_{it},C_{i}\right)$.

\renewcommand{\theassumptionx}{LH}
\begin{assumptionx}\label{assu:Homogeneity}\textup{(Linearity and Time Homogeneity)}
	For any $(k,t)$ with $1\le t<t+k\le T$, the conditional expectation
	of $\Delta_{k}D_{it}$ given $\left(\Delta_{k}W_{it},C_{i}\right)$
	can be expressed as
	\[
	E\left[\Delta_{k}D_{it}|\Delta_{k}W_{it},C_{i}\right]=\Delta_{k}W_{it}'\delta^{W}+C_{i}'\delta_{k,t}^{C}+\delta_{k,t}^{I},
	\]
	where the coefficient $\delta^{W}$ does not vary across $(k,t)$.
\end{assumptionx}

Assumption \ref{assu:Homogeneity} imposes two restrictions: linearity
of the conditional mean function, and time homogeneity of the coefficient
$\delta^{W}$ across $(k,t)$. The linearity component can be made
less restrictive through flexible covariate specifications (\citealp{angrist1999empirical}),
while the time homogeneity component imposes a fundamental restriction
that cannot be relaxed within the TWFE framework. Under this assumption, the TWFE and $k$-period FD regressions of $D_{it}$ on $W_{it}$ share the same population coefficient, so the adjustment term in Theorem \ref{theorem:FE_not_equal_FD} vanishes.

The following theorem characterizes $\beta_{\text{FE}}^{D}$ under
the full set of assumptions. Appendix \ref{subsec:Generalization} shows that Assumption \ref{assu:PT} and both linearity and time homogeneity
components of Assumption \ref{assu:Homogeneity} are necessary: relaxing
any of them introduces bias terms that preclude causal interpretation.
\begin{theorem}
	\label{thm:Causal_TWFE}Under Assumptions \ref{assu:PO}, \ref{assu:PT},
	\ref{assu:Variation}, and \ref{assu:Homogeneity},
	\[
	\beta_{\text{FE}}^{D}=\sum_{t=1}^{T}E\left[\tau_{it}\omega_{it}\right],
	\]
	where $\tau_{it}$ is the per-unit effect defined in \eqref{eq:tau_it}. The weights satisfy $\sum_{t=1}^{T}E\left[\omega_{it}\right]=1$
	and
	\[
	\omega_{it}\propto\left(D_{it}-d^{0}\right)\left\{ \left(\widetilde{D}_{it}-\overline{\widetilde{D}}_{i}\right)-\left(\widetilde{W}_{it}-\overline{\widetilde{W}}_{i}\right)'\delta^{W}\right\} .
	\]
\end{theorem}
Theorem \ref{thm:Causal_TWFE} interprets $\beta_{\text{FE}}^{D}$
as a weighted average of per-unit treatment effects, with weights
summing to one but possibly negative. The weight function $\omega_{it}$
interacts a deviation of $D_{it}$ from the baseline level $d^{0}$
with a TWFE residual of $D_{it}$.

However, this weighted-average interpretation faces multiple challenges.
Three issues concern the assumptions needed for the weighted-average
interpretation, while a fourth issue arises even when this interpretation
holds.

First, the common trends assumption (Assumption \ref{assu:PT}) must
hold across all difference lengths $k$. This requirement extends
far beyond the canonical two-period DID setting, and becomes increasingly
difficult to justify as the number of time periods grows. The diagnostics
in Section \ref{subsec:Diagnosing-Common-Trends} assess this requirement
horizon by horizon.

Second, the common trends assumption conditions on covariate changes $\Delta_{k}W_{it}$, unlike in the canonical two-period DID \citep{heckman1997matching,heckman1998matching,abadie2005semiparametric}, which conditions on pretreatment levels. \citet{caetano2022difference} formalize why conditioning on changes is theoretically less attractive; if the
treatment change $\Delta_{k}D_{it}$ influences the covariate change
$\Delta_{k}W_{it}$, then variation in $\Delta_{k}D_{it}$ with $\Delta_{k}W_{it}$
being fixed does not correspond to the \emph{ceteris paribus} variation
that defines the causal effect of the treatment.

Third, the linearity and time homogeneity assumption (Assumption \ref{assu:Homogeneity})
can be restrictive unless all covariates are discrete and time-invariant.
This assumption demands that the relationship between treatment changes
and covariate changes be not only linear but also constant across
all time periods and difference lengths. Despite being essential for
causal interpretation, these requirements are typically assumed implicitly
rather than explicitly defended, unlike common trends assumptions.

Fourth, even when the weighted-average interpretation holds, the weights
$\omega_{it}$ can be negative, as similarly highlighted in binary
treatment cases (e.g., \citealp{Borusyak2018,Chaisemartin2020,goodman2018difference}).
Whether this poses a problem depends on the correlation between weights
and treatment effects. If some weights are negative and weights are
strongly correlated with per-unit treatment effects, then $\beta_{\text{FE}}^{D}$
can even lie outside the support of per-unit treatment effects $\tau_{it}$.
Nevertheless, negative weights alone are not necessarily problematic
if treatment effects are uncorrelated with weights, as noted by \citet{Chaisemartin2020}
in their Corollary 2. Appendix \ref{subsec:Causal-Interpretation-without} shows that the weight function $\omega_{it}$ contains redundant variation
that is orthogonal to treatment effects under a stricter common
trends assumption, which may alleviate the concern about negative
weights in some cases.

Overall, these issues highlight that causal interpretation of TWFE
estimators involves more restrictive assumptions than commonly recognized,
particularly regarding the length over which common trends can be
justified, the role of time-varying covariates, and the need for linearity
and time homogeneity conditions, all of which are rarely scrutinized
in practice.

\subsubsection*{Remarks}

\citet{Chaisemartin2020} provide, in their online appendix, a weighted-average interpretation
of the TWFE coefficient that parallels Theorem \ref{thm:Causal_TWFE}. In
the current setup, their counterpart to Assumption \ref{assu:PT}
can be expressed as
\begin{equation}
	E\left[\Delta_{1}Y_{it}(d^{0})|D_{i1},\ldots,D_{iT},W_{i1},\ldots,W_{iT},C_{i}\right]=E\left[\Delta_{1}Y_{it}(d^{0})|\Delta_{1}W_{it},C_{i}\right].\label{eq:CD_cond}
\end{equation}
They also assume linearity and time homogeneity of $E\left[\Delta_{1}Y_{it}(d^{0})|\Delta_{1}W_{it},C_{i}\right]$, analogous in structure to Assumption \ref{assu:Homogeneity}, but applied to the outcome rather than the treatment.

Their assumption \eqref{eq:CD_cond} conditions on entire treatment
and covariate histories. Thus, this single-period ($k=1$) assumption
implies the corresponding assumption for all difference lengths $k=1,\ldots,T-1$
when combined with their linearity and time homogeneity assumption.
By contrast, Assumption \ref{assu:PT} conditions only on concurrent
changes $(\Delta_{k}D_{it},\Delta_{k}W_{it})$, and the assumption
for $k=1$ does not imply the corresponding ones for longer differences.
For example, $\Delta_{2}Y_{it}(d^{0})$ may correlate with $\Delta_{2}D_{it}$
even when $\Delta_{1}$ terms show no correlation, due to feedback
between outcome changes in one period and treatment changes in another.

The contribution of Theorem \ref{thm:Causal_TWFE} relative to their
result lies in showing that Assumption \ref{assu:PT}, which relies
solely on concurrent changes rather than full histories, is central
to causal interpretation of the TWFE coefficient. The importance of
covariate changes rather than levels for TWFE interpretation is well recognized
in canonical two-period settings, stemming from the known equivalence
between TWFE and FD. Theorem \ref{thm:Causal_TWFE} shows that concurrent
changes remain the fundamental basis for causal interpretation even
in multiperiod settings. \citet{caetano2024difference} also emphasize the role of covariate changes for TWFE interpretation but focus on the two-period case; in their multiperiod extension, they instead condition on full covariate histories.

While this emphasis on concurrent changes clarifies the role of covariates,
the implications for treatment dynamics warrant closer examination.
Exogeneity of an entire treatment path $\left(D_{i1},\ldots,D_{iT}\right)$
as in \eqref{eq:CD_cond}, namely a strict exogeneity condition,
is a common assumption in panel data models. This assumption rules
out the possibility that a shock to the current outcome influences
future treatment status or is correlated with past treatment status.
Although the constraints that Assumption \ref{assu:PT} imposes on treatment
dynamics appear weaker than strict exogeneity when considered for
each $k=1,\ldots,T-1$ individually, similar restrictions arise when
considered collectively. Specifically, the common trends assumption
for periods $t$ to $t+2k$ inherently makes it unlikely for potential
outcome changes $\Delta_{k}Y_{it}(d^{0})$ in the first $k$ periods
to influence treatment changes $\Delta_{k}D_{i,t+k}$ in the subsequent
$k$ periods. It also makes correlations between prior treatment changes
$\Delta_{k}D_{it}$ and subsequent potential outcome changes $\Delta_{k}Y_{i,t+k}(d^{0})$
practically implausible. Without excluding such scenarios, correlations
between $\Delta_{2k}Y_{it}(d^{0})$ and $\Delta_{2k}D_{it}$ would
arise, even in the absence of correlations between $\Delta_{k}Y_{it}(d^{0})$
and $\Delta_{k}D_{it}$ or between $\Delta_{k}Y_{i,t+k}(d^{0})$ and
$\Delta_{k}D_{i,t+k}$. 

Yet a critical distinction of Assumption \ref{assu:PT} from strict
exogeneity is that these restrictions manifest as testable departures
from common trends at specific horizons rather than as a failure of
a monolithic condition that must be maintained or abandoned entirely.
This horizon-specific structure is precisely what the diagnostic
approach in Section \ref{subsec:Diagnosing-Common-Trends} exploits.

\subsection{FD Interpretation}\label{subsec:FD-Interpretation}

The causal interpretation of the $k$-period FD coefficient follows
directly from the TWFE framework, but invokes assumptions only for
the specific difference length $k$: the common trends assumption for
the chosen $k$ alone (Assumption \ref{assu:PT-k}), together with the
following $k$-specific counterparts of Assumptions \ref{assu:Variation}
and \ref{assu:Homogeneity}:

\renewcommand{\theassumptionx}{V--k}
\begin{assumptionx}\label{assu:Variation-k}\textup{(Variation, $k$-period FD)}
	$\sum_{t=1}^{T-k}E\left[Var\left(\Delta_{k}D_{it}|\Delta_{k}W_{it},C_{i}\right)\right]>0$.
\end{assumptionx}

\renewcommand{\theassumptionx}{LH--k}
\begin{assumptionx}\label{assu:Homogeneity-k}\textup{(Linearity and Time Homogeneity, $k$-period FD)}
	For $t\in\{1,\ldots,T-k\}$, the conditional expectation of $\Delta_{k}D_{it}$
	given $\left(\Delta_{k}W_{it},C_{i}\right)$ can be expressed as
	\[
	E\left[\Delta_{k}D_{it}|\Delta_{k}W_{it},C_{i}\right]=\Delta_{k}W_{it}'\delta_{k}^{W}+C_{i}'\delta_{k,t}^{C}+\delta_{k,t}^{I},
	\]
	where the coefficient $\delta_{k}^{W}$ does not vary across $t$.
\end{assumptionx}

These assumptions need only hold for the specific $k$ rather than
across all $k=1,\ldots,T-1$. As noted in Section \ref{subsec:TWFE-Interpretation},
Assumption \ref{assu:PT-k} for smaller $k$ does not imply it holds
for larger $k$.

The population $k$-period FD coefficient $\beta_{\text{FD},k}^{D}$
is obtained from a regression of $\Delta_{k}Y_{it}$ on $\Delta_{k}D_{it}$
controlling for $\Delta_{k}W_{it}$ and $C_{i}$ interacted with period
indicators. Under the stated assumptions, this coefficient has the
following causal interpretation.
\begin{theorem}
	\label{thm:Causal_FD}Under Assumptions \ref{assu:PO}, \ref{assu:PT-k},
	\ref{assu:Variation-k}, and \ref{assu:Homogeneity-k},
	\[
	\beta_{\text{FD},k}^{D}=\sum_{t=1}^{T}E\left[\tau_{it}\omega_{it}^{(k)}\right],
	\]
	where $\tau_{it}$ is the per-unit effect defined in \eqref{eq:tau_it}.
	The weights satisfy $\sum_{t=1}^{T}E\left[\omega_{it}^{(k)}\right]=1$
	and 
	\[
	\omega_{it}^{(k)}\propto(D_{it}-d^{0})\sum_{s=1}^{T}\mathbbm{1}_{|t-s|=k}\left(\widetilde{D}_{it}-\widetilde{D}_{is}-\left(\widetilde{W}_{it}-\widetilde{W}_{is}\right)'\delta_{k}^{W}\right).
	\]
\end{theorem}
Theorem \ref{thm:Causal_FD} shows that the $k$-period FD coefficient
admits a weighted-average interpretation of per-unit treatment effects,
with weights that depend on the unpredictable component of $k$-period
treatment changes.

The FD interpretation inherits several problems from the TWFE framework.
The common trends assumption remains difficult to justify for large
$k$, though researchers can mitigate this concern by choosing smaller
values of $k$. The identification strategy still relies on covariate
changes rather than predetermined covariate levels, creating the same
theoretical concerns about the \emph{ceteris paribus} variation. Covariate impacts
must be homogeneous across time periods $t$, though homogeneity
across difference lengths $k$ is no longer required. Finally, the
possibility of negative weights persists, complicating the interpretation
of the weighted average.

Among these inherited problems, those associated with covariates can
be readily addressed within the FD framework. Rather than controlling
for covariate changes $\Delta_{k}W_{it}$, FD regressions can be modified
to control for the initial covariate levels $W_{it}$ or other predetermined
variables. Moreover, the homogeneity restriction on covariate impacts
across time periods can be relaxed by allowing period-specific coefficients.

\subsection{Discussion: When Should We Use TWFE?}\label{subsec:Discussion}

Theorems \ref{thm:Causal_TWFE} and \ref{thm:Causal_FD} establish an
asymmetry: the FD interpretation invokes assumptions only for the specific
difference length $k$, whereas the TWFE interpretation demands these assumptions
hold simultaneously across all possible difference lengths. This asymmetry
raises a natural question: why aggregate across all available FD estimators
to form a TWFE estimator, when each individual FD estimator may provide
valid causal identification under weaker conditions?

In traditional panel data econometrics, the answer centers on efficiency
gains. Within the textbook fixed effects model, TWFE and all $k$-period
FD estimators identify the same population parameter, and the TWFE
estimator achieves greater efficiency under additional distributional
assumptions about the error term. Indeed, adding the constant effects
assumption (Assumption \ref{assu:CE}) to the assumptions in Section
\ref{subsec:TWFE-Interpretation} yields this equivalence:
\begin{corollary}
	If Assumptions \ref{assu:PO}, \ref{assu:PT}, \ref{assu:Variation},
	\ref{assu:Homogeneity}, and \ref{assu:CE} hold, then $\beta_{\text{FE}}^{D}=\tau=\beta_{\text{FD},k}^{D}$
	for all $k=1,\ldots,T-1$.
\end{corollary}
However, this efficiency justification loses its force when the estimators
identify different parameters. The possibility of heterogeneous treatment
effects alone undermines this foundation, even when all other identifying
assumptions hold. Moreover, potential bias from violated common trends
assumptions at longer horizons can be too large to justify the
gains from reduced standard errors. Long differences may be less reliable
for causal identification if unobserved heterogeneity in trends grows
over time.\footnote{\citet{millimet2025mis} formalize a related mechanism within the
	linear model: when unit-specific heterogeneity drifts over time, longer
	differences introduce greater bias than shorter ones.} In addition, as noted in Section \ref{subsec:TWFE-Interpretation},
the common trends assumption essentially rules out feedback from past
outcome changes to future treatment changes within its time frame.
Over longer horizons, this assumption becomes harder to justify, as
the likelihood of such feedback increases.

One might alternatively justify TWFE aggregation by appealing to interest
in long-run effects, since TWFE regressions exploit long-run differences
as well as short-run differences. Identifying effects over $k$-period
horizons inevitably requires assuming common trends over at least
$k$ periods, making the assumption a necessary compromise on these
grounds. However, even $k$-period FD coefficients have little connection
to $k$-period treatment effects when dynamic effects are present,
as formalized in Appendix \ref{subsec:Dynamic-Effects} under an assumption
of homogeneous treatment effects. The TWFE coefficient would therefore
represent a difficult-to-interpret weighted average of these already
complex effect measures, even under the strong homogeneous effect
assumption.

These considerations suggest that TWFE maintains clear advantages
only under three joint conditions: (1) common trends assumptions hold
for all difference lengths $k$, (2) treatment effects are static
rather than dynamic, and (3) treatment effects are linear and homogeneous
across units and time periods. Substantial differences in FD estimates across horizons suggest that at least one of these conditions is violated. Common trends diagnostics developed in Section \ref{subsec:Diagnosing-Common-Trends} provide evidence on the first condition at specific horizons, while also providing information about the second. The weight diagnostics established in the recent literature (e.g., \citealp{Chaisemartin2020}) are useful for evaluating the third condition, though with important caveats that arise from assuming common trends at only one treatment level, as highlighted by \citet{fabre2022robustness} in binary treatment settings and by Appendix \ref{subsec:Causal-Interpretation-without} in broader settings.

When any of these conditions fails, the case for TWFE becomes weaker. Researchers may then focus on FD regressions with specific $k$, which inherit some limitations of
TWFE but allow more targeted identifying assumptions. Alternatively,
they may employ heterogeneity-robust estimators tailored to their
specific setting (e.g., \citealt{Chaisemartin2020,de2020difference};
\citealt{callaway2020difference,wooldridge2021two,de2022difference,Borusyak2018,callaway2021difference}), though such methods may be more challenging to apply in broader settings
with nonbinary, nonstaggered treatments and time-varying covariates,
and some of them still rely on common trends assumptions across the
entire panel duration.

\section{Empirical Illustration}\label{sec:Empirical-Illustration}

This section illustrates insights from the econometric results by
studying the impact of minimum wages on employment using TWFE regressions.
The TWFE regression analysis based on state-level panel data in the
US has attracted considerable attention for decades. The econometric
framework of this paper, applicable to a broad class of multiperiod
panel settings, is well-suited for this empirical context.
The treatment variable, the state minimum wage, is continuous and
changes many times in a given state. The existing insights from the
TWFE regression literature, typically addressing binary treatments
and staggered adoption designs, do not directly apply to this setting.

To present the estimates, I use a state--year panel of employment
outcomes and minimum wage laws in 50 US states and the District of
Columbia from 1979 to 2019. As a dependent variable in a TWFE regression,
I use the log employment rate of teens (ages 16--19) from the Current
Population Survey (CPS). Another dependent variable is the net job creation rate from the
Business Dynamics Statistics (BDS), using four major sectors with
high shares of workers without college education.\footnote{The four sectors are: Construction, Manufacturing, Retail Trade, and
	Accommodation and Food Services. Using the data from all sectors does
	not qualitatively change the estimates presented below, but the estimated
	effects become smaller in magnitude.} These employment measures, as stock and flow variables, capture different
aspects of labor market adjustment.

The sample restriction and the teen employment outcome follow
\citet{manning2021elusive}, who aggregates CPS data from 1979 to 2019 into
quarterly state-level observations. My analysis is at annual frequency
instead, because the BDS net job creation series is annual. The focus on
age groups or sectors
more likely affected by the minimum wage is motivated by the fact that the majority
of workers in the US are not directly affected by the minimum wage.
The minimum wage data are from \citet{Vaghul2019}, and I use the
log of the annual average of the state minimum wage as the treatment
variable in each regression. I use census region indicators as covariates.

\begin{figure}[t]
\caption{The TWFE and FD Coefficients on Log Minimum Wage}

\label{Fig:TWFE_FD}

\vspace{0.5em}
\centering
\begin{threeparttable}
\vspace{0.25em}

\begin{minipage}[t]{0.5\columnwidth}%
(a) Log Employment Rate of Teens

\includegraphics[width=1\columnwidth]{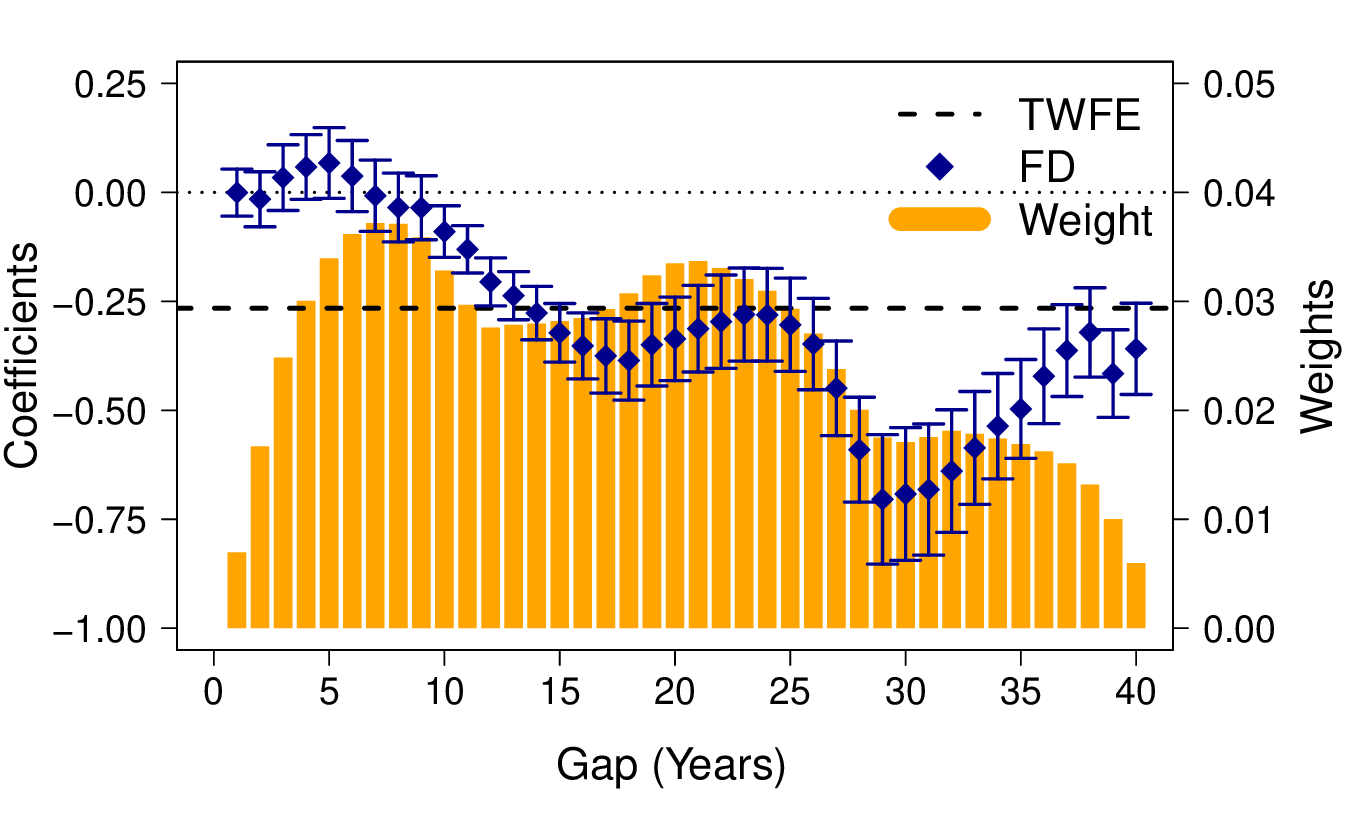}%
\end{minipage}\,\,\,\,\,\,\,\,%
\begin{minipage}[t]{0.5\columnwidth}%
(b) Net Job Creation Rate (\%)

\includegraphics[width=1\columnwidth]{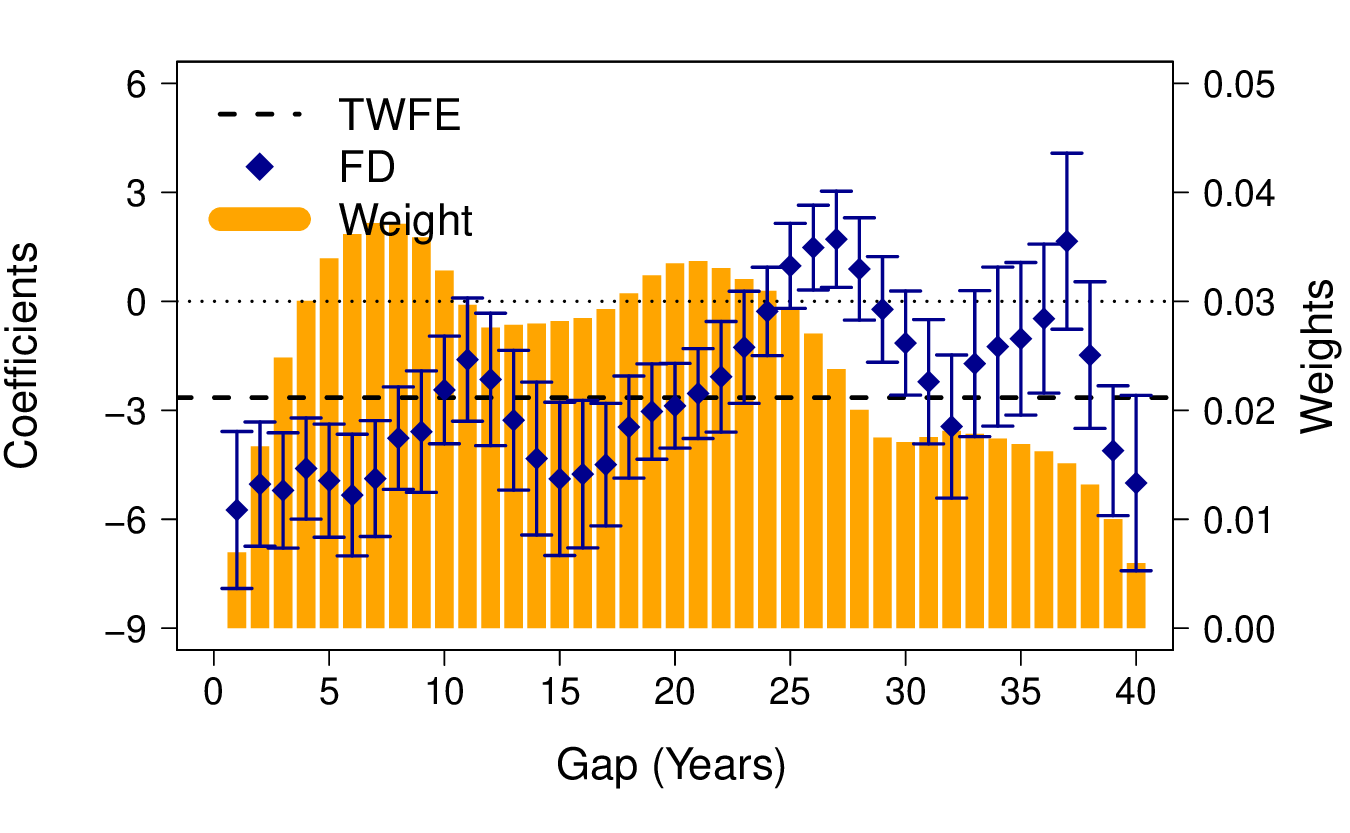}%
\end{minipage}

\vspace{0.5em}
\begin{tablenotes}
\small

\item Notes: The FD coefficients are illustrated with standard error
bars. Covariates are census region indicators. The standard errors
are clustered at the state level.

\end{tablenotes}
\end{threeparttable}
\end{figure}

I begin by illustrating the weighted-average characterization suggested
in Section \ref{subsec:Weighted-Average-Relationship}. Figure \ref{Fig:TWFE_FD}
presents the FD coefficients $\widehat{\beta}_{\text{FD},k}$ for
$k=1,\ldots,40$ as dots accompanied by standard error bars, the TWFE
coefficient $\widehat{\beta}_{\text{FE}}$ as a horizontal dashed
line, and the weights given by Theorem \ref{theorem:FE_not_equal_FD}
as bars. Since all covariates are time-invariant, the adjustment term
in Theorem \ref{theorem:FE_not_equal_FD} can be ignored. While the
inclusion of time-varying covariates would result in a deviation from
the exact weighted-average interpretation, Appendix \ref{subsec:Covariates} finds that the deviation is minimal in this setting even after including
time-varying covariates such as log per-capita income and log teen
population.

In Panel (a) of Figure \ref{Fig:TWFE_FD}, the TWFE coefficient is
$\widehat{\beta}_{\text{FE}}=-0.266$, whereas the FD coefficient
is $\widehat{\beta}_{\text{FD},1}=-0.001$ with a one-year gap and
becomes larger in magnitude as the gap increases. While the difference
among ``short'' FD, ``long'' FD, and TWFE estimates itself has
long been recognized in the minimum wage effects literature (e.g., \citealp{neumark1992employment}),
an explicit numerical relationship among these estimates has not been
previously established. The theoretical results in Section \ref{sec:Numerical-Equivalence}
demonstrate that a TWFE regression aggregates these heterogeneous
FD coefficients into a single coefficient. The weighted average of
all 40 FD coefficients is indeed $-0.266$. Panel (b) of Figure \ref{Fig:TWFE_FD}
presents analogous plots for a regression of the net job creation
rate. Even though the TWFE coefficient is $\widehat{\beta}_{\text{FE}}=-2.65$,
the one-year FD coefficient $\widehat{\beta}_{\text{FD},1}=-5.74$ is
larger in magnitude and some long-run FD coefficients have different
signs. The weighted average of the FD coefficients is again confirmed
to be identical to the TWFE coefficient.

The weights depicted in Figure \ref{Fig:TWFE_FD} are identical across
the two panels. This consistency arises because the weights depend
only on the explanatory variables, not on the dependent variables.
As observed in \eqref{eq:weight_cov}, these weights are associated
with the unexplained variation of treatment changes. Consequently,
the weights exhibit a distinct pattern: they are small for both short-
and long-run FD coefficients, albeit for different reasons. For the
short-run FD coefficients, the weights are small because minimum wage
adjustments tend to be incremental over short periods. By contrast,
for the long-run FD coefficients, the weights are small due to the
limited number of observations of long-run changes.

The substantial differences among FD coefficients documented in Figure
\ref{Fig:TWFE_FD} raise fundamental questions about causal interpretation.
Traditional panel data econometrics justifies TWFE's aggregation of
these estimates on efficiency grounds, but this efficiency justification
is valid only when all FD estimators identify the same population
parameter. The patterns observed here suggest that different FD estimators
likely identify distinct parameters rather than providing noisy estimates of
a common effect. Indeed, the hypothesis that these coefficients are
all identical can be rejected even at a significance level of 0.01\%.
More concerning, some of these estimates may not identify causal parameters
at all if the common trends assumption is violated. With a 41-year
panel spanning 1979--2019, it seems unlikely that this assumption
holds uniformly across all $k=1,\ldots,40$. The procedures developed
in Section \ref{subsec:Diagnosing-Common-Trends} provide a systematic
way to evaluate these concerns across different time horizons.

Table \ref{Table:Reg1} presents regressions based on equations \eqref{eq:lead_reg}
and \eqref{eq:lag_reg}. These diagnostics test whether past and future
treatment changes have zero coefficients when regressing outcome changes
on them, controlling for concurrent treatment changes and covariates.
For practical presentation, I focus on even values of $k\le10$ with
lead/lag length $\ell=k/2$.

\begin{table}[t]
\caption{Diagnosing Common Trends Violation for Minimum Wage Changes}
\label{Table:Reg1}

\centering
\begin{threeparttable}
\setlength\tabcolsep{0.15em}

\begin{tabular}{llcccccccccccccc}
 &  &  &  &  &  &  &  &  &  &  &  &  &  &  & \tabularnewline
\hline 
\multirow{2}{*}{} &  & \multicolumn{2}{c}{$k=2$} &  & \multicolumn{2}{c}{$k=4$} &  & \multicolumn{2}{c}{$k=6$} &  & \multicolumn{2}{c}{$k=8$} &  & \multicolumn{2}{c}{$k=10$}\tabularnewline
\cline{3-4}\cline{6-7}\cline{9-10}\cline{12-13}\cline{15-16}
 &  & Lead & Lag &  & Lead & Lag &  & Lead & Lag &  & Lead & Lag &  & Lead & Lag\tabularnewline
\hline 
\multicolumn{16}{l}{\textbf{Panel A. Log Teen Employment Rate}}\tabularnewline
Future change &  & 0.101 &  &  & 0.081 &  &  & --0.013 &  &  & --0.094 &  &  & --0.092 & \tabularnewline
 &  & (0.061) &  &  & (0.056) &  &  & (0.056) &  &  & (0.061) &  &  & (0.050) & \tabularnewline
Past change &  &  & --0.128 &  &  & 0.017 &  &  & 0.024 &  &  & --0.021 &  &  & --0.121\tabularnewline
 &  &  & (0.070) &  &  & (0.040) &  &  & (0.076) &  &  & (0.085) &  &  & (0.116)\tabularnewline
Concurrent &  & --0.034 & 0.027 &  & 0.025 & --0.008 &  & 0.032 & 0.036 &  & --0.002 & 0.052 &  & 0.009 & 0.013\tabularnewline
change &  & (0.061) & (0.058) &  & (0.067) & (0.066) &  & (0.092) & (0.067) &  & (0.107) & (0.062) &  & (0.118) & (0.054)\tabularnewline
P-value &  & \multicolumn{2}{c}{0.019} &  & \multicolumn{2}{c}{0.288} &  & \multicolumn{2}{c}{0.936} &  & \multicolumn{2}{c}{0.048} &  & \multicolumn{2}{c}{0.002}\tabularnewline
\hline 
\multicolumn{16}{l}{\textbf{Panel B. Net Job Creation Rate (\%)}}\tabularnewline
Future change &  & 1.88 &  &  & 4.24 &  &  & 2.99 &  &  & 2.79 &  &  & 4.18 & \tabularnewline
 &  & (1.23) &  &  & (1.53) &  &  & (1.16) &  &  & (1.28) &  &  & (1.48) & \tabularnewline
Past change &  &  & 0.19 &  &  & --4.24 &  &  & --3.58 &  &  & --1.45 &  &  & --1.32\tabularnewline
 &  &  & (2.60) &  &  & (1.55) &  &  & (1.53) &  &  & (2.12) &  &  & (2.27)\tabularnewline
Concurrent &  & --6.17 & --5.98 &  & --4.48 & --5.10 &  & --5.42 & --5.55 &  & --4.87 & --5.00 &  & --4.13 & --5.16\tabularnewline
change &  & (2.35) & (2.72) &  & (1.78) & (1.64) &  & (1.87) & (1.46) &  & (1.88) & (1.22) &  & (2.16) & (1.31)\tabularnewline
P-value &  & \multicolumn{2}{c}{0.280} &  & \multicolumn{2}{c}{0.004} &  & \multicolumn{2}{c}{0.017} &  & \multicolumn{2}{c}{0.037} &  & \multicolumn{2}{c}{0.005}\tabularnewline
\hline 
\end{tabular}

\vspace{0.2em}
\begin{tablenotes}
\footnotesize

\item Note: Each column reports a regression of outcome changes on
future or past minimum wage changes using equation \eqref{eq:lead_reg}
or \eqref{eq:lag_reg} with $\ell=k/2$, controlling for concurrent
minimum wage changes and census region indicators. Standard errors
are in parentheses and clustered at the state level. P-values are from a Wald test of the joint null
that future and past change coefficients equal zero.

\end{tablenotes}
\end{threeparttable}
\end{table}

For teen employment, the hypothesis that both past and future treatment
change coefficients equal zero is rejected at the 5\% significance
level for $k=2$, 8, and 10. For $k=2$, the future treatment change coefficient
(0.101) and the past treatment change coefficient (--0.128) exceed
the concurrent change coefficient in magnitude, suggesting short-term
feedback mechanisms or dynamic treatment effects. For $k=8$ and
10, both past and future minimum wage changes have negative coefficients,
which raises the possibility of unobserved long-term factors that
simultaneously drive minimum wage policies and teen employment trends
over extended periods.

The job creation rate estimates also exhibit patterns that are hard
to reconcile with the common trends assumption. The hypothesis of
zero coefficients is rejected at the 5\% level for $k=4$, 6, 8,
and 10. Most notably, the future treatment change coefficients are
consistently positive and significant across these horizons, indicating
that increases in net job creation systematically precede minimum
wage increases. This pattern may reflect intertemporal outcome-to-treatment
feedback, where policymakers respond to improving labor market conditions
by raising minimum wages.

Overall, the diagnostics indicate that the common trends assumption
is credible only at short horizons. Thus, in this setting, estimators
relying exclusively on short-run changes may be more appropriate than
the TWFE estimator for identifying causal effects.
Yet, because employment is a stock variable that may adjust only gradually,
an exclusive focus on short-run variation may never recover the economically
relevant long-run effect, as emphasized in the literature (e.g., \citealp{meer2016effects}).
This tension underscores the difficulty of designing empirical strategies
that are both credible and substantively informative, and highlights
the value of diagnostics in making such trade-offs explicit.

\section{Conclusion}\label{sec:Conclusion}
Building on a long-overlooked numerical equivalence between TWFE and
pooled FD regressions, this paper clarifies how TWFE regressions translate
panel data into the coefficient of interest: they reflect associations
among changes in the variables, pooling all possible short- and long-run
comparisons. This property has a direct implication for causal
interpretation. Common trends must hold simultaneously across all time
horizons, restrictions that collectively parallel the familiar strict
exogeneity condition but arrive decomposed by horizon. This structure
opens the requirement to empirical scrutiny: unlike a monolithic exogeneity
condition that must be maintained or abandoned as a whole, common trends
at each horizon can be assessed separately, using the diagnostic procedures
proposed in this paper together with the comparison of FD estimates
across horizons.

This paper focuses on diagnosing problems with TWFE regressions rather
than proposing solutions. While recent literature has developed various
alternative estimators in specific contexts such as binary treatments
or staggered adoption designs, extending robust causal inference methods
to broader settings with general treatment paths and time-varying
covariates remains an important challenge for future research.


\bibliographystyle{econ-econometrica}
\bibliography{DID}

\newpage{}


\appendix
\counterwithin{equation}{section}
\renewcommand{\theequation}{\thesection.\arabic{equation}}
\counterwithin{theorem}{section}
\renewcommand{\thetheorem}{\thesection\arabic{theorem}}
\counterwithin{corollary}{section}
\renewcommand{\thecorollary}{\thesection\arabic{corollary}}
\setcounter{secnumdepth}{2}

\section{Proofs of Results}\label{sec:Proofs}
The proofs use the following algebraic identities.
The sample covariance between any two real sequences $\{a_{t}\}^{T}_{t=1}$ and
$\{b_{t}\}^{T}_{t=1}$ admits the U-statistics representation:
\begin{equation}
\frac{1}{T}\sum^{T}_{t=1}\left(a_{t}-\overline{a}\right)\left(b_{t}-\overline{b}\right)=\frac{1}{T^{2}}\sum^{T-1}_{k=1}\sum^{T-k}_{t=1}\Delta_{k}a_{t}\Delta_{k}b_{t}.\label{eq:U-stat}
\end{equation}
For any real sequence $\{b_{t}\}_{t=1}^{T}$ and any $A(s,t)$ with
$A(s,t)=-A(t,s)$ and $A(t,t)=0$: 
\begin{equation}
\sum_{s>t}(b_{s}-b_{t})A(s,t)=\sum_{t=1}^{T}b_{t}\sum_{s=1}^{T}A(t,s).
\label{eq:sym_id}
\end{equation}

\medskip

\noindent\textbf{Proof of Theorem \ref{thm:FE_FD_DID}:}
Define $e_{it}\equiv Y_{it}-X_{it}'\beta-\alpha_{i}-C_{i}'\gamma_{t}-\mu_{t}$. Using the U-statistics identity \eqref{eq:U-stat}, the TWFE objective \eqref{eq:FE_reg} can be expressed as
\begin{equation}
\frac{1}{T}\sum_{i=1}^{N}\sum_{k=1}^{T-1}\sum_{t=1}^{T-k}\left(\Delta_{k}e_{it}\right)^{2}+T\sum_{i=1}^{N}\overline{e}_{i}^{2}.\label{eq:e_identity}
\end{equation}
The first component is identical to \eqref{eq:PFD_reg} up to a $1/T$ scaling. For any $\left(\beta,\{\gamma_{t},\mu_{t}\}_{t=1}^{T}\right)$,
setting $\alpha_{i}=\overline{Y}_{i}-\overline{X}_{i}'\beta-C_{i}'\overline{\gamma}-\overline{\mu}$ forces $\overline{e}_{i}=0$ for each $i=1,\ldots,N$. Hence the minimizer of the TWFE objective \eqref{eq:FE_reg} also minimizes the pooled FD objective \eqref{eq:PFD_reg}.\hfill$\square$\\

\noindent \textbf{Proof of Theorem \ref{theorem:FE_not_equal_FD}:}
By the Frisch--Waugh--Lovell (FWL) theorem, each $\widehat{\beta}_{\text{FD},k}$ satisfies
\begin{eqnarray}
\left(\sum^{N}_{i=1}\sum^{T-k}_{t=1}\Delta_{k}\widetilde{X}_{it}\Delta_{k}\widetilde{X}_{it}'\right)\widehat{\beta}_{\text{FD},k} & = & \left(\sum^{N}_{i=1}\sum^{T-k}_{t=1}\Delta_{k}\widetilde{X}_{it}\Delta_{k}\widetilde{Y}_{it}\right).\label{eq:FD_FWL}
\end{eqnarray}
Applying \eqref{eq:U-stat} to the normal equation of \eqref{eq:FE_reg} yields
\begin{equation}
\left(\sum^{T-1}_{k=1}\sum^{N}_{i=1}\sum^{T-k}_{t=1}\Delta_{k}\widetilde{X}_{it}\Delta_{k}\widetilde{X}_{it}'\right)\widehat{\beta}_{\text{FE}}=\left(\sum^{T-1}_{k=1}\sum^{N}_{i=1}\sum^{T-k}_{t=1}\Delta_{k}\widetilde{X}_{it}\Delta_{k}\widetilde{Y}_{it}\right).\label{eq:FE_FWL}
\end{equation}
If \eqref{eq:FE_FWL} has a unique solution, then for any $\{\widehat{\beta}_{\text{FD},k}\}_{k=1}^{T-1}$ solving \eqref{eq:FD_FWL},
\begin{equation}
\widehat{\beta}_{\text{FE}}=\sum^{T-1}_{k=1}\widehat{\varOmega}_{k}\widehat{\beta}_{\text{FD},k},\label{eq:FE_FD_multi}
\end{equation}
where $\widehat{\varOmega}_{k}=\left(\sum^{T-1}_{\ell=1}\sum^{N}_{i=1}\sum^{T-\ell}_{t=1}\Delta_{\ell}\widetilde{X}_{it}\Delta_{\ell}\widetilde{X}_{it}'\right)^{-1}\left(\sum^{N}_{i=1}\sum^{T-k}_{t=1}\Delta_{k}\widetilde{X}_{it}\Delta_{k}\widetilde{X}_{it}'\right)$. The same derivation applied to the regression of $D_{it}$ on $W_{it}$ yields
\begin{equation}
\sum^{T-1}_{k=1}\left(\widehat{\delta}^{W}_{\text{FD},k}-\widehat{\delta}^{W}_{\text{FE}}\right)'\left(\sum^{N}_{i=1}\sum^{T-k}_{t=1}\Delta_{k}\widetilde{W}_{it}\Delta_{k}\widetilde{W}_{it}'\right)=0.\label{eq:W_agg}
\end{equation}

Then,
\begin{flalign*}
 & \sum^{T-1}_{k=1}\sum^{N}_{i=1}\sum^{T-k}_{t=1}\Delta_{k}Y_{it}\left(\Delta_{k}\widetilde{D}_{it}-\Delta_{k}\widetilde{W}_{it}'\widehat{\delta}^{W}_{\text{FE}}\right)\\
= & \sum^{T-1}_{k=1}\sum^{N}_{i=1}\sum^{T-k}_{t=1}\left\{ \widehat{\beta}^{D}_{\text{FD},k}\Delta_{k}D_{it}\left(\Delta_{k}\widetilde{D}_{it}-\Delta_{k}\widetilde{W}_{it}'\widehat{\delta}^{W}_{\text{FD},k}\right)+\Delta_{k}Y_{it}\Delta_{k}\widetilde{W}_{it}'\left(\widehat{\delta}^{W}_{\text{FD},k}-\widehat{\delta}^{W}_{\text{FE}}\right)\right\} \\
= & \sum^{T-1}_{k=1}\sum^{N}_{i=1}\sum^{T-k}_{t=1}\left\{ \widehat{\beta}^{D}_{\text{FD},k}\Delta_{k}D_{it}\left(\Delta_{k}\widetilde{D}_{it}-\Delta_{k}\widetilde{W}_{it}'\widehat{\delta}^{W}_{\text{FE}}\right)+\left(\Delta_{k}Y_{it}-\widehat{\beta}^{D}_{\text{FD},k}\Delta_{k}D_{it}\right)\Delta_{k}\widetilde{W}_{it}'\left(\widehat{\delta}^{W}_{\text{FD},k}-\widehat{\delta}^{W}_{\text{FE}}\right)\right\} \\
= & \sum^{T-1}_{k=1}\sum^{N}_{i=1}\sum^{T-k}_{t=1}\left\{ \widehat{\beta}^{D}_{\text{FD},k}\Delta_{k}D_{it}\left(\Delta_{k}\widetilde{D}_{it}-\Delta_{k}\widetilde{W}_{it}'\widehat{\delta}^{W}_{\text{FE}}\right)+\left(\widehat{\delta}^{W}_{\text{FD},k}-\widehat{\delta}^{W}_{\text{FE}}\right)'\left(\Delta_{k}\widetilde{W}_{it}\Delta_{k}\widetilde{W}_{it}'\right)\widehat{\beta}^{W}_{\text{FD},k}\right\} \\
= & \sum^{T-1}_{k=1}\sum^{N}_{i=1}\sum^{T-k}_{t=1}\left\{ \widehat{\beta}^{D}_{\text{FD},k}\Delta_{k}D_{it}\left(\Delta_{k}\widetilde{D}_{it}-\Delta_{k}\widetilde{W}_{it}'\widehat{\delta}^{W}_{\text{FE}}\right)+\left(\widehat{\delta}^{W}_{\text{FD},k}-\widehat{\delta}^{W}_{\text{FE}}\right)'\left(\Delta_{k}\widetilde{W}_{it}\Delta_{k}\widetilde{W}_{it}'\right)\left(\widehat{\beta}^{W}_{\text{FD},k}-\widehat{\beta}^{W}_{\text{FE}}\right)\right\}.
\end{flalign*}
The first and third equalities follow from the FWL theorem; $\widehat{\beta}^{D}_{\text{FD},k}$ is obtained by regressing $\Delta_{k}Y_{it}$ on $\Delta_{k}\widetilde{D}_{it}-\Delta_{k}\widetilde{W}_{it}'\widehat{\delta}^{W}_{\text{FD},k}$, and $\widehat{\beta}^{W}_{\text{FD},k}$ is given by regressing $\Delta_{k}Y_{it}-\widehat{\beta}^{D}_{\text{FD},k}\Delta_{k}D_{it}$ on $\Delta_{k}\widetilde{W}_{it}$. The second equality uses addition and subtraction. The last equality uses \eqref{eq:W_agg}.\hfill$\square$\\

\noindent \textbf{Proof of Theorem \ref{thm:Causal_TWFE}:}  Assumption~\ref{assu:Homogeneity} implies $E\!\left[\Delta_{k}\widetilde{D}_{it}\mid\Delta_{k}W_{it},C_{i}\right]=\Delta_{k}\widetilde{W}_{it}'\delta^{W}$. Then, using the population analogue of \eqref{eq:FE_FWL} and the law of iterated expectations,
\begin{align*}
&\delta^{W}_{\text{FE}}=\left(\sum^{T-1}_{k=1}\sum^{T-k}_{t=1}E\left[\Delta_{k}\widetilde{W}_{it}\Delta_{k}\widetilde{W}_{it}'\right]\right)^{-1}\left(\sum^{T-1}_{k=1}\sum^{T-k}_{t=1}E\left[\Delta_{k}\widetilde{W}_{it}\Delta_{k}\widetilde{D}_{it}\right]\right) \nonumber \\
&=\left(\sum^{T-1}_{k=1}\sum^{T-k}_{t=1}E\left[\Delta_{k}\widetilde{W}_{it}\Delta_{k}\widetilde{W}_{it}'\right]\right)^{-1}\left(\sum^{T-1}_{k=1}\sum^{T-k}_{t=1}E\left[\Delta_{k}\widetilde{W}_{it}E\left[\Delta_{k}\widetilde{D}_{it}|\Delta_{k}W_{it},C_{i}\right]\right]\right)=\delta^{W}.
\end{align*}
Thus,
\begin{equation}
\Delta_{k}\widetilde{D}_{it}-\Delta_{k}\widetilde{W}_{it}'\delta_{\text{FE}}^{W} = \Delta_{k}\widetilde{D}_{it}-E[\Delta_{k}\widetilde{D}_{it}|\Delta_{k}W_{it},C_{i}] = \Delta_{k}D_{it}-E[\Delta_{k}D_{it}|\Delta_{k}W_{it},C_{i}].\label{resid_FE}
\end{equation}

The denominator of \eqref{eq:pop_FE} then equals $\sum_{k=1}^{T-1}\sum_{t=1}^{T-k}E\!\left[\bigl(\Delta_{k}D_{it}-E[\Delta_{k}D_{it}\mid\Delta_{k}W_{it},C_{i}]\bigr)^{2}\right]$, which is strictly positive under Assumption~\ref{assu:Variation}.

Write $Y_{it}=\tau_{it}(D_{it}-d^{0})+Y_{it}(d^{0})$. The contribution of $Y_{it}(d^{0})$ to the numerator of  \eqref{eq:pop_FE}  satisfies
\begin{align*}
&E\!\left[\Delta_{k}Y_{it}(d^{0})(\Delta_{k}\widetilde{D}_{it}-\Delta_{k}\widetilde{W}_{it}'\delta_{\text{FE}}^{W})\right] = E\!\left[E\!\left[\Delta_{k}Y_{it}(d^{0})|\Delta_{k}D_{it},\Delta_{k}W_{it},C_{i}\right](\Delta_{k}\widetilde{D}_{it}-\Delta_{k}\widetilde{W}_{it}'\delta_{\text{FE}}^{W})\right]\\
&=E\!\left[E\!\left[\Delta_{k}Y_{it}(d^{0})|\Delta_{k}W_{it},C_{i}\right]\left(\Delta_{k}D_{it}-E\!\left[\Delta_{k}D_{it}\mid\Delta_{k}W_{it},C_{i}\right]\right)\right]=0,
\end{align*}
where the first and third equalities follow from the law of iterated expectations, and the second equality uses Assumption~\ref{assu:PT} and \eqref{resid_FE}.

By setting $A(s,t)=\widetilde{D}_{is}-\widetilde{D}_{it}-(\widetilde{W}_{is}-\widetilde{W}_{it})'\delta^{W}$ and $b_{t}=\tau_{it}(D_{it}-d^{0})$ or $b_{t}=D_{it}-d^{0}$  in \eqref{eq:sym_id}, the remaining part of the numerator and the denominator in \eqref{eq:pop_FE} can each be rewritten as sums over $t$, yielding
\begin{equation}
\beta_{\mathrm{FE}}^{D}=
\frac{\sum_{t=1}^{T}E\!\left[\tau_{it}(D_{it}-d^{0})\sum_{s=1}^{T}\left(\widetilde{D}_{it}-\widetilde{D}_{is}-\left(\widetilde{W}_{it}-\widetilde{W}_{is}\right)'\delta^{W}\right)\right]}{\sum_{t=1}^{T}E\!\left[(D_{it}-d^{0})\sum_{s=1}^{T}\left(\widetilde{D}_{it}-\widetilde{D}_{is}-\left(\widetilde{W}_{it}-\widetilde{W}_{is}\right)'\delta^{W}\right)\right]}.\label{eq:FE_prf}
\end{equation}
The expression for $\omega_{it}$ follows by comparing the numerator and denominator of \eqref{eq:FE_prf}.\hfill$\square$\\

\noindent \textbf{Proof of Theorem \ref{thm:Causal_FD}:} The population $k$-period FD coefficient is given by
\begin{equation}
\beta_{\text{FD},k}^{D}=\frac{\sum_{t=1}^{T-k}E\left[\Delta_{k}Y_{it}\left(\Delta_{k}\widetilde{D}_{it}-\Delta_{k}\widetilde{W}_{it}'\delta_{\text{FD},k}^{W}\right)\right]}{\sum_{t=1}^{T-k}E\left[\Delta_{k}D_{it}\left(\Delta_{k}\widetilde{D}_{it}-\Delta_{k}\widetilde{W}_{it}'\delta_{\text{FD},k}^{W}\right)\right]},\label{eq:pop_FD}
\end{equation}
where $\delta_{\text{FD},k}^{W}=\left(\sum_{t=1}^{T-k}E\left[\Delta_{k}\widetilde{W}_{it}\Delta_{k}\widetilde{W}_{it}'\right]\right)^{-1}\left(\sum_{t=1}^{T-k}E\left[\Delta_{k}\widetilde{W}_{it}\Delta_{k}\widetilde{D}_{it}\right]\right)$. Assumption~\ref{assu:Homogeneity-k} implies $E\!\left[\Delta_{k}\widetilde{D}_{it}\mid\Delta_{k}W_{it},C_{i}\right]=\Delta_{k}\widetilde{W}_{it}'\delta_{k}^{W}$ and $\delta_{\mathrm{FD},k}^{W}=\delta_{k}^{W}$, which yield
\begin{equation}
\Delta_{k}\widetilde{D}_{it}-\Delta_{k}\widetilde{W}_{it}'\delta_{\text{FD},k}^{W} = \Delta_{k}D_{it}-E[\Delta_{k}D_{it}|\Delta_{k}W_{it},C_{i}].\label{eq:resid_FD}
\end{equation}
The denominator of \eqref{eq:pop_FD} then equals $\sum_{t=1}^{T-k}E\!\left[\bigl(\Delta_{k}D_{it}-E[\Delta_{k}D_{it}\mid\Delta_{k}W_{it},C_{i}]\bigr)^{2}\right]$, which is strictly positive under Assumption~\ref{assu:Variation-k}.

Write $Y_{it}=\tau_{it}(D_{it}-d^{0})+Y_{it}(d^{0})$. The contribution of $Y_{it}(d^{0})$ to the numerator of  \eqref{eq:pop_FD}  satisfies
\begin{align*}
&E\!\left[\Delta_{k}Y_{it}(d^{0})\left(\Delta_{k}\widetilde{D}_{it}-\Delta_{k}\widetilde{W}_{it}'\delta_{\text{FD},k}^{W}\right)\right]
= E\!\left[E\!\left[\Delta_{k}Y_{it}(d^{0})|\Delta_{k}D_{it},\Delta_{k}W_{it},C_{i}\right](\Delta_{k}\widetilde{D}_{it}-\Delta_{k}\widetilde{W}_{it}'\delta_{\text{FD},k}^{W})\right]\\
&=E\!\left[E\!\left[\Delta_{k}Y_{it}(d^{0})\mid\Delta_{k}W_{it},C_{i}\right]\left(\Delta_{k}D_{it}-E\!\left[\Delta_{k}D_{it}\mid\Delta_{k}W_{it},C_{i}\right]\right)\right]=0,
\end{align*}
where the first and third equalities follow from the law of iterated expectations, and the second equality uses Assumption~\ref{assu:PT-k} and \eqref{eq:resid_FD}.

By setting $A(s,t)=\mathbbm{1}_{|s-t|=k}\bigl(\widetilde{D}_{is}-\widetilde{D}_{it}-(\widetilde{W}_{is}-\widetilde{W}_{it})'\delta_{k}^{W}\bigr)$ and $b_{t}=\tau_{it}(D_{it}-d^{0})$ or $b_{t}=D_{it}-d^{0}$ in \eqref{eq:sym_id}, the remaining part of the numerator and the denominator can each be rewritten as sums over $t$, yielding
\begin{equation}
\beta_{\mathrm{FD},k}^{D}=
\frac{\sum_{t=1}^{T}E\!\left[\tau_{it}(D_{it}-d^{0})\sum_{s:\,|s-t|=k}\left(\widetilde{D}_{it}-\widetilde{D}_{is}-\left(\widetilde{W}_{it}-\widetilde{W}_{is}\right)'\delta_{k}^{W}\right)\right]}{\sum_{t=1}^{T}E\!\left[(D_{it}-d^{0})\sum_{s:\,|s-t|=k}\left(\widetilde{D}_{it}-\widetilde{D}_{is}-\left(\widetilde{W}_{it}-\widetilde{W}_{is}\right)'\delta_{k}^{W}\right)\right]}.\label{eq:FD_k_prf}
\end{equation}
The expression for $\omega_{it}^{(k)}$ follows by comparing the numerator and denominator of \eqref{eq:FD_k_prf}.\hfill$\square$\\

\section{Key Extensions}\label{sec:Extension}

This section extends Section \ref{sec:Causal-Interpretation} in three
key directions. Proofs of additional theoretical results are provided
at the end of this section.

\subsection{Generalization of Theorem \ref{thm:Causal_TWFE}}\label{subsec:Generalization}
Theorem \ref{thm:Causal_TWFE} follows from a more general result
that characterizes the TWFE coefficient using a causal effect component
and bias components associated with violations of Assumptions \ref{assu:PT}
and \ref{assu:Homogeneity}. To demonstrate that both linearity and
time homogeneity components of Assumption \ref{assu:Homogeneity}
are essential, I specify the following intermediate assumption. 

\renewcommand{\theassumptionx}{L}
\begin{assumptionx}\label{assu:Linearity}\textup{(Linearity)}
	For any $(k,t)$ with $1\le t<t+k\le T$, the conditional expectation of $\Delta_{k}D_{it}$
	given $\left(\Delta_{k}W_{it},C_{i}\right)$ can be expressed as
	\[
	E\left[\Delta_{k}D_{it}|\Delta_{k}W_{it},C_{i}\right]=\Delta_{k}W_{it}'\delta^{W}_{k,t}+C_{i}'\delta^{C}_{k,t}+\delta^{I}_{k,t}.
	\]
\end{assumptionx}

Assumption \ref{assu:Linearity} requires the conditional mean function
$E\left[\Delta_{k}D_{it}|\Delta_{k}W_{it},C_{i}\right]$ to be linear,
mirroring a conventional linearity assumption for causal interpretation
in cross-sectional regressions (\citealp{angrist1999empirical}).
\begin{theorem}
	\label{lem:Causal_TWFE-g}Suppose that Assumptions \ref{assu:PO}
	and \ref{assu:Variation} hold. Then,
	\[
	\beta^{D}_{\text{FE}}=\sum^{T}_{t=1}E\left[\tau_{it}\omega_{it}\right]+B_{1}+B_{2}+B_{3},
	\]
	where $\tau_{it}$ is defined in \eqref{eq:tau_it} and $\omega_{it}$ in Theorem \ref{thm:Causal_TWFE}.
	The first bias term $B_{1}$ is given by 
	\[
	B_{1}=\frac{\sum^{T-1}_{k=1}\sum^{T-k}_{t=1}E\left[Cov\left(\Delta_{k}Y_{it}(d^{0}),\Delta_{k}D_{it}|\Delta_{k}W_{it},C_{i}\right)\right]}{\sum^{T-1}_{k=1}\sum^{T-k}_{t=1}E\left[\Delta_{k}D_{it}\left(\Delta_{k}\widetilde{D}_{it}-\Delta_{k}\widetilde{W}_{it}'\delta^{W}_{\text{FE}}\right)\right]},
	\]
	which satisfies $B_{1}=0$ under Assumption \ref{assu:PT}. The second
	and the third bias terms are given by
	\begin{eqnarray*}
		B_{2} & = & \frac{\sum^{T-1}_{k=1}\sum^{T-k}_{t=1}E\left[\Delta_{k}Y_{it}(d^{0})\left(E\left[\Delta_{k}\widetilde{D}_{it}|\Delta_{k}W_{it},C_{i}\right]-\Delta_{k}\widetilde{W}_{it}'\overline{\delta}^{W}_{k,t}\right)\right]}{\sum^{T-1}_{k=1}\sum^{T-k}_{t=1}E\left[\Delta_{k}D_{it}\left(\Delta_{k}\widetilde{D}_{it}-\Delta_{k}\widetilde{W}_{it}'\delta^{W}_{\text{FE}}\right)\right]},\\
		B_{3} & = & \frac{\sum^{T-1}_{k=1}\sum^{T-k}_{t=1}E\left[\Delta_{k}Y_{it}(d^{0})\Delta_{k}\widetilde{W}_{it}'\left(\overline{\delta}^{W}_{k,t}-\delta^{W}_{\text{FE}}\right)\right]}{\sum^{T-1}_{k=1}\sum^{T-k}_{t=1}E\left[\Delta_{k}D_{it}\left(\Delta_{k}\widetilde{D}_{it}-\Delta_{k}\widetilde{W}_{it}'\delta^{W}_{\text{FE}}\right)\right]},
	\end{eqnarray*}
	where $\overline{\delta}^{W}_{k,t}$ is the population coefficient
	on $\Delta_{k}W_{it}$ from a regression of $\Delta_{k}D_{it}$ on
	$(\Delta_{k}W_{it},C_{i})$. Under Assumption \ref{assu:Linearity},
	$\overline{\delta}^{W}_{k,t}=\delta^{W}_{k,t}$ and $B_{2}=0$. Under
	Assumption \ref{assu:Homogeneity}, $B_{2}=B_{3}=0$.
\end{theorem}
Theorem \ref{lem:Causal_TWFE-g} demonstrates that the TWFE coefficient
$\beta^{D}_{\text{FE}}$ deviates from the weighted-average causal
interpretation through three channels: (i) the correlation between
potential outcome trends $\Delta_{k}Y_{it}(d^{0})$ and treatment
trends $\Delta_{k}D_{it}$ conditional on covariates; (ii) nonlinearity
of the conditional mean function of treatment trends $\Delta_{k}D_{it}$;
and (iii) variation in the conditional mean function across $(k,t)$.
Assumption \ref{assu:PT} eliminates the first channel, Assumption
\ref{assu:Linearity} eliminates the second, and Assumption \ref{assu:Homogeneity}
eliminates both the second and third channels.

\subsection{Dynamic Effects\label{subsec:Dynamic-Effects}}

The framework in Section \ref{sec:Causal-Interpretation} restricts
potential outcomes to depend only on current treatment status, ruling
out dynamic effects where past treatments influence current outcomes.
This section relaxes this restriction to allow for dynamic treatment
effects, though under a constant effects assumption to maintain tractability.
While this assumption is restrictive, the analysis serves to illustrate
that even under this strong assumption, $k$-period FD coefficients
lack clear interpretation as $k$-period treatment effects, and the
TWFE coefficient is even more difficult to interpret.

\renewcommand{\theassumptionx}{PO--D}
\begin{assumptionx}
\label{assu:PO-D}\textup{(Potential Outcome, Dynamic)} For each
$t=1,\ldots,T$, $\{Y_{it}(d_{t},\ldots,d_{t-L}):(d_{t},\ldots,d_{t-L})\in(\underline{d},\overline{d})^{L+1}\}$
is a stochastic process that defines a potential outcome associated
with each possible treatment level $(d_{t},\ldots,d_{t-L})\in(\underline{d},\overline{d})^{L+1}$,
where $-\infty\le\underline{d}<\overline{d}\le\infty$. The observed
outcome is given by $Y_{it}=Y_{it}(D_{it},\ldots,D_{i,t-L})$.
\end{assumptionx}

\renewcommand{\theassumptionx}{CT--D}
\begin{assumptionx}
\label{assu:PT-D}\textup{(Conditional Common Trends, Dynamic)} There
exists $d^{0}\in(\underline{d},\overline{d})$ such that 
\[
E\left[\Delta_{k}Y_{it}(d^{0},\ldots,d^{0})|\Delta_{k}D_{it},\Delta_{k}W_{it},C_{i}\right]=E\left[\Delta_{k}Y_{it}(d^{0},\ldots,d^{0})|\Delta_{k}W_{it},C_{i}\right]
\]
 for any $(k,t)$ with $1\le t<t+k\le T$.
\end{assumptionx}

\renewcommand{\theassumptionx}{CE--D}
\begin{assumptionx}
\label{assu:CE-D}\textup{(Constant Effects, Dynamic)} The potential
outcome function satisfies $Y_{it}(d_{t}',\ldots,d_{t-L}')-Y_{it}(d_{t},\ldots,d_{t-L})=\sum^{L}_{\ell=0}\gamma_{\ell}(d_{t-\ell}'-d_{t-\ell})$
for any $(d_{t}',\ldots,d_{t-L}'),(d_{t},\ldots,d_{t-L})\in(\underline{d},\overline{d})^{L+1}$.
\end{assumptionx}

Assumption \ref{assu:PO-D} extends the potential outcome framework
to allow outcomes to depend on treatment history up to $L$ periods
in the past, including the treatment status before the beginning of
the panel. This encompasses a broad class of dynamic effects, from
immediate impacts to persistent effects that decay or accumulate over
time. Assumption \ref{assu:PT-D} maintains the common trends requirement
but now applies to potential outcomes under a constant treatment path
at the baseline level $d^{0}$. Assumption \ref{assu:CE-D} requires
that the impact of treatment $\ell$ periods ago be constant across
units, time periods, and treatment histories.

Under these assumptions, the parameters $\{\gamma_{\ell}\}^{L}_{\ell=0}$
capture the dynamic structure of treatment effects. The contemporaneous
effect is $\gamma_{0}$, while $\gamma_{\ell}$ for $\ell>0$ represents
the additional impact on current outcomes from treatment received
$\ell$ periods ago, holding all other treatment history constant.
Partial sums of $\{\gamma_{\ell}\}^{L}_{\ell=0}$ represent cumulative
treatment effects at different horizons. For example, if the treatment
increases by one unit once and for all, its impact on the outcome
$K$ periods later is $\sum^{\min\{K,L\}}_{\ell=0}\gamma_{\ell}$.
\begin{theorem}
\label{thm:Causal_FD-Dyn}Under Assumptions \ref{assu:PO-D}, \ref{assu:PT-D}, \ref{assu:CE-D},
\ref{assu:Variation}, and \ref{assu:Homogeneity},
\[
\beta^{D}_{\text{FD},k}=\gamma_{0}+\sum^{L}_{\ell=1}\frac{\sum^{T-k}_{t=1}E\left[Cov\left(\Delta_{k}D_{it},\Delta_{k}D_{i,t-\ell}|\Delta_{k}W_{it},C_{i}\right)\right]}{\sum^{T-k}_{t=1}E\left[Var\left(\Delta_{k}D_{it}|\Delta_{k}W_{it},C_{i}\right)\right]}\gamma_{\ell},
\]

\[
\beta^{D}_{\text{FE}}=\gamma_{0}+\sum^{L}_{\ell=1}\frac{\sum^{T-1}_{k=1}\sum^{T-k}_{t=1}E\left[Cov\left(\Delta_{k}D_{it},\Delta_{k}D_{i,t-\ell}|\Delta_{k}W_{it},C_{i}\right)\right]}{\sum^{T-1}_{k=1}\sum^{T-k}_{t=1}E\left[Var\left(\Delta_{k}D_{it}|\Delta_{k}W_{it},C_{i}\right)\right]}\gamma_{\ell}.
\]
\end{theorem}
Theorem \ref{thm:Causal_FD-Dyn} characterizes how FD and TWFE coefficients
relate to the underlying dynamic effect parameters. While both $\beta^{D}_{\text{FD},k}$
and $\beta^{D}_{\text{FE}}$ fully capture the contemporaneous effect
parameter $\gamma_{0}$, the extent to which each coefficient captures
dynamic effects $\gamma_{\ell}$ for $\ell>0$ depends on how much
current treatment changes $\Delta_{k}D_{it}$ predict lagged treatment
changes $\Delta_{k}D_{i,t-\ell}$. 

With $\ell=k$, there is no mechanical correlation between $\Delta_{k}D_{it}$
and $\Delta_{k}D_{i,t-k}$ since these represent treatment changes
over nonoverlapping time periods. While these treatment changes can
still be correlated through underlying treatment dynamics, there is
no inherent connection between $k$-period FD coefficients and treatment
effects of $k$-period horizon, contrary to what might be intuitively
expected. The TWFE coefficient becomes even more difficult to interpret,
as it represents a complex weighted average across all difference
lengths $k$ and lag lengths $\ell$, with weights determined by the
full correlation structure of treatment paths across the entire panel.

The presence of dynamic effects affects the diagnostic procedures
developed in Section \ref{subsec:Diagnosing-Common-Trends}. Under
Assumptions \ref{assu:PO-D} and \ref{assu:CE-D}, observed outcome
changes decompose as:
\[
\Delta_{k}Y_{it}=\Delta_{k}Y_{it}(d^{0},\ldots,d^{0})+\sum^{L}_{\ell=0}\gamma_{\ell}\Delta_{k}D_{i,t-\ell}.
\]
The decomposition shows that outcome changes $\Delta_{k}Y_{it}$ depend
on lagged treatment changes $\Delta_{k}D_{i,t-\ell}$ for $\ell>0$.
In lag diagnostics \eqref{eq:lag_reg}, some of these lagged changes directly appear as
regressors. Even in lead diagnostics \eqref{eq:lead_reg}, future treatment changes may
correlate with these lagged changes. Therefore, diagnostic procedures
in Section \ref{subsec:Diagnosing-Common-Trends} flag dynamic effects
in addition to violations of common trends assumptions.

\subsection{No Baseline Treatment Level}\label{subsec:Causal-Interpretation-without}

While a natural baseline treatment level such as ``no treatment''
exists in some applications, in other applications making Assumption
\ref{assu:PT} only for a particular treatment level $d^{0}$ can
be too arbitrary. Moreover, even when a natural baseline exists, there
is typically no clear theoretical justification for why common trends
should hold exclusively at this level while being allowed to fail
at all others, absent dynamic effects. This assumption affects weight
diagnostics commonly used to assess TWFE sensitivity to treatment
effect heterogeneity. The prominence of negative weights in recent
literature depends crucially on assuming common trends at only one
level; \citet{fabre2022robustness} shows that in binary treatment
settings with static effects, assuming common trends for both untreated
and treated potential outcomes allows the TWFE coefficient to be expressed
as a convex combination of treatment effects.

The following analysis addresses these issues by requiring common
trends to hold at all treatment levels.

\renewcommand{\theassumptionx}{CTA}
\begin{assumptionx}
\label{assu:PTE}\textup{(Conditional Common Trends at Any Level of Treatment)}
For any $d\in(\underline{d},\overline{d})$ and any $(k,t)$ with
$1\le t<t+k\le T$, 
\[
E\left[\Delta_{k}Y_{it}(d)|\Delta_{k}D_{it},\Delta_{k}W_{it},C_{i}\right]=E\left[\Delta_{k}Y_{it}(d)|\Delta_{k}W_{it},C_{i}\right].
\]
\end{assumptionx}

\renewcommand{\theassumptionx}{LC}
\begin{assumptionx}
\label{assu:Diff}\textup{(Lipschitz Continuity)} For each $t=1,\ldots,T$,
there exists a random variable $M_{it}$ that satisfies $|Y_{it}(d')-Y_{it}(d)|\le M_{it}|d'-d|$
for any $d',d\in(\underline{d},\overline{d})$, $E\left[M^{2}_{it}\right]<\infty$,
and $E\left[M^{2}_{it}D^{2}_{it}\right]<\infty$.
\end{assumptionx}

Assumption \ref{assu:PTE} extends Assumption \ref{assu:PT} to every
possible treatment level. Even though Assumption \ref{assu:PTE} is
technically stronger than Assumption \ref{assu:PT}, it can be practically
more reasonable, particularly in settings where no natural baseline
treatment level exists. For instance, with minimum wage policy, assuming
common trends only at \$10/hour but not at \$9 or \$11 lacks theoretical
justification. In such cases, Assumption \ref{assu:PTE} avoids the
need to defend why common trends should hold exclusively at a single,
arbitrarily chosen treatment level while being allowed to fail at
all others. An alternative baseline-free approach may assume common
trends for the status-quo outcome $\Delta_{k}Y_{it}(D_{it})$ conditional
on the same initial treatment status $D_{it}$, akin to a strategy
employed by \citet{de2024difference}. This alternative assumption
can support causal interpretation for estimators that control for
initial treatment status, but it cannot for the standard TWFE estimator,
which does not control for initial treatment status.

Assumption \ref{assu:Diff} is a regularity condition which ensures
that a potential outcome process $Y_{it}(d)$ is differentiable almost
everywhere and is sufficiently smooth. Under this differentiability,
the derivative $Y_{it}'(d)$ represents the marginal causal effect
of treatment at level $d$.\footnote{This definition does not preclude discrete treatment settings, since
potential outcomes $Y_{it}(d)$ defined for discrete $d$ can be extended
to the real line without loss of generality. While my exposition uses
derivatives and integrals for continuous treatments, the results apply
to discrete treatments by replacing them with differences and summations.} Given this definition, Assumption \ref{assu:PTE} also results in
conditional common trends for treatment effects $Y_{it}'(d)$, implying
that the treatment effects are additive in unit-specific and period-specific
terms. This contrasts with Assumption \ref{assu:PT}, which applies
only to a baseline treatment level and allows arbitrary treatment
effect heterogeneity.

I maintain Assumptions \ref{assu:PO}, \ref{assu:Variation}, and
\ref{assu:Homogeneity}, which are also used in Section \ref{subsec:TWFE-Interpretation}.
It turns out that these assumptions are sufficient for providing a
weighted-average interpretation of a TWFE coefficient, but not sufficient
for an unambiguous weighted-average interpretation.
\begin{theorem}
\label{thm:Causal_TWFE_d}Suppose Assumptions \ref{assu:PO}, \ref{assu:Variation},
\ref{assu:Homogeneity}, \ref{assu:PTE}, and \ref{assu:Diff} hold.
Then, for any $d^{0}\in(\underline{d},\overline{d})$ and any $g:(\underline{d},\overline{d})\to\mathbb{R}$
with $\int^{\overline{d}}_{\underline{d}}|g(x)|dx<\infty$, 
\[
\beta^{D}_{\text{FE}}=\int^{\overline{d}}_{\underline{d}}\sum^{T}_{t=1}E\left[Y_{it}'(x)\left(\psi_{it}(x;d^{0})+g(x)R_{it}\right)\right]dx,
\]
where $R_{it}=\left(\widetilde{D}_{it}-\overline{\widetilde{D}}_{i}\right)-\left(\widetilde{W}_{it}-\overline{\widetilde{W}}_{i}\right)'\delta^{W}_{\text{FE}}$
is a TWFE residual and the weight function
\[
\psi_{it}(d;d^{0})\equiv\frac{\left(\ind_{D_{it}\ge d}-\ind_{d^{0}\ge d}\right)R_{it}}{\sum^{T}_{s=1}E\left[D_{is}R_{is}\right]}
\]
satisfies $\int^{\overline{d}}_{\underline{d}}\sum^{T}_{t=1}E\left[\psi_{it}(x;d^{0})\right]dx=1$.
\end{theorem}
Theorem \ref{thm:Causal_TWFE_d} expresses the population TWFE coefficient
$\beta^{D}_{\text{FE}}$ as a weighted average of treatment effects
$Y_{it}'(d)$, where weights sum to one but may be negative. However,
there are infinitely many weight functions that can yield this weighted-average
expression. This multiplicity stems from two sources: the choice of
baseline treatment $d^{0}$, and more broadly, the possibility to
add any function proportional to the TWFE residual $R_{it}$. This
issue arises because Assumption \ref{assu:PTE} implies that treatment
effects $Y_{it}'(d)$ are additive in unit and period effects. Due
to this assumption, $Y_{it}'(d)$ is orthogonal to the TWFE residual
$R_{it}$, that is, $\sum^{T}_{t=1}E\left[Y_{it}'(d)R_{it}\right]=0$
for any $d\in(\underline{d},\overline{d})$. As a result, any function
proportional to $R_{it}$, such as $\ind_{d^{0}\ge d}R_{it}$, can
be added to or subtracted from the weight function without affecting
the weighted average.

This property poses challenges beyond ambiguity in the weighted-average
interpretation: it can also exaggerate the sensitivity of the TWFE
coefficient to treatment effect heterogeneity. This sensitivity is
often assessed through the variance of weights or the proportion of
negative weights (e.g., \citealp{Chaisemartin2020}). While these
measures remain valid when common trends hold only for a baseline
treatment level, they inflate under Assumption \ref{assu:PTE} due
to extraneous variation unrelated to treatment effects, making the
TWFE coefficient appear more sensitive to heterogeneity than it may
actually be.

To address this issue, I introduce additional conditions that characterize
the exogeneity requirements more explicitly than Assumption \ref{assu:PTE}.
These conditions help eliminate the weight variation orthogonal to
$Y_{it}'(d)$, yielding a more interpretable weighted-average representation.

\renewcommand{\theassumptionx}{SE}
\begin{assumptionx}
\label{assu:SE}\textup{(Strict Exogeneity)} For any $d\in(\underline{d},\overline{d})$
and $(k,t)$ with $1\le t<t+k\le T$, 
\[
E\left[\Delta_{k}Y_{it}(d)|D_{i,t+k},D_{it},\overline{D}_{i},W_{i,t+k},W_{it},\overline{W}_{i},C_{i}\right]=E\left[\Delta_{k}Y_{it}(d)|W_{i,t+k},W_{it},\overline{W}_{i},C_{i}\right].
\]
\end{assumptionx}

\renewcommand{\theassumptionx}{S}
\begin{assumptionx}
\label{assu:Suff}\textup{(Sufficiency)} For any $d\in(\underline{d},\overline{d})$
and $(k,t)$ with $1\le t<t+k\le T$, 
\[
E\left[\Delta_{k}Y_{it}(d)|W_{i,t+k},W_{it},\overline{W}_{i},C_{i}\right]=E\left[\Delta_{k}Y_{it}(d)|\Delta_{k}W_{it},C_{i}\right].
\]
\end{assumptionx}

Assumption \ref{assu:SE} is a nonparametric and weaker version of
a strict exogeneity condition that is common in panel data models.
Unlike traditional strict exogeneity, which typically requires the
entire treatment path $(D_{i1},\ldots,D_{iT})$ to be exogenous---as
seen in condition (\ref{eq:CD_cond}) used by \citet{Chaisemartin2020}---Assumption
\ref{assu:SE} requires exogeneity of only the pair $(D_{i,t+k},D_{it})$
and the time-series average $\overline{D}_{i}$. Similarly, this
assumption conditions only on $\left(W_{i,t+k},W_{it},\overline{W}_{i}\right)$
for time-varying covariates, avoiding reliance on the full covariate
history. Nevertheless, the time-series treatment average $\overline{D}_{i}$
depends on past and future treatment statuses beyond the specified
time frame $[t,t+k]$. As noted in Section \ref{subsec:TWFE-Interpretation},
Assumption \ref{assu:PT} (and \ref{assu:PTE}) collectively impose
an exogeneity requirement that rules out the possibility that potential
outcome changes $\Delta_{k}Y_{it}(d)$ influence future treatment
assignments or correlate with past treatment assignments, mitigating
concerns about this constraint. Including $\overline{D}_{i}$ and
$\overline{W}_{i}$ in the time-invariant covariates $C_{i}$, though
uncommon in practice, would further relax these requirements.

Assumption \ref{assu:Suff} ensures that concurrent covariate changes
$\Delta_{k}W_{it}$ are sufficient for predicting potential outcome
changes $\Delta_{k}Y_{it}(d)$. This assumption is necessary because
Assumption \ref{assu:SE} alone does not imply Assumption \ref{assu:PTE}.
While Assumption \ref{assu:SE} allows $\Delta_{k}Y_{it}(d)$ to depend
on $(W_{i,t+k},W_{it},\overline{W}_{i})$, a TWFE regression predicts
outcome changes using only $\Delta_{k}W_{it}$. Assumption \ref{assu:Suff}
eliminates potential omitted variable bias from this discrepancy.
Together, Assumptions \ref{assu:SE} and \ref{assu:Suff} imply Assumption
\ref{assu:PTE}.
\begin{theorem}
\label{thm:Causal_TWFE_d2}Under Assumptions \ref{assu:PO}, \ref{assu:Variation},
\ref{assu:Homogeneity}, \ref{assu:Diff}, \ref{assu:SE}, and \ref{assu:Suff},
\[
\beta^{D}_{\text{FE}}=\int^{\overline{d}}_{\underline{d}}\sum^{T}_{t=1}E\left[Y_{it}'(x)\psi^{*}_{it}(x)\right]dx,
\]
where the weight function is given by:
\begin{equation}
\psi^{*}_{it}(d)\equiv\frac{\frac{1}{T}\sum^{T}_{s=1}\ind_{D_{is}\ge d}R_{is}+\frac{1}{T}\sum^{T}_{s=1}E\left[\ind_{D_{it}\ge d}R_{it}-\ind_{D_{is}\ge d}R_{is}|W_{it}-W_{is},C_{i}\right]}{\sum^{T}_{s=1}E\left[D_{is}R_{is}\right]}\label{eq:psi_star_def}
\end{equation}
and satisfies $\int^{\overline{d}}_{\underline{d}}\sum^{T}_{t=1}E\left[\psi^{*}_{it}(x)\right]dx=1$.
In addition,
\begin{equation}
\sum^{T}_{t=1}E\left[\psi^{*}_{it}(d)^{2}\right]\le\sum^{T}_{t=1}E\left[\left(\psi^{*}_{it}(d)+g(d)R_{it}\right)^{2}\right]\thinspace\thinspace\thinspace\text{for all } d\in(\underline{d},\overline{d})\label{eq:psi_optimal}
\end{equation}
holds for any $g:(\underline{d},\overline{d})\to\mathbb{R}$.
$\psi^{*}_{it}(d)$ can also be expressed as
\begin{equation}
\psi^{*}_{it}(d)=\frac{1}{T}\sum^{T}_{s=1}\psi_{is}(d;d^{0})+\frac{1}{T}\sum^{T}_{s=1}E\left[\psi_{it}(d;d^{0})-\psi_{is}(d;d^{0})|W_{it}-W_{is},C_{i}\right],\label{eq:psi_star_alt}
\end{equation}
where $\psi_{it}(d;d^{0})$ is the weight function defined in Theorem
\ref{thm:Causal_TWFE_d} for any $d^{0}\in(\underline{d},\overline{d})$.
\end{theorem}
Theorem \ref{thm:Causal_TWFE_d2} expresses the coefficient $\beta^{D}_{\text{FE}}$
as a weighted average of treatment effects $Y_{it}'(d)$, with weights
summing to one but still possibly negative, much like Theorem \ref{thm:Causal_TWFE_d}.
However, its weight function $\psi^{*}_{it}(d)$ does not depend on
an arbitrary baseline level $d^{0}$. While any function proportional
to $R_{it}$ can still be added to $\psi^{*}_{it}(d)$ without affecting
the weighted average, doing so always increases the weight's square
norm as shown in (\ref{eq:psi_optimal}).

As observed in (\ref{eq:psi_star_def}), the weight function $\psi^{*}_{it}(d)$
is additive in two components, where the first component is constant
within the same unit $i$ and the second component is constant within
the same period $t$ conditional on covariates. Moreover, as expressed
in (\ref{eq:psi_star_alt}), $\psi^{*}_{it}(d)$ can be obtained by
collapsing any possible weight function $\psi_{it}(d;d^{0})$ into
the unit-specific and the period-specific terms in exactly the same
manner.

Considering the simplest case with no covariates is helpful for offering
an intuition for the weight function. In this case, the weight function
is given by
\[
\psi^{*}_{it}(d)=\frac{\frac{1}{T}\sum^{T}_{s=1}\left(\ind_{D_{is}\ge d}R_{is}-E\left[\ind_{D_{is}\ge d}R_{is}\right]\right)+E\left[\ind_{D_{it}\ge d}R_{it}\right]}{\sum^{T}_{s=1}E\left[D_{is}R_{is}\right]},
\]
which is a predicted value from a regression of $\psi_{it}(d;d^{0})$
on additive unit and period fixed effects. Given that a common trends
assumption relies on treatment effects $Y_{it}'(d)$ being additive
in unit and period effects, any weight variation orthogonal to unit
and period effects is orthogonal to $Y_{it}'(d)$. Projecting the
weight function onto unit and period effects eliminates this redundant
weight variation.\footnote{See \citet{ishimaru2021empirical} for a discussion
in a related context, in which an instrumental variables regression
employs a DID-type identification strategy.}

While the set of assumptions in Theorem \ref{thm:Causal_TWFE} also
holds in Theorem \ref{thm:Causal_TWFE_d2}, using the weight function
$\omega_{it}$ in Theorem \ref{thm:Causal_TWFE} can be misleading
for defining the weight on each observation. In particular, the total
weight on an observation ($i,t$) aggregated over $d\in(\underline{d},\overline{d})$
in Theorem \ref{thm:Causal_TWFE_d2} is given by:
\begin{multline}
\omega^{*}_{it}\equiv\int^{\overline{d}}_{\underline{d}}\psi^{*}_{it}(x)dx=\frac{\frac{1}{T}\sum^{T}_{s=1}D_{is}R_{is}+\frac{1}{T}\sum^{T}_{s=1}E\left[D_{it}R_{it}-D_{is}R_{is}|W_{it}-W_{is},C_{i}\right]}{\sum^{T}_{s=1}E\left[D_{is}R_{is}\right]}\\
=\frac{1}{T}\sum^{T}_{s=1}\omega_{is}+\frac{1}{T}\sum^{T}_{s=1}E\left[\omega_{it}-\omega_{is}|W_{it}-W_{is},C_{i}\right].\label{eq:weight_agg}
\end{multline}
This weight projection eliminates redundant variation in $\omega_{it}$
exactly in the same manner as in (\ref{eq:psi_star_alt}). This projection
holds for any baseline level $d^{0}$ chosen for defining $\omega_{it}$.
For example, in a binary treatment setting under Assumption \ref{assu:PTE},
it is well known that selecting $d^{0}=0$ gives treated observations
nonzero (potentially negative) weights, while $d^{0}=1$ gives untreated
observations such weights. Applying the projection in equation (\ref{eq:weight_agg})
to either weighting scheme yields $\omega^{*}_{it}$.

\subsection{Proofs}
\subsubsection{Proof of Theorem \ref{lem:Causal_TWFE-g}}
The denominator in (\ref{eq:pop_FE}) is given by
\begin{multline*}
	\sum^{T-1}_{k=1}\sum^{T-k}_{t=1}E\left[\Delta_{k}D_{it}\left(\Delta_{k}\widetilde{D}_{it}-\Delta_{k}\widetilde{W}_{it}'\delta^{W}_{\text{FE}}\right)\right]\\
	=\sum^{T-1}_{k=1}\sum^{T-k}_{t=1}E\left[\left(\Delta_{k}D_{it}-E\left[\Delta_{k}D_{it}|\Delta_{k}W_{it},C_{i}\right]\right)^{2}+\left(E\left[\Delta_{k}D_{it}|\Delta_{k}W_{it},C_{i}\right]-\Delta_{k}\widetilde{W}_{it}'\delta^{W}_{\text{FE}}\right)^{2}\right],
\end{multline*}
which is strictly positive under Assumption \ref{assu:Variation}.

Using identity (\ref{eq:sym_id}), the denominator in (\ref{eq:pop_FE})
can be alternatively expressed as
\begin{multline}
	\sum^{T-1}_{k=1}\sum^{T-k}_{t=1}E\left[\Delta_{k}D_{it}\left(\Delta_{k}\widetilde{D}_{it}-\Delta_{k}\widetilde{W}_{it}'\delta^{W}_{\text{FE}}\right)\right]\\
	=\sum_{s>t}E\left[\left((D_{is}-d^{0})-(D_{it}-d^{0})\right)\left(\widetilde{D}_{is}-\widetilde{D}_{it}-\left(\widetilde{W}_{is}-\widetilde{W}_{it}\right)'\delta^{W}_{\text{FE}}\right)\right]\\
	=\sum^{T}_{t=1}E\left[(D_{it}-d^{0})\sum^{T}_{s=1}\left(\widetilde{D}_{it}-\widetilde{D}_{is}-\left(\widetilde{W}_{it}-\widetilde{W}_{is}\right)'\delta^{W}_{\text{FE}}\right)\right].\label{eq:denom-ab}
\end{multline}

Using the definition of $\tau_{it}$, decompose the observed outcome
as $Y_{it}=\tau_{it}(D_{it}-d^{0})+Y_{it}(d^{0})$ and substitute
this into the numerator in (\ref{eq:pop_FE}). The treatment effect
component yields
\begin{multline}
	\sum_{s>t}E\left[\left(\tau_{is}(D_{is}-d^{0})-\tau_{it}(D_{it}-d^{0})\right)\left(\widetilde{D}_{is}-\widetilde{D}_{it}-\left(\widetilde{W}_{is}-\widetilde{W}_{it}\right)'\delta^{W}_{\text{FE}}\right)\right]\\
	=\sum^{T}_{t=1}E\left[\tau_{it}(D_{it}-d^{0})\sum^{T}_{s=1}\left(\widetilde{D}_{it}-\widetilde{D}_{is}-\left(\widetilde{W}_{it}-\widetilde{W}_{is}\right)'\delta^{W}_{\text{FE}}\right)\right].\label{eq:numer-a}
\end{multline}
The baseline outcome component yields
\begin{multline}
	E\left[\Delta_{k}Y_{it}(d^{0})\left(\Delta_{k}\widetilde{D}_{it}-\Delta_{k}\widetilde{W}_{it}'\delta^{W}_{\text{FE}}\right)\right]\\
	=E\left[\left(\Delta_{k}Y_{it}(d^{0})-E\left[\Delta_{k}Y_{it}(d^{0})|\Delta_{k}W_{it},C_{i}\right]\right)\left(\Delta_{k}\widetilde{D}_{it}-\Delta_{k}\widetilde{W}_{it}'\delta^{W}_{\text{FE}}\right)\right]\\
	+E\left[E\left[\Delta_{k}Y_{it}(d^{0})|\Delta_{k}W_{it},C_{i}\right]\left(\Delta_{k}\widetilde{D}_{it}-\Delta_{k}\widetilde{W}_{it}'\delta^{W}_{\text{FE}}\right)\right]\\
	=E\left[Cov\left(\Delta_{k}Y_{it}(d^{0}),\Delta_{k}D_{it}|\Delta_{k}W_{it},C_{i}\right)\right]+E\left[\Delta_{k}Y_{it}(d^{0})\left(E\left[\Delta_{k}\widetilde{D}_{it}|\Delta_{k}W_{it},C_{i}\right]-\Delta_{k}\widetilde{W}_{it}'\overline{\delta}^{W}_{k,t}\right)\right]\\
	+E\left[\Delta_{k}Y_{it}(d^{0})\Delta_{k}\widetilde{W}_{it}'\left(\overline{\delta}^{W}_{k,t}-\delta^{W}_{\text{FE}}\right)\right].\label{eq:numer-b}
\end{multline}

Combining (\ref{eq:denom-ab}), (\ref{eq:numer-a}) and (\ref{eq:numer-b})
yields $\beta^{D}_{\text{FE}}=\sum^{T}_{t=1}E\left[\omega_{it}\tau_{it}\right]+B_{1}+B_{2}+B_{3},$
where the weights satisfy $\sum^{T}_{t=1}E[\omega_{it}]=1$ and 
\begin{eqnarray*}
	\omega_{it} & \propto & (D_{it}-d^{0})\sum^{T}_{s=1}\left(\widetilde{D}_{it}-\widetilde{D}_{is}-\left(\widetilde{W}_{it}-\widetilde{W}_{is}\right)'\delta^{W}_{\text{FE}}\right)\\
	& \propto & (D_{it}-d^{0})\left(\widetilde{D}_{it}-\overline{\widetilde{D}}_{i}-\left(\widetilde{W}_{it}-\overline{\widetilde{W}}_{i}\right)'\delta^{W}_{\text{FE}}\right).
\end{eqnarray*}

Assumption \ref{assu:PT} implies $Cov\left(\Delta_{k}Y_{it}(d^{0}),\Delta_{k}D_{it}|\Delta_{k}W_{it},C_{i}\right)=0$,
which in turn implies $B_{1}=0$. Assumption \ref{assu:Linearity}
implies $\overline{\delta}^{W}_{k,t}=\delta^{W}_{k,t}$ and
\[
E\left[\Delta_{k}\widetilde{D}_{it}|\Delta_{k}W_{it},C_{i}\right]=\Delta_{k}\widetilde{W}_{it}'\delta^{W}_{k,t},
\]
which in turn imply $B_{2}=0$. As shown in the proof of Theorem \ref{thm:Causal_TWFE},
Assumption \ref{assu:Homogeneity} implies $\delta^{W}_{\text{FE}}=\delta^{W}$
and
\[
E\left[\Delta_{k}\widetilde{D}_{it}|\Delta_{k}W_{it},C_{i}\right]=\Delta_{k}\widetilde{W}_{it}'\delta^{W},
\]
which in turn imply $B_{2}=B_{3}=0$.\hfill$\square$\\

\subsubsection{Proof of Theorem \ref{thm:Causal_FD-Dyn}}\label{subsec:Proof_Causal_FD-Dyn}
Under Assumptions \ref{assu:PO-D} and \ref{assu:CE-D}, observed
outcome changes decompose as:
\begin{equation}
\Delta_{k}Y_{it}=\Delta_{k}Y_{it}(d^{0},\ldots,d^{0})+\sum^{L}_{\ell=0}\gamma_{\ell}\Delta_{k}D_{i,t-\ell}.\label{eq:decom-D}
\end{equation}
Then, the numerator in (\ref{eq:pop_FE}) is given by
\begin{multline*}
\sum^{T-1}_{k=1}\sum^{T-k}_{t=1}E\left[\Delta_{k}Y_{it}\left(\Delta_{k}\widetilde{D}_{it}-\Delta_{k}\widetilde{W}_{it}'\delta^{W}_{\text{FE}}\right)\right]\\
=\sum^{T-1}_{k=1}\sum^{T-k}_{t=1}E\left[\Delta_{k}Y_{it}(d^{0},\ldots,d^{0})\left(\Delta_{k}\widetilde{D}_{it}-\Delta_{k}\widetilde{W}_{it}'\delta^{W}_{\text{FE}}\right)\right]\\
+\sum^{L}_{\ell=0}\gamma_{\ell}\sum^{T-1}_{k=1}\sum^{T-k}_{t=1}E\left[\Delta_{k}D_{i,t-\ell}\left(\Delta_{k}\widetilde{D}_{it}-\Delta_{k}\widetilde{W}_{it}'\delta^{W}_{\text{FE}}\right)\right]\\
=\sum^{T-1}_{k=1}\sum^{T-k}_{t=1}E\left[\Delta_{k}Y_{it}(d^{0},\ldots,d^{0})\left(\Delta_{k}D_{it}-E[\Delta_{k}D_{it}|\Delta_{k}W_{it},C_{i}]\right)\right]\\
+\sum^{L}_{\ell=0}\gamma_{\ell}\sum^{T-1}_{k=1}\sum^{T-k}_{t=1}E\left[\Delta_{k}D_{i,t-\ell}\left(\Delta_{k}D_{it}-E[\Delta_{k}D_{it}|\Delta_{k}W_{it},C_{i}]\right)\right]\\
=\sum^{L}_{\ell=0}\gamma_{\ell}\sum^{T-1}_{k=1}\sum^{T-k}_{t=1}E\left[Cov\left(\Delta_{k}D_{i,t-\ell},\Delta_{k}D_{it}|\Delta_{k}W_{it},C_{i}\right)\right],
\end{multline*}
where the first equality uses \eqref{eq:decom-D}, the second equality uses Assumption \ref{assu:Homogeneity},
and the third equality uses Assumption \ref{assu:PT-D}. In addition,
the denominator in (\ref{eq:pop_FE}) is given by
\begin{multline*}
\sum^{T-1}_{k=1}\sum^{T-k}_{t=1}E\left[\Delta_{k}D_{it}\left(\Delta_{k}\widetilde{D}_{it}-\Delta_{k}\widetilde{W}_{it}'\delta^{W}_{\text{FE}}\right)\right]=\sum^{T-1}_{k=1}\sum^{T-k}_{t=1}E\left[\Delta_{k}D_{it}\left(\Delta_{k}D_{it}-E[\Delta_{k}D_{it}|\Delta_{k}W_{it},C_{i}]\right)\right]\\
=\sum^{T-1}_{k=1}\sum^{T-k}_{t=1}E\left[Var\left(\Delta_{k}D_{it}|\Delta_{k}W_{it},C_{i}\right)\right],
\end{multline*}
where the first equality uses Assumption \ref{assu:Homogeneity}.

The numerator in (\ref{eq:pop_FD}) is given by
\begin{multline*}
\sum^{T-k}_{t=1}E\left[\Delta_{k}Y_{it}\left(\Delta_{k}\widetilde{D}_{it}-\Delta_{k}\widetilde{W}_{it}'\delta^{W}_{\text{FD},k}\right)\right]\\
=\sum^{T-k}_{t=1}E\left[\Delta_{k}Y_{it}(d^{0},\ldots,d^{0})\left(\Delta_{k}\widetilde{D}_{it}-\Delta_{k}\widetilde{W}_{it}'\delta^{W}_{\text{FD},k}\right)\right]+\sum^{L}_{\ell=0}\gamma_{\ell}\sum^{T-k}_{t=1}E\left[\Delta_{k}D_{i,t-\ell}\left(\Delta_{k}\widetilde{D}_{it}-\Delta_{k}\widetilde{W}_{it}'\delta^{W}_{\text{FD},k}\right)\right]\\
=\sum^{T-k}_{t=1}E\left[\Delta_{k}Y_{it}(d^{0},\ldots,d^{0})\left(\Delta_{k}D_{it}-E[\Delta_{k}D_{it}|\Delta_{k}W_{it},C_{i}]\right)\right]\\
+\sum^{L}_{\ell=0}\gamma_{\ell}\sum^{T-k}_{t=1}E\left[\Delta_{k}D_{i,t-\ell}\left(\Delta_{k}D_{it}-E[\Delta_{k}D_{it}|\Delta_{k}W_{it},C_{i}]\right)\right]\\
=\sum^{L}_{\ell=0}\gamma_{\ell}\sum^{T-k}_{t=1}E\left[Cov\left(\Delta_{k}D_{i,t-\ell},\Delta_{k}D_{it}|\Delta_{k}W_{it},C_{i}\right)\right],
\end{multline*}
where the second equality uses Assumption \ref{assu:Homogeneity}
and the third equality uses Assumption \ref{assu:PT-D}. In addition,
the denominator in (\ref{eq:pop_FD}) is given by
\begin{multline*}
\sum^{T-k}_{t=1}E\left[\Delta_{k}D_{it}\left(\Delta_{k}\widetilde{D}_{it}-\Delta_{k}\widetilde{W}_{it}'\delta^{W}_{\text{FD},k}\right)\right]=\sum^{T-k}_{t=1}E\left[\Delta_{k}D_{it}\left(\Delta_{k}D_{it}-E[\Delta_{k}D_{it}|\Delta_{k}W_{it},C_{i}]\right)\right]\\
=\sum^{T-k}_{t=1}E\left[Var\left(\Delta_{k}D_{it}|\Delta_{k}W_{it},C_{i}\right)\right],
\end{multline*}
where the first equality uses Assumption \ref{assu:Homogeneity}.\hfill$\square$\\

\subsubsection{Proof of Theorem \ref{thm:Causal_TWFE_d}\label{subsec:Proof_Causal_TWFE_d}}

Given the definition of $R_{it}$ and Assumption \ref{assu:Homogeneity},
\begin{align}
\Delta_{k}R_{it} & =\Delta_{k}\widetilde{D}_{it}-\Delta_{k}\widetilde{W}_{it}'\delta^{W}_{\text{FE}}\nonumber \\
 & =\Delta_{k}D_{it}-E\left[\Delta_{k}D_{it}|\Delta_{k}W_{it},C_{i}\right].\label{eq:def_dRH}
\end{align}

For any $d\in(\underline{d},\overline{d})$,
\begin{eqnarray}
\sum^{T}_{t=1}E\left[Y_{it}(d)R_{it}\right] & = & \frac{1}{T}\sum^{T-1}_{k=1}\sum^{T-k}_{t=1}E\left[\Delta_{k}Y_{it}(d)\Delta_{k}R_{it}\right]\nonumber \\
 & = & \frac{1}{T}\sum^{T-1}_{k=1}\sum^{T-k}_{t=1}E\left[\Delta_{k}Y_{it}(d)\left(\Delta_{k}D_{it}-E\left[\Delta_{k}D_{it}|\Delta_{k}W_{it},C_{i}\right]\right)\right]\nonumber \\
 & = & 0.\label{eq:orth}
\end{eqnarray}
The first equality uses \eqref{eq:U-stat} and $\overline{R}_{i}=0$.
The second equality uses \eqref{eq:def_dRH}. The third equality follows
from Assumption \ref{assu:PTE}. Under Assumption \ref{assu:Diff},
(\ref{eq:orth}) also implies
\begin{equation}
\sum^{T}_{t=1}E\left[Y_{it}'(d)R_{it}\right]=0\label{eq:orth_d}
\end{equation}
for any $d\in(\underline{d},\overline{d})$.

The numerator of \eqref{eq:pop_FE} is given by
\begin{eqnarray}
\sum^{T-1}_{k=1}\sum^{T-k}_{t=1}E\left[\Delta_{k}Y_{it}\Delta_{k}R_{it}\right] & = & T\sum^{T}_{t=1}E\left[Y_{it}R_{it}\right]\nonumber \\
 & = & T\sum^{T}_{t=1}E\left[\left(Y_{it}(D_{it})-Y_{it}(d^{0})\right)R_{it}\right]\nonumber \\
 & = & T\sum^{T}_{t=1}E\left[\int^{\overline{d}}_{\underline{d}}Y_{it}'(x)\left(\ind_{D_{it}\ge x}-\ind_{d^{0}\ge x}\right)dxR_{it}\right]\nonumber \\
 & = & T\int^{\overline{d}}_{\underline{d}}\sum^{T}_{t=1}E\left[Y_{it}'(x)\left(\ind_{D_{it}\ge x}-\ind_{d^{0}\ge x}\right)R_{it}\right]dx.\label{eq:numer_TWFE_d}
\end{eqnarray}
The first equality uses \eqref{eq:U-stat} and $\overline{R}_{i}=0$,
and the second equality uses \eqref{eq:orth} and Assumption \ref{assu:PO}.
The third equality follows from the fundamental theorem of calculus
and an identity that $\ind_{D_{it}\ge x}-\ind_{d^{0}\ge x}$ is equal
to $1$ if $d^{0}<x\le D_{it}$, to $-1$ if $D_{it}<x\le d^{0}$,
and to $0$ otherwise. The fourth equality follows from Fubini's theorem.
Note that the absolute integrability condition required for using
the theorem is satisfied because Assumption \ref{assu:Diff} and the
Cauchy--Schwarz inequality imply
\begin{eqnarray*}
E\left[\int^{\overline{d}}_{\underline{d}}\left|Y_{it}'(x)\left(\ind_{D_{it}\ge x}-\ind_{d^{0}\ge x}\right)R_{it}\right|dx\right] & \le & E\left[M_{it}\cdot\left|D_{it}-d^{0}\right|\cdot\left|R_{it}\right|\right]\\
 & \le & \sqrt{E\left[M^{2}_{it}\left(D_{it}-d^{0}\right)^{2}\right]E\left[R^{2}_{it}\right]}\\
 & < & \infty.
\end{eqnarray*}
Combining \eqref{eq:orth_d} and \eqref{eq:numer_TWFE_d} yields the
weighted average expression in Theorem \ref{thm:Causal_TWFE_d}.

In addition, 
\begin{align*}
\int^{\overline{d}}_{\underline{d}}\sum^{T}_{t=1}E\left[\psi_{it}(x;d^{0})\right]dx & =\sum^{T}_{t=1}E\left[\int^{\overline{d}}_{\underline{d}}\psi_{it}(x;d^{0})dx\right]\\
 & =\frac{\sum^{T}_{t=1}E\left[\left(D_{it}-d^{0}\right)R_{it}\right]}{\sum^{T}_{t=1}E\left[D_{it}R_{it}\right]}\\
 & =1,
\end{align*}
where the first equality uses Fubini's theorem and the last equality
follows from $E[R_{it}]=0$.\hfill$\square$\\

\subsubsection{Proof of Theorem \ref{thm:Causal_TWFE_d2}\label{subsec:Proof_Causal_TWFE_d2}}

Since the set of assumptions in Theorem \ref{thm:Causal_TWFE_d} is
still satisfied, 
\[
\beta^{D}_{\text{FE}}=\int^{\overline{d}}_{\underline{d}}\sum^{T}_{t=1}E\left[Y_{it}'(x)\psi_{it}(x;d^{0})\right]dx.
\]
For any $d\in(\underline{d},\overline{d})$,
\begin{flalign}
 & \sum^{T}_{t=1}E\left[Y_{it}'(d)\left(\psi_{it}(d;d^{0})-\frac{1}{T}\sum^{T}_{s=1}\psi_{is}(d;d^{0})\right)\right]\nonumber \\
= & \frac{1}{T}\sum_{s>t}E\left[\left(Y_{is}'(d)-Y_{it}'(d)\right)\left(\psi_{is}(d;d^{0})-\psi_{it}(d;d^{0})\right)\right]\nonumber \\
= & \frac{1}{T}\sum_{s>t}E\left[E\left[Y_{is}'(d)-Y_{it}'(d)|D_{is},D_{it},\overline{D}_{i},W_{is},W_{it},\overline{W}_{i},C_{i}\right]\left(\psi_{is}(d;d^{0})-\psi_{it}(d;d^{0})\right)\right]\nonumber \\
= & \frac{1}{T}\sum_{s>t}E\left[E\left[Y_{is}'(d)-Y_{it}'(d)|W_{is}-W_{it},C_{i}\right]\left(\psi_{is}(d;d^{0})-\psi_{it}(d;d^{0})\right)\right]\nonumber \\
= & \frac{1}{T}\sum_{s>t}E\left[\left(Y_{is}'(d)-Y_{it}'(d)\right)E\left[\psi_{is}(d;d^{0})-\psi_{it}(d;d^{0})|W_{is}-W_{it},C_{i}\right]\right].\label{eq:DY_Dpsi}
\end{flalign}
The first equality uses \eqref{eq:U-stat}. The second equality follows
from the law of iterated expectations, since $\psi_{it}(d;d^{0})$
is determined by $(D_{it},\overline{D}_{i},W_{it},\overline{W}_{i},C_{i})$.
The third equality uses Assumptions \ref{assu:Diff}, \ref{assu:SE}, and \ref{assu:Suff}.
The last equality follows from the law of iterated expectations.

Letting $A(s,t)=E\left[\psi_{is}(d;d^{0})-\psi_{it}(d;d^{0})|W_{is}-W_{it},C_{i}\right]$
and $b_{t}=Y_{it}'(d)$ in \eqref{eq:sym_id} and applying it to \eqref{eq:DY_Dpsi} yields
\[
\sum^{T}_{t=1}E\left[Y_{it}'(d)\psi_{it}(d;d^{0})\right]=\sum^{T}_{t=1}E\left[Y_{it}'(d)\frac{1}{T}\sum^{T}_{s=1}\left(\psi_{is}(d;d^{0})+E\left[\psi_{it}(d;d^{0})-\psi_{is}(d;d^{0})|W_{it}-W_{is},C_{i}\right]\right)\right].
\]
This proves \eqref{eq:psi_star_alt}.

In addition,
\[
\frac{1}{T}\sum^{T}_{s=1}\psi_{is}(d;d^{0})=\frac{\frac{1}{T}\sum^{T}_{s=1}\left(\ind_{D_{is}\ge d}-\ind_{d^{0}\ge d}\right)R_{is}}{\sum^{T}_{s=1}E\left[D_{is}R_{is}\right]}=\frac{\frac{1}{T}\sum^{T}_{s=1}\ind_{D_{is}\ge d}R_{is}}{\sum^{T}_{s=1}E\left[D_{is}R_{is}\right]}
\]
follows from $\frac{1}{T}\sum^{T}_{s=1}R_{is}=0$ and
\begin{flalign*}
 & E\left[\psi_{it}(d;d^{0})-\psi_{is}(d;d^{0})|W_{it}-W_{is},C_{i}\right]\\
= & \frac{E\left[\ind_{D_{it}\ge d}R_{it}-\ind_{D_{is}\ge d}R_{is}-(R_{it}-R_{is})\ind_{d^{0}\ge d}|W_{it}-W_{is},C_{i}\right]}{\sum^{T}_{t'=1}E\left[D_{it'}R_{it'}\right]}\\
= & \frac{E\left[\ind_{D_{it}\ge d}R_{it}-\ind_{D_{is}\ge d}R_{is}|W_{it}-W_{is},C_{i}\right]}{\sum^{T}_{t'=1}E\left[D_{it'}R_{it'}\right]}
\end{flalign*}
follows from (\ref{eq:def_dRH}). This proves (\ref{eq:psi_star_def}).

Finally,
\begin{align*}
\sum^{T}_{t=1}E\left[\psi^{*}_{it}(d)R_{it}\right] & =\frac{1}{T}\sum^{T}_{s=1}\sum^{T}_{t=1}E\left[R_{it}\left(\psi_{is}(d;d^{0})+E\left[\psi_{it}(d;d^{0})-\psi_{is}(d;d^{0})|W_{it}-W_{is},C_{i}\right]\right)\right]\\
 & =\frac{1}{T}\sum^{T}_{s=1}\sum^{T}_{t=1}E\left[R_{it}E\left[\psi_{it}(d;d^{0})-\psi_{is}(d;d^{0})|W_{it}-W_{is},C_{i}\right]\right]\\
 & =\frac{1}{T}\sum_{s>t}E\left[\left(R_{is}-R_{it}\right)E\left[\psi_{is}(d;d^{0})-\psi_{it}(d;d^{0})|W_{is}-W_{it},C_{i}\right]\right]\\
 & =\frac{1}{T}\sum_{s>t}E\left[E\left[R_{is}-R_{it}|W_{is}-W_{it},C_{i}\right]\left(\psi_{is}(d;d^{0})-\psi_{it}(d;d^{0})\right)\right]\\
 & =0,
\end{align*}
where the first equality uses \eqref{eq:psi_star_alt}, the second
equality follows from $\sum^{T}_{t=1}R_{it}=0$, the third equality
uses \eqref{eq:sym_id} by letting $A(s,t)=E\left[\psi_{is}(d;d^{0})-\psi_{it}(d;d^{0})|W_{is}-W_{it},C_{i}\right]$
and $b_{t}=R_{it}$, the fourth equality uses the law of iterated
expectations, and the last equality uses \eqref{eq:def_dRH}.
This proves \eqref{eq:psi_optimal}.\hfill$\square$\\

\section{Results in Unbalanced Panels}\label{Asec:Unbalanced}
The main paper assumes a balanced panel setting, in which each unit--period
combination has exactly one observation. This section discusses how
the results in the main paper extend to, or differ in, a more general
setting. Proofs of additional theoretical results are provided at
the end of this section.

\subsection{Equivalence of Least-Squares Objectives}
Without assuming a balanced panel, the least-squares objective in
(\ref{eq:FE_reg}) should be rewritten as
\begin{equation}
\underset{\beta,\{\alpha_{i}\}_{i=1}^{N},\{\gamma_{t},\mu_{t}\}_{t=1}^{T}}{\min}\sum_{i=1}^{N}\sum_{t=1}^{T}B_{it}\left(Y_{it}-X_{it}'\beta-\alpha_{i}-C_{i}'\gamma_{t}-\mu_{t}\right)^{2}.\label{eq:FE_LS_w}
\end{equation}
In an unbalanced panel, $B_{it}$ is an observability indicator that
takes a value of 1 if an observation $(i,t)$ is available and a value
of $0$ if the observation is missing. More generally, $B_{it}$ can
be any real number. For example, a TWFE regression may use repeated
cross-section data with $B_{it}$ observations in group $i$ at period
$t$. The equivalence between TWFE and pooled FD regressions naturally
extends to these cases.
\begin{theorem}
\label{thm:TWFE_FD_ub}The coefficient on $X_{it}$ given by (\ref{eq:FE_LS_w})
is algebraically identical to the coefficient on $\Delta_{k}X_{it}$
given by the following least-squares problem.
\begin{align}
\underset{\beta,\left\{ \gamma_{t},\mu_{t}\right\} _{t=1}^{T}}{\min} & \sum_{k=1}^{T-1}\sum_{i=1}^{N}\sum_{t=1}^{T-k}\frac{B_{it}B_{i,t+k}}{\overline{B}_{i}}\left(\Delta_{k}Y_{it}-\Delta_{k}X_{it}'\beta-C_{i}'\Delta_{k}\gamma_{t}-\Delta_{k}\mu_{t}\right)^{2}.\label{eq:FD_equiv_w}
\end{align}
\end{theorem}

\subsection{Weighted-Average Relationship}

The least-squares problem \eqref{eq:FD_equiv_w} is distinct from 
\begin{equation}
\underset{\beta,\left\{ \gamma_{k,t}^{*}\right\} _{k,t}}{\min}\sum_{i=1}^{N}\sum_{k=1}^{T-1}\sum_{t=1}^{T-k}\frac{B_{it}B_{i,t+k}}{\overline{B}_{i}}\left(\Delta_{k}Y_{it}-\Delta_{k}X_{it}'\beta-C_{i}'\gamma_{k,t}^{*}-\mu_{k,t}^{*}\right)^{2}.\label{eq:FD_equiv_flex}
\end{equation}
While (\ref{eq:FD_equiv_w}) includes fixed effects for starting and
ending periods in an additive manner, (\ref{eq:FD_equiv_flex}) accounts
for fixed effects associated with all possible combinations of starting
and ending periods. These separability restrictions, $\gamma_{k,t}^{*}=\gamma_{t+k}-\gamma_{t}$
and $\mu_{k,t}^{*}=\mu_{t+k}-\mu_{t}$, do not bind in a balanced
panel. However,
this property does not extend to unbalanced panels. 

As a result, even when $X_{it}$ is univariate, unlike in Theorem
\ref{theorem:FE_not_equal_FD}, the TWFE coefficient $\widehat{\beta}_{\text{FE}}$
given by (\ref{eq:FE_LS_w}) or (\ref{eq:FD_equiv_w}) is not exactly
identical to a convex combination of FD coefficients $\widehat{\beta}_{\text{FD},k}$
that are given by solving
\begin{equation}
\underset{\beta,\left\{ \gamma_{t}^{*},\mu_{t}^{*}\right\} _{t=1}^{T-k}}{\min}\sum_{i=1}^{N}\sum_{t=1}^{T-k}\frac{B_{it}B_{i,t+k}}{\overline{B}_{i}}\left(\Delta_{k}Y_{it}-\Delta_{k}X_{it}'\beta-C_{i}'\gamma_{t}^{*}-\mu_{t}^{*}\right)^{2}\label{eq:FD_UR}
\end{equation}
for $k=1,\ldots,T-1$. One possible way to restore a weighted-average
relationship is to use (\ref{eq:FD_equiv_flex}) instead of (\ref{eq:FE_LS_w})
or (\ref{eq:FD_equiv_w}) for producing the TWFE coefficient.

\subsection{Causal Interpretation}

Now, I consider causal interpretation of the population TWFE coefficient
that arises from the regression (\ref{eq:FE_LS_w}). For simplicity,
I focus on an unbalanced panel in which $B_{it}$ is either $0$ or
$1$ and $\overline{B}_{i}>0$. Furthermore, I focus on the setting
with a baseline treatment level, which corresponds to the one in Section
\ref{subsec:TWFE-Interpretation}. The required assumptions for causal
interpretation of the TWFE coefficient should be modified as follows.

\renewcommand{\theassumptionx}{CT-u}
\begin{assumptionx}
\label{assu:PT_ub}\textup{(Conditional Common Trends)} There exists
$d^{0}\in(\underline{d},\overline{d})$ such that 
\[
E\left[\Delta_{k}Y_{it}(d^{0})|\Delta_{k}D_{it},\Delta_{k}W_{it},C_{i},\overline{B}_{i},B_{it}=B_{i,t+k}=1\right]=E\left[\Delta_{k}Y_{it}(d^{0})|\Delta_{k}W_{it},C_{i},B_{it}=B_{i,t+k}=1\right]
\]
 for any $(k,t)$ with $1\le t<t+k\le T$.
\end{assumptionx}

\renewcommand{\theassumptionx}{V-u}
\begin{assumptionx}
\label{assu:Variation_ub}\textup{(Variation)} $\sum_{k=1}^{T-1}\sum_{t=1}^{T-k}E\left[Var\left(\Delta_{k}D_{it}|\Delta_{k}W_{it},C_{i},\overline{B}_{i},B_{it}=B_{i,t+k}=1\right)\right]>0$.
\end{assumptionx}

\renewcommand{\theassumptionx}{LH-u}
\begin{assumptionx}
\label{assu:Homogeneity_ub}\textup{(Linearity and Time Homogeneity)}
For any $(k,t)$ with $1\le t<t+k\le T$,
\[
E\left[\Delta_{k}D_{it}|\Delta_{k}W_{it},C_{i},\overline{B}_{i},B_{it}=B_{i,t+k}=1\right]=\Delta_{k}W_{it}'\delta^{W}+C_{i}'\left(\delta_{t+k}^{C}-\delta_{t}^{C}\right)+\left(\delta_{t+k}^{I}-\delta_{t}^{I}\right),
\]
where the coefficient $\delta^{W}$ does not vary across $(k,t)$.
\end{assumptionx}

These assumptions differ from the corresponding assumptions (\ref{assu:PT},
\ref{assu:Variation}, and \ref{assu:Homogeneity}) in Section \ref{sec:Causal-Interpretation}
in three ways. First, the assumptions about the potential outcome
change $\Delta_{k}Y_{it}(d^{0})$ and the treatment change $\Delta_{k}D_{it}$
are now conditional on the changes being observed, $B_{it}=B_{i,t+k}=1$.
Second, the observation rate $\overline{B}_{i}$ has to be exogenous,
although this requirement can be removed when time-invariant covariates
$C_{i}$ include $\overline{B}_{i}$. Third, the coefficient on $C_{i}$
and the intercept coefficient in Assumption \ref{assu:Homogeneity_ub}
are less flexible than in Assumption \ref{assu:Homogeneity}, where
the coefficients can depend on $(t,k)$ in an arbitrary manner.
\begin{theorem}
\label{thm:Causal_TWFE-u}Under Assumptions \ref{assu:PO}, \ref{assu:PT_ub},
\ref{assu:Variation_ub}, and \ref{assu:Homogeneity_ub},
\[
\beta_{\text{FE}}^{D}=\sum_{t=1}^{T}E\left[\tau_{it}\omega_{it}\right],
\]
where $\tau_{it}$ is the per-unit effect defined in \eqref{eq:tau_it}.
The weights satisfy $\sum_{t=1}^{T}E\left[\omega_{it}\right]=1$ and
\begin{equation}
\omega_{it}\propto\frac{B_{it}}{\overline{B}_{i}}\left(D_{it}-d^{0}\right)\sum_{s=1}^{T}B_{is}\left(D_{it}-D_{is}-E\left[D_{it}-D_{is}|W_{it}-W_{is},C_{i},\overline{B}_{i},B_{it}=B_{is}=1\right]\right).\label{eq:weight_ub}
\end{equation}
\end{theorem}
This theorem, much like Theorem \ref{thm:Causal_TWFE}, expresses
the coefficient $\beta_{\text{FE}}^{D}$ as a weighted average of
per-unit treatment effects $\tau_{it}$, with weights summing to one
but possibly negative. The weight function is similar to Theorem \ref{thm:Causal_TWFE}
but is adjusted for observability $B_{it}$.

\subsection{Proofs}

\subsubsection{Proof of Theorem \ref{thm:TWFE_FD_ub}}

Suppose a sequence $\{w_{t}\}_{t=1}^{T}$ satisfies $\sum_{t=1}^{T}w_{t}=T$
and define $\overline{a}^{w}\equiv\frac{1}{T}\sum_{t=1}^{T}w_{t}a_{t}$
for another sequence $\{a_{t}\}_{t=1}^{T}$. Basic algebra yields
\begin{multline}
\frac{1}{T}\sum_{k=1}^{T-1}\sum_{t=1}^{T-k}w_{t+k}w_{t}\left(\Delta_{k}a_{t}\right)^{2}=\frac{1}{2T}\sum_{s=1}^{T}\sum_{t=1}^{T}w_{s}w_{t}\left(a_{s}-a_{t}\right)^{2}=\frac{1}{2T}\sum_{s=1}^{T}\sum_{t=1}^{T}w_{s}w_{t}\left(a_{s}-\overline{a}^{w}+\overline{a}^{w}-a_{t}\right)^{2}\\
=\frac{1}{2T}\sum_{s=1}^{T}\sum_{t=1}^{T}w_{s}w_{t}\left\{ \left(a_{s}-\overline{a}^{w}\right)^{2}+\left(a_{t}-\overline{a}^{w}\right)^{2}-2\left(a_{s}-\overline{a}^{w}\right)\left(a_{t}-\overline{a}^{w}\right)\right\} =\sum_{t=1}^{T}w_{t}\left(a_{t}-\overline{a}^{w}\right)^{2}\\
=\sum_{t=1}^{T}w_{t}a_{t}^{2}-\left(\overline{a}^{w}\right)^{2}.\label{eq:lm1_g}
\end{multline}
Exploiting the algebraic identity (\ref{eq:lm1_g}) by letting $w_{t}=\frac{B_{it}}{\overline{B}_{i}}$
and $a_{t}=Y_{it}-X_{it}'\beta-\alpha_{i}-C_{i}'\gamma_{t}-\mu_{t}$
for each $i=1,\ldots,N$, the objective function (\ref{eq:FE_LS_w})
can be expressed as
\begin{multline}
\sum_{i=1}^{N}\sum_{t=1}^{T}B_{it}\left(Y_{it}-X_{it}'\beta-\alpha_{i}-C_{i}'\gamma_{t}-\mu_{t}\right)^{2}\\
=\frac{1}{T}\sum_{i=1}^{N}\sum_{k=1}^{T-1}\sum_{t=1}^{T-k}\frac{B_{i,t+k}B_{it}}{\overline{B}_{i}}\left(\Delta_{k}Y_{it}-\Delta_{k}X_{it}'\beta-C_{i}'\Delta_{k}\gamma_{t}-\Delta_{k}\mu_{t}\right)^{2}\\
+\sum_{i=1}^{N}\overline{B}_{i}\left(\sum_{t=1}^{T}\frac{B_{it}}{T\overline{B}_{i}}\left(Y_{it}-X_{it}'\beta-C_{i}'\gamma_{t}-\mu_{t}\right)-\alpha_{i}\right)^{2}.\label{eq:FE_LS_wA}
\end{multline}
Since choosing $\beta$ and $\{\gamma_{t},\mu_{t}\}_{t=1}^{T}$ that
solve (\ref{eq:FD_equiv_w}) and letting $\alpha_{i}=\sum_{t=1}^{T}\frac{B_{it}}{T\overline{B}_{i}}\left(Y_{it}-X_{it}'\beta-C_{i}'\gamma_{t}-\mu_{t}\right)$
for each $i=1,\ldots,N$ results in minimizing (\ref{eq:FE_LS_wA}),
the two objectives (\ref{eq:FE_LS_w}) and (\ref{eq:FD_equiv_w})
yield the same estimates of $\beta$.\hfill$\square$\\

\subsubsection{Proof of Theorem \ref{thm:Causal_TWFE-u}}

Suppose $X_{it}$ consists of a scalar treatment variable $D_{it}$
and a vector of time-varying covariates $W_{it}$ in (\ref{eq:FE_LS_w}).
Due to the equivalence between (\ref{eq:FE_LS_w}) and (\ref{eq:FD_equiv_w}),
the population TWFE coefficient on $D_{it}$ is given by solving
\begin{align}
\underset{\beta^{D},\beta^{W},\left\{ \gamma_{t},\mu_{t}\right\} _{t=1}^{T}}{\min} & \sum_{k=1}^{T-1}\sum_{t=1}^{T-k}E\left[\frac{B_{it}B_{i,t+k}}{\overline{B}_{i}}\left(\Delta_{k}Y_{it}-\beta^{D}\Delta_{k}D_{it}-\Delta_{k}W_{it}'\beta^{W}-C_{i}'\Delta_{k}\gamma_{t}-\Delta_{k}\mu_{t}\right)^{2}\right].\label{eq:FD_equiv_w_pop}
\end{align}

Assumption \ref{assu:Homogeneity_ub} and the FWL theorem imply that
the same $\beta^{D}$ as (\ref{eq:FD_equiv_w_pop}) can be recovered
by solving
\begin{align}
\underset{\beta^{D}}{\min} & \sum_{k=1}^{T-1}\sum_{t=1}^{T-k}E\left[\frac{B_{it}B_{i,t+k}}{\overline{B}_{i}}\left\{ \Delta_{k}Y_{it}-\beta^{D}\left(\Delta_{k}D_{it}-E\left[\Delta_{k}D_{it}|\Delta_{k}W_{it},C_{i},\overline{B}_{i},B_{it}=B_{i,t+k}=1\right]\right)\right\} ^{2}\right].\label{eq:FD_equiv_w_popr}
\end{align}
This regression yields
\begin{equation}
\beta^{D}=\frac{\sum_{k=1}^{T-1}\sum_{t=1}^{T-k}E\left[\frac{B_{it}B_{i,t+k}}{\overline{B}_{i}}\Delta_{k}Y_{it}\left(\Delta_{k}D_{it}-E\left[\Delta_{k}D_{it}|\Delta_{k}W_{it},C_{i},\overline{B}_{i},B_{it}=B_{i,t+k}=1\right]\right)\right]}{\sum_{k=1}^{T-1}\sum_{t=1}^{T-k}E\left[\frac{B_{it}B_{i,t+k}}{\overline{B}_{i}}\Delta_{k}D_{it}\left(\Delta_{k}D_{it}-E\left[\Delta_{k}D_{it}|\Delta_{k}W_{it},C_{i},\overline{B}_{i},B_{it}=B_{i,t+k}=1\right]\right)\right]}.\label{eq:TWFE_pop_ub}
\end{equation}

The numerator in (\ref{eq:TWFE_pop_ub}) can be expressed as
\begin{align}
 & \sum_{k=1}^{T-1}\sum_{t=1}^{T-k}E\left[\frac{B_{it}B_{i,t+k}}{\overline{B}_{i}}\Delta_{k}Y_{it}\left(\Delta_{k}D_{it}-E\left[\Delta_{k}D_{it}|\Delta_{k}W_{it},C_{i},\overline{B}_{i},B_{it}=B_{i,t+k}=1\right]\right)\right]\nonumber \\
= & \sum_{k=1}^{T-1}\sum_{t=1}^{T-k}E\left[\frac{B_{it}B_{i,t+k}}{\overline{B}_{i}}\left(\Delta_{k}Y_{it}-\Delta_{k}Y_{it}(d^{0})\right)\left(\Delta_{k}D_{it}-E\left[\Delta_{k}D_{it}|\Delta_{k}W_{it},C_{i},\overline{B}_{i},B_{it}=B_{i,t+k}=1\right]\right)\right]\nonumber \\
= & \sum_{s=1}^{T}\sum_{t=1}^{T}E\left[\frac{B_{it}B_{is}}{\overline{B}_{i}}\left(Y_{it}(D_{it})-Y_{it}(d^{0})\right)\left(D_{is}-D_{it}-E\left[D_{is}-D_{it}|W_{is}-W_{it},C_{i},\overline{B}_{i},B_{it}=B_{is}=1\right]\right)\right]\nonumber \\
= & \sum_{t=1}^{T}E\left[\tau_{it}\frac{B_{it}}{\overline{B}_{i}}\left(D_{it}-d^{0}\right)\sum_{s=1}^{T}B_{is}\left(D_{it}-D_{is}-E\left[D_{it}-D_{is}|W_{it}-W_{is},C_{i},\overline{B}_{i},B_{it}=B_{is}=1\right]\right)\right].\label{eq:numer_ub}
\end{align}
Note that the first equality uses Assumption \ref{assu:PT_ub}, the
second equality follows from Assumption \ref{assu:PO} and \eqref{eq:sym_id},
and the last equality uses the definition of $\tau_{it}$.

The denominator in (\ref{eq:TWFE_pop_ub}) is strictly positive due
to Assumption \ref{assu:Variation_ub}. Also, using the transformation
applied to the numerator by letting $Y_{it}(d)=d$, the denominator
in (\ref{eq:TWFE_pop_ub}) can be expressed as 
\begin{align}
 & \sum_{k=1}^{T-1}\sum_{t=1}^{T-k}E\left[\frac{B_{it}B_{i,t+k}}{\overline{B}_{i}}\Delta_{k}D_{it}\left(\Delta_{k}D_{it}-E\left[\Delta_{k}D_{it}|\Delta_{k}W_{it},C_{i},\overline{B}_{i},B_{it}=B_{i,t+k}=1\right]\right)\right]\nonumber \\
= & \sum_{t=1}^{T}E\left[\frac{B_{it}}{\overline{B}_{i}}\left(D_{it}-d^{0}\right)\sum_{s=1}^{T}B_{is}\left(D_{it}-D_{is}-E\left[D_{it}-D_{is}|W_{it}-W_{is},C_{i},\overline{B}_{i},B_{it}=B_{is}=1\right]\right)\right].\label{eq:denom_ub}
\end{align}
Combining (\ref{eq:numer_ub}) and (\ref{eq:denom_ub}) yields the
weight expression (\ref{eq:weight_ub}) and $\sum_{t=1}^{T}E\left[\omega_{it}\right]=1$.\hfill$\square$\\

\edef\temp{\noexpand\ref{thm:FE_FD_DID}}
\section{Connection of Theorem \temp\ with Prior Work}\label{sec:Comparison-GB}
This section explores how Theorem \ref{thm:FE_FD_DID}, which links
TWFE and pooled FD regressions via a U-statistics identity, connects
to prior DID decomposition results, notably \citet{goodman2018difference}
and \citet{strezhnev2018semiparametric}. It extends the period-based
expansion in Theorem \ref{thm:FE_FD_DID} to cross-unit differences
and situates earlier findings within a broader numerical framework.

\subsubsection{Cross-Unit Expansion of TWFE}

Theorem \ref{thm:FE_FD_DID} derives TWFE coefficients from a least-squares
problem pooling $k$-period differences over time. Given TWFE’s symmetry
across units and periods, this approach extends to differences across
units, as shown below.
\begin{theorem}
\label{thm:TWFE-DID} Let $\widehat{\beta}_{\text{FE}}$ be the coefficient
on $X_{it}$ from the TWFE least-squares problem defined in (\ref{eq:FE_reg}).
Then $\beta=\widehat{\beta}_{\text{FE}}$ also solves:

\begin{align}
\underset{\beta,\{\gamma_{t}\}_{t=1}^{T}}{\min} & \sum_{i<j}\sum_{k=1}^{T-1}\sum_{t=1}^{T-k}\left\{ \left(\Delta_{k}Y_{jt}-\Delta_{k}Y_{it}\right)-\left(\Delta_{k}X_{jt}-\Delta_{k}X_{it}\right)'\beta-\left(C_{j}-C_{i}\right)'\Delta_{k}\gamma_{t}\right\} ^{2}.\label{eq:DID_reg}
\end{align}
\end{theorem}
\begin{proof}
This result builds on the proof of Theorem \ref{thm:FE_FD_DID}. Applying the U-statistics identity
\eqref{eq:U-stat} to the cross-sectional summation across $i=1,\ldots,N$
in \eqref{eq:e_identity} yields:
\begin{multline}
\sum_{k=1}^{T-1}\sum_{t=1}^{T-k}\sum_{i=1}^{N}\left(\Delta_{k}e_{it}\right)^{2}=\sum_{k=1}^{T-1}\sum_{t=1}^{T-k}\left\{ \sum_{i=1}^{N}\left(\Delta_{k}e_{it}-\frac{1}{N}\sum_{j=1}^{N}\Delta_{k}e_{jt}\right)^{2}+N\left(\frac{1}{N}\sum_{i=1}^{N}\Delta_{k}e_{it}\right)^{2}\right\} \\
=\frac{1}{N}\sum_{k=1}^{T-1}\sum_{t=1}^{T-k}\sum_{1\le i<j\le N}\left(\Delta_{k}e_{jt}-\Delta_{k}e_{it}\right)^{2}+N\sum_{k=1}^{T-1}\sum_{t=1}^{T-k}\left(\frac{1}{N}\sum_{i=1}^{N}\Delta_{k}e_{it}\right)^{2}.\label{eq:FD_DID}
\end{multline}
The first term in \eqref{eq:FD_DID} is identical to \eqref{eq:DID_reg}
and does not depend on $\{\mu_{t}\}_{t=1}^{T}$. For any $\left(\beta,\{\gamma_{t}\}_{t=1}^{T}\right)$,
setting $\mu_{t}=\frac{1}{N}\sum_{i=1}^{N}\left(Y_{it}-X_{it}'\beta-C_{i}'\gamma_{t}\right)$
ensures that the second term vanishes. Therefore, $\left(\beta,\{\gamma_{t}\}_{t=1}^{T}\right)$
minimizing (\ref{eq:DID_reg}) solves (\ref{eq:FE_reg}) as well.\hfill$\square$\\
\end{proof}
The least-squares problem (\ref{eq:DID_reg}) can be interpreted as
a regression of a DID in $Y_{it}$ across units and periods on the
corresponding DID in $X_{it}$, controlling for a difference in time-invariant
covariates $C_{i}$. For a scalar $X_{it}$ without covariates, Theorem
\ref{thm:TWFE-DID} implies:
\begin{align}
\widehat{\beta}_{\text{FE}}= & \frac{\sum_{s>t,j>i}\left\{ (Y_{js}-Y_{jt})-(Y_{is}-Y_{it})\right\} \left\{ (X_{js}-X_{jt})-(X_{is}-X_{it})\right\} }{\sum_{s>t,j>i}\left\{ (X_{js}-X_{jt})-(X_{is}-X_{it})\right\} ^{2}}.\label{eq:DID_elem}
\end{align}
In a staggered adoption design with binary $X_{it}$, each nonzero
term in the numerator corresponds to a two-period, two-unit DID: it
equals $(Y_{is}-Y_{it})-(Y_{js}-Y_{jt})$ if only unit $i$ starts
treatment between periods $t$ and $s$, or the reverse if only $j$
does. Thus, $\widehat{\beta}_{\text{FE}}$ averages all such DID comparisons.

\subsubsection{Relation to Prior Work}

This representation aligns with \citet{goodman2018difference} and
\citet{strezhnev2018semiparametric}, who decompose TWFE estimators
in binary treatment settings into averages of two-unit, two-period
DID comparisons. Their approaches rely on algebraic methods tailored
to binary treatment settings. While \citet{goodman2018difference}
employs a similar identity as a lemma, he uses it for an expansion
across units but not periods, and does not present it as a U-statistics
result. By contrast, my approach explicitly leverages the U-statistics
identity to uncover a general structure that applies to binary, discrete,
or continuous regressors, with or without covariates, revealing their
results as special cases.

To further illustrate the link, I demonstrate that the weighted-average
expression in \citet{goodman2018difference} results from aggregating
(\ref{eq:DID_elem}) by treatment timing. Define the treatment as
$X_{it}=\ind_{t>g_{i}}$, where $g_{i}\in\{1,\ldots,T\}$ represents
the last period in which the unit $i$ is untreated, with $g_{i}=T$
for the never-treated group (denoted $U$ in \citealp{goodman2018difference}).
Without loss of generality, suppose that units $i=1,\ldots,N$ are
ordered so that $g_{j}\le g_{i}$ for any $j>i$. Let $n_{g}$ be
the number of units that belong to group $g$, where $\sum_{g=1}^{T}n_{g}=N$.
Define
\[
\overline{Y}_{gt}=\frac{1}{n_{g}}\sum_{i:g_{i}=g}Y_{it}
\]
to be the group mean of the outcome. Given these,
\begin{align*}
 & \sum_{s>t,j>i}\left\{ (Y_{js}-Y_{jt})-(Y_{is}-Y_{it})\right\} \left\{ (X_{js}-X_{jt})-(X_{is}-X_{it})\right\} \\
= & \sum_{s>t,h>g}n_{g}n_{h}\left\{ (\overline{Y}_{gs}-\overline{Y}_{gt})-(\overline{Y}_{hs}-\overline{Y}_{ht})\right\} \left(\ind_{t\le g<s}-\ind_{t\le h<s}\right)\\
= & \frac{1}{N}\sum_{t\le g<s\le h}n_{g}n_{h}\left\{ (\overline{Y}_{gs}-\overline{Y}_{gt})-(\overline{Y}_{hs}-\overline{Y}_{ht})\right\} -\frac{1}{N}\sum_{g<t\le h<s}n_{g}n_{h}\left\{ (\overline{Y}_{gs}-\overline{Y}_{gt})-(\overline{Y}_{hs}-\overline{Y}_{ht})\right\} ,
\end{align*}
where the last equality follows from the identity that $\ind_{t\le g<s}-\ind_{t\le h<s}$
is equal to $1$ if $t\le g<s\le h$, to $-1$ if $g<t\le h<s$, and
to $0$ otherwise.

Then, following the notation in \citet{goodman2018difference}, for
any two groups $g$ and $h$, let
\[
\overline{Y}_{g}^{POST(h)}=\frac{1}{T-h}\sum_{t=h+1}^{T}\overline{Y}_{gt}
\]
be the mean outcome of the group $g$ during periods in which the
group $h<T$ is treated, and let
\[
\overline{Y}_{g}^{PRE(h)}=\frac{1}{h}\sum_{t=1}^{h}\overline{Y}_{gt}
\]
be the mean outcome of the group $g$ during periods in which the
group $h$ is untreated. In addition, for $g<h$, let
\[
\overline{Y}_{m}^{MID(g,h)}=\frac{1}{h-g}\sum_{t=g+1}^{h}\overline{Y}_{mt}
\]
be the mean outcome of a group $m\in\{g,h\}$ during periods in which
the group $g$ is treated and the group $h$ is untreated.

Using these definitions,
\begin{align}
 & \sum_{s>t,j>i}\left\{ (Y_{js}-Y_{jt})-(Y_{is}-Y_{it})\right\} \left\{ (X_{js}-X_{jt})-(X_{is}-X_{it})\right\} \nonumber \\
= & \frac{1}{N}\sum_{g=1}^{T-1}\sum_{h=g+1}^{T}\sum_{t=1}^{g}\sum_{s=g+1}^{h}n_{g}n_{h}\left\{ (\overline{Y}_{gs}-\overline{Y}_{gt})-(\overline{Y}_{hs}-\overline{Y}_{ht})\right\} \nonumber \\
 & -\frac{1}{N}\sum_{g=1}^{T-2}\sum_{h=g+1}^{T-1}\sum_{t=g+1}^{h}\sum_{s=h+1}^{T}n_{g}n_{h}\left\{ (\overline{Y}_{gs}-\overline{Y}_{gt})-(\overline{Y}_{hs}-\overline{Y}_{ht})\right\} \nonumber \\
= & \frac{1}{N}\sum_{g=1}^{T-1}\sum_{h=g+1}^{T}g(h-g)n_{g}n_{h}\left\{ (\overline{Y}_{g}^{MID(g,h)}-\overline{Y}_{g}^{PRE(g)})-(\overline{Y}_{h}^{MID(g,h)}-\overline{Y}_{h}^{PRE(g)})\right\} \nonumber \\
 & +\frac{1}{N}\sum_{g=1}^{T-2}\sum_{h=g+1}^{T-1}(h-g)(T-h)n_{g}n_{h}\left\{ (\overline{Y}_{h}^{POST(h)}-\overline{Y}_{h}^{MID(g,h)})-(\overline{Y}_{g}^{POST(h)}-\overline{Y}_{g}^{MID(g,h)})\right\} .\label{eq:numer}
\end{align}
Note that $(\overline{Y}_{g}^{MID(g,h)}-\overline{Y}_{g}^{PRE(g)})-(\overline{Y}_{h}^{MID(g,h)}-\overline{Y}_{h}^{PRE(g)})$
is a DID comparison between an early-treated group $g$ and a late-treated
(or never-treated) group $h$ across the periods in which only $g$
is treated and the periods in which both are untreated, the not-yet-treated
$h$ serving as the control group. In addition, $(\overline{Y}_{h}^{POST(h)}-\overline{Y}_{h}^{MID(g,h)})-(\overline{Y}_{g}^{POST(h)}-\overline{Y}_{g}^{MID(g,h)})$
is a DID comparison between a late-treated group $h$ and an early-treated
group $g$ across the periods in which both $g$ and $h$ are treated
and the periods in which only $g$ is treated, the already-treated
$g$ serving as the control group.

Similarly, the denominator in (\ref{eq:DID_elem}) satisfies
\begin{equation}
\sum_{s>t,j>i}\left\{ (X_{js}-X_{jt})-(X_{is}-X_{it})\right\} ^{2}=\frac{1}{N}\sum_{g=1}^{T-1}\sum_{h=g+1}^{T}g(h-g)n_{g}n_{h}+\frac{1}{N}\sum_{g=1}^{T-2}\sum_{h=g+1}^{T-1}(h-g)(T-h)n_{g}n_{h}.\label{eq:denom}
\end{equation}

Combining (\ref{eq:numer}) and (\ref{eq:denom}) yields
\begin{align}
\widehat{\beta}_{\text{FE}}= & \sum_{g=1}^{T-1}w_{gT}\left\{ (\overline{Y}_{g}^{POST(g)}-\overline{Y}_{g}^{PRE(g)})-(\overline{Y}_{T}^{POST(g)}-\overline{Y}_{T}^{PRE(g)})\right\} \nonumber \\
+ & \sum_{g=1}^{T-2}\sum_{h=g+1}^{T-1}w_{gh}\left[\frac{g}{T-h+g}\left\{ (\overline{Y}_{g}^{MID(g,h)}-\overline{Y}_{g}^{PRE(g)})-(\overline{Y}_{h}^{MID(g,h)}-\overline{Y}_{h}^{PRE(g)})\right\} \right.\nonumber \\
 & \left.+\frac{T-h}{T-h+g}\left\{ (\overline{Y}_{h}^{POST(h)}-\overline{Y}_{h}^{MID(g,h)})-(\overline{Y}_{g}^{POST(h)}-\overline{Y}_{g}^{MID(g,h)})\right\} \right],\label{eq:GBdecom}
\end{align}
where the weights are defined as
\[
w_{gh}=\frac{n_{g}n_{h}(h-g)(T-h+g)}{\sum_{g'=1}^{T-1}\sum_{h'=g'+1}^{T}n_{g'}n_{h'}(h'-g')(T-h'+g')}.
\]
This weighted-average expression is consistent with the one derived
in \citet{goodman2018difference}.

\section{Additional Empirical Results}

\subsection{Impact of Covariates on the Weighted-Average Interpretation}\label{subsec:Covariates}

\begin{figure}
\caption{The TWFE and FD Coefficients with Covariates}

\label{Fig:TWFE_FD_c}

\vspace{0.5em}
\centering
\begin{threeparttable}
\vspace{0.25em}

\begin{minipage}[t]{0.7\columnwidth}%
(a) Log Employment Rate of Teens

\includegraphics[width=1\columnwidth]{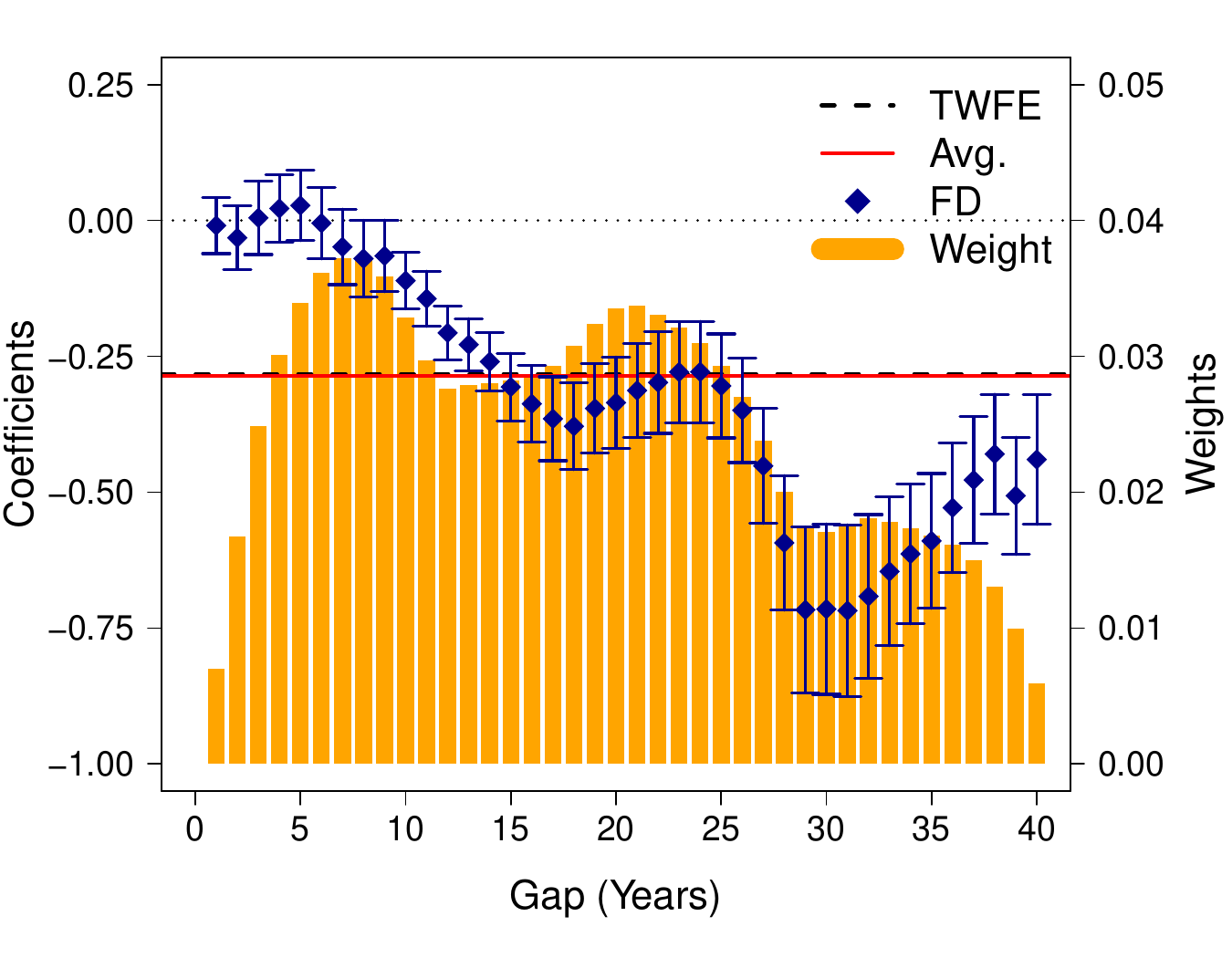}%
\end{minipage}

\vspace{0.75em}

\begin{minipage}[t]{0.7\columnwidth}%
(b) Net Job Creation Rate (\%)

\includegraphics[width=1\columnwidth]{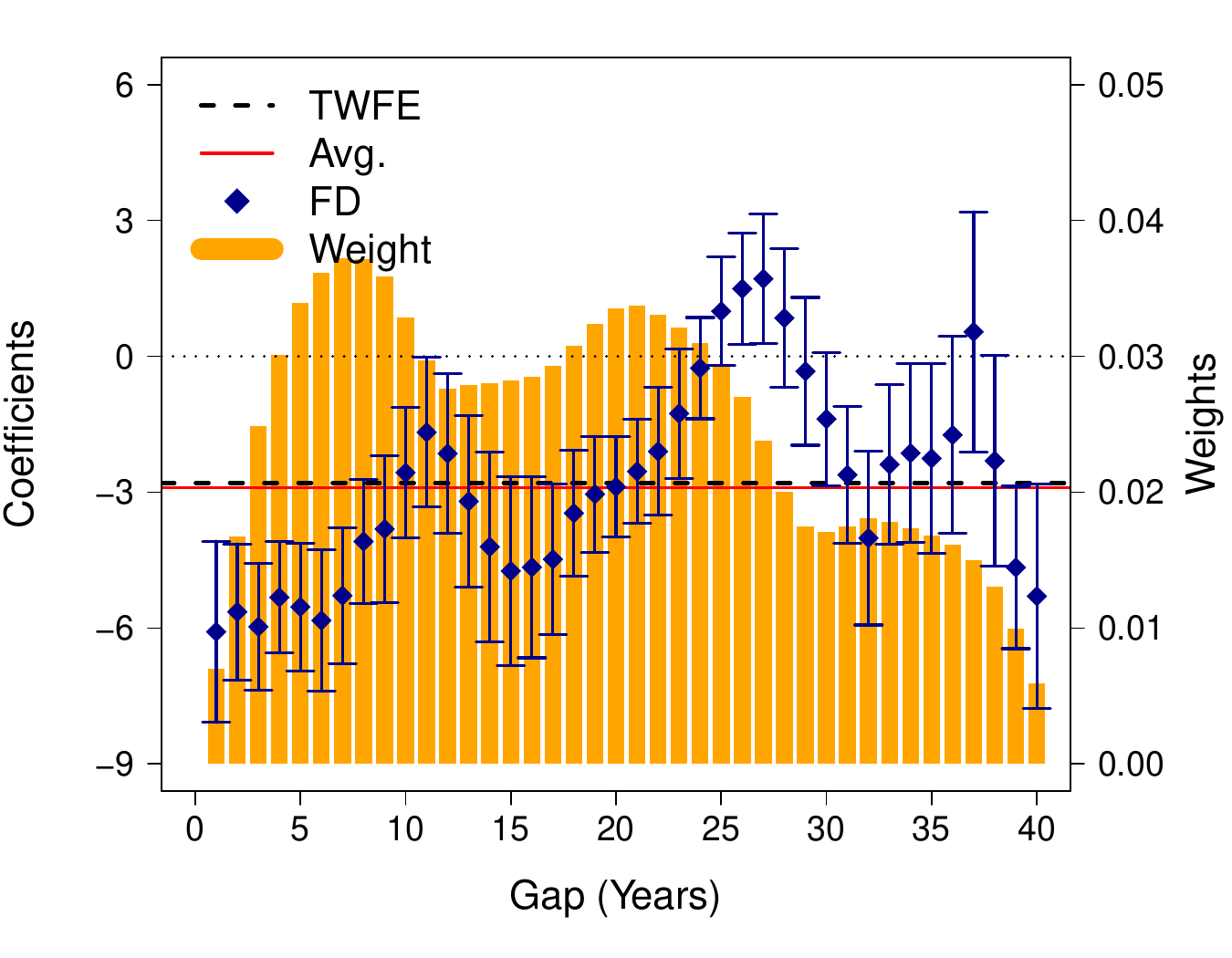}%
\end{minipage}

\vspace{0.75em}
\begin{tablenotes}
\small

\item Notes: The FD coefficients are illustrated with standard error
bars. The standard errors are robust to heteroskedasticity and
correlation across observations on the same state. Covariates are
census region dummies, log per capita income, and log teen population.

\end{tablenotes}
\end{threeparttable}
\end{figure}

Figure \ref{Fig:TWFE_FD_c} replicates the analysis from Figure \ref{Fig:TWFE_FD}
with the inclusion of covariates. Specifically, I include census region dummies,
log per capita income, and log teen population as covariates. The
latter two variables are time-varying, necessitating an adjustment
to the weighted-average interpretation. As in Figure \ref{Fig:TWFE_FD},
the FD coefficients $\widehat{\beta}_{\text{FD},k}^{D}$ for $k=1,\ldots,40$
are presented as dots with standard error bars. The TWFE coefficient
$\widehat{\beta}_{\text{FE}}^{D}$ is indicated by a dotted horizontal
line, while the weighted average of the FD coefficients defined in
Theorem \ref{theorem:FE_not_equal_FD} is represented by a solid horizontal
line. The minimal difference between these two lines suggests that
the inclusion of time-varying covariates does not substantially alter
the weighted-average interpretation of the TWFE coefficient. In panel
(a) of Figure \ref{Fig:TWFE_FD_c}, presenting the analysis for the
log teen employment rate, the TWFE coefficient $\widehat{\beta}_{\text{FE}}^{D}=-0.282$,
while the weighted average of the FD coefficients is $-0.286$. Panel
(b) presents the corresponding analysis for the net job creation rate.
The TWFE coefficient $\widehat{\beta}_{\text{FE}}^{D}=-2.80$ remains
close to the weighted average of the FD coefficients, which is $-2.90$.

\subsection{Weights on Causal Effect Parameters}\label{subsec:Weights-on-Causal}

Section \ref{sec:Causal-Interpretation} provides causal interpretation
of the TWFE coefficient, demonstrating that the population TWFE coefficient
can be understood as a weighted average of causal effects. For this
interpretation to hold, the common trends assumption must be valid
for any $k$-period gap, a condition that appears questionable in
the minimum wage setting. Even if the common trends assumption is
satisfied, a further challenge arises: if some weights are negative
and these weights are systematically related to the causal effects
of minimum wages, causal interpretation of the TWFE coefficient may
be compromised. Therefore, I examine how the weight function derived
in Theorem \ref{thm:Causal_TWFE_d2} is apportioned across calendar
years, states, and minimum wage levels.

\subsubsection*{Weights on Calendar Years}

\begin{figure}
\caption{Weights on Calendar Years}

\label{Fig:Causal_weight_years}

\vspace{-0.5em}
\centering
\begin{threeparttable}

\includegraphics[width=0.8\columnwidth]{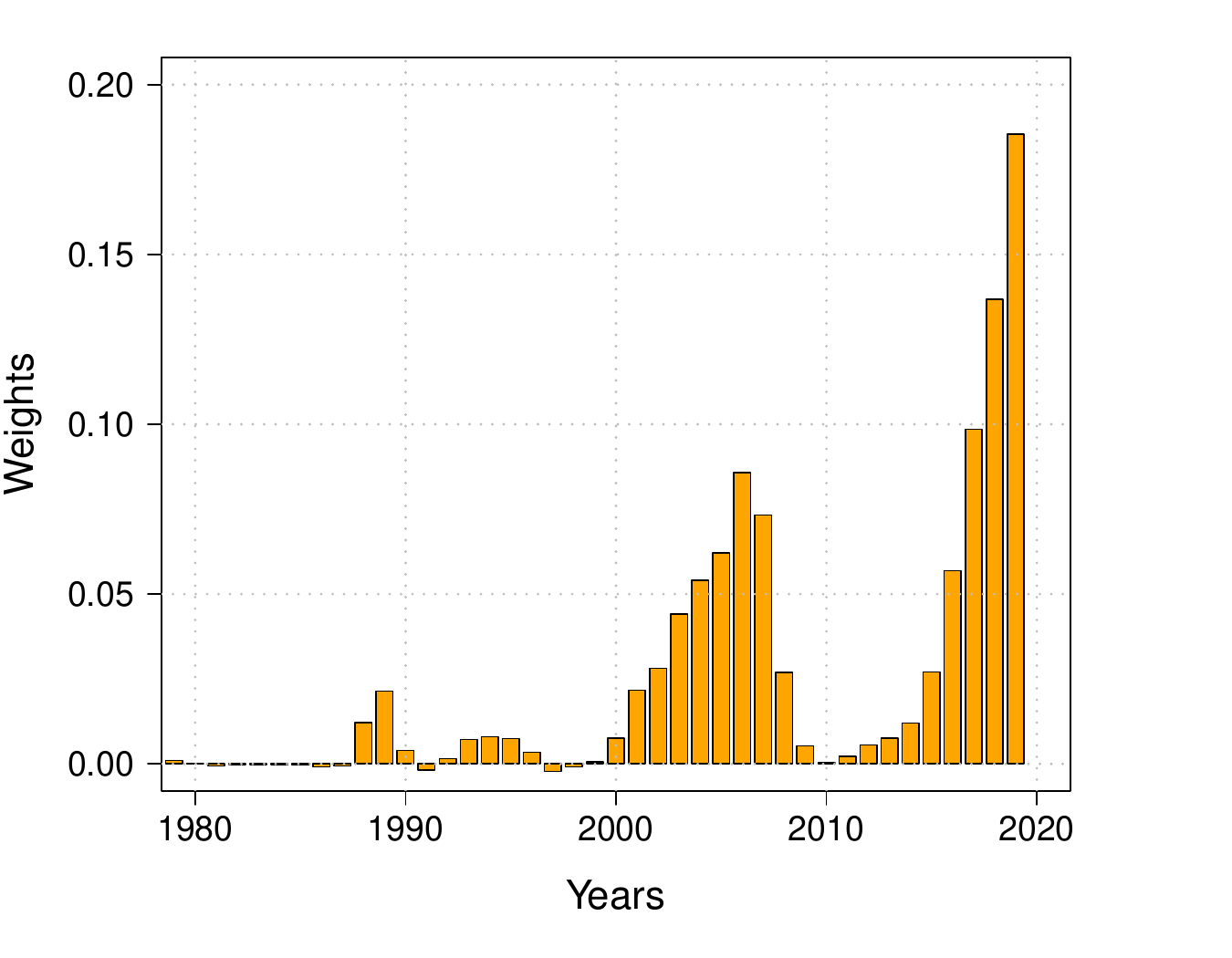}

\vspace{0.5em}
\begin{tablenotes}
\footnotesize

\item Note: The weights on years $t=1,\ldots,T$ are given by $E\left[\omega_{it}^{*}\right]$,
defined in equation \eqref{eq:weight_agg}, and sum to $1$.

\end{tablenotes}
\end{threeparttable}
\end{figure}

Figure \ref{Fig:Causal_weight_years} illustrates the pattern of weights
assigned to different calendar years in the state--year panel data
spanning 1979 to 2019. The weights are heavily concentrated in years
when the federal minimum wage remained unchanged for extended periods,
particularly in the mid-2000s and late 2010s. This pattern aligns
with the expectation that years with greater variation in minimum
wages contribute more to the estimation. Although some years are assigned
negative weights, their magnitudes are negligible.

\subsubsection*{Weights on States}

\begin{figure}
\caption{The Relationship Between Weights and State Policies}

\label{Fig:Causal_weight_state}

\centering
\begin{threeparttable}
\vspace{0.25em}

\includegraphics[width=0.7\columnwidth]{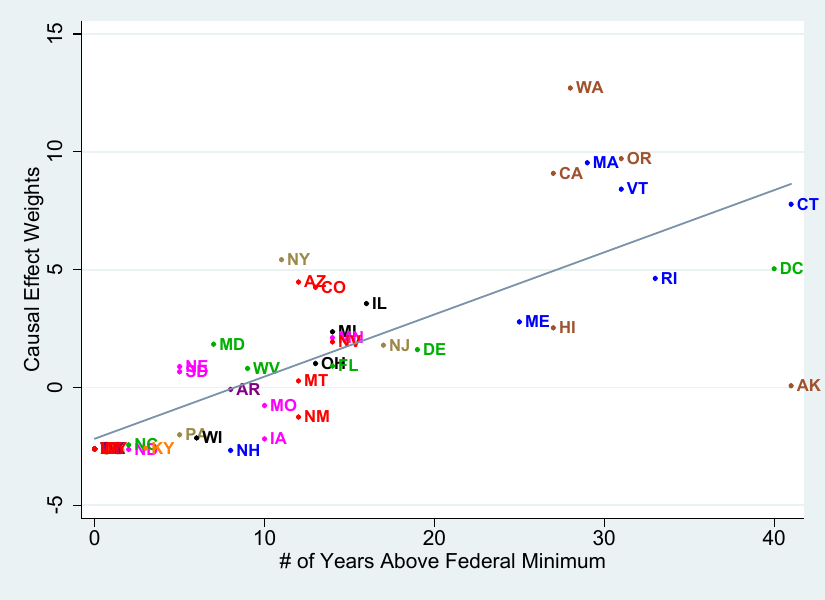}

\vspace{0.5em}
\begin{tablenotes}
\footnotesize

\item Note: The weights on states $i=1,\ldots,N$ are given by $\sum_{t=1}^{T}\omega_{it}^{*}$,
defined in equation \eqref{eq:weight_agg}, and have a mean of $1$.

\end{tablenotes}
\end{threeparttable}
\end{figure}

Figure \ref{Fig:Causal_weight_state} presents the distribution of
weights across states, with the vertical axis showing the weight assigned
to each state and the horizontal axis indicating the number of years
the state's minimum wage exceeded the federal minimum wage. The analysis
reveals a strong positive correlation ($\rho=0.798$) between the
state weights and the number of years a state's minimum wage was above
the federal level. This suggests the TWFE coefficient places greater
emphasis on states that have experienced more prolonged deviations
from the federal minimum wage.

Notably, 24 states are assigned negative weights, most of which closely
tracked the federal minimum wage for the majority of the years analyzed.
This raises a potential concern if the treatment effects of minimum
wage changes differ systematically across states with positive versus
negative weights. A common argument in the minimum wage literature
posits that minimum wage increases may have negative employment effects
only after exceeding the productivity of the marginal job. This threshold
is not necessarily reached immediately, especially in labor markets
characterized by monopsony power or frictions. If labor productivity
differs across states, it could imply heterogeneity in the minimum
wage effects. However, this variation would reflect nonlinearity in
the treatment effects rather than true effect heterogeneity across
states.

\subsubsection*{Weights on Minimum Wage Levels}

\begin{figure}
\caption{Weights on Treatment Levels, with Two Definitions of Treatment}

\label{Fig:Causal_weight_levels}

\vspace{0.5em}
\centering
\begin{threeparttable}
\vspace{0.25em}

\begin{minipage}[t]{0.75\columnwidth}%
(a) Log Minimum Wage

\includegraphics[width=1\columnwidth]{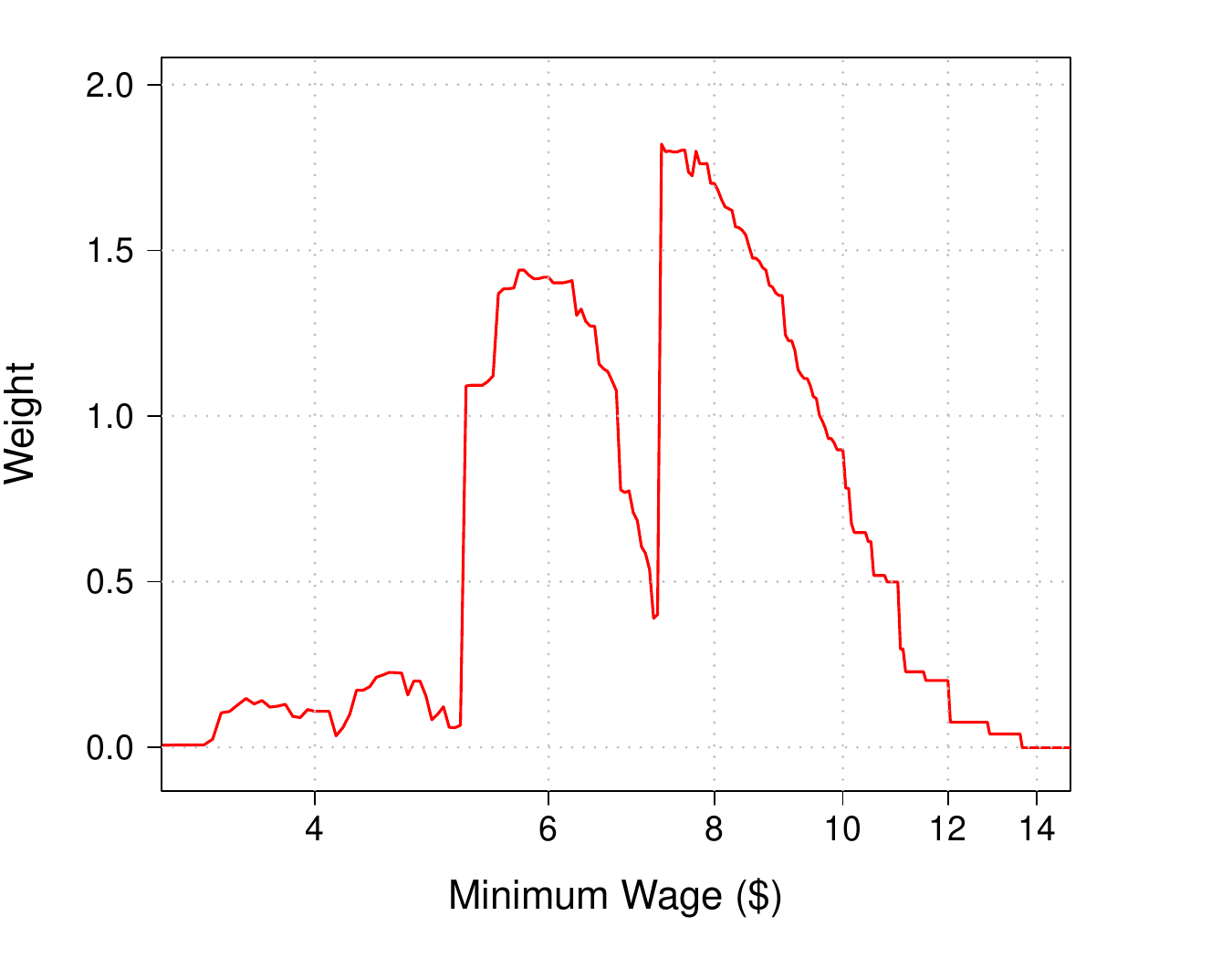}%
\end{minipage}

\vspace{0.75em}

\begin{minipage}[t]{0.75\columnwidth}%
(b) Gap Between Minimum Wage and Prime-age 25th Percentile

\includegraphics[width=1\columnwidth]{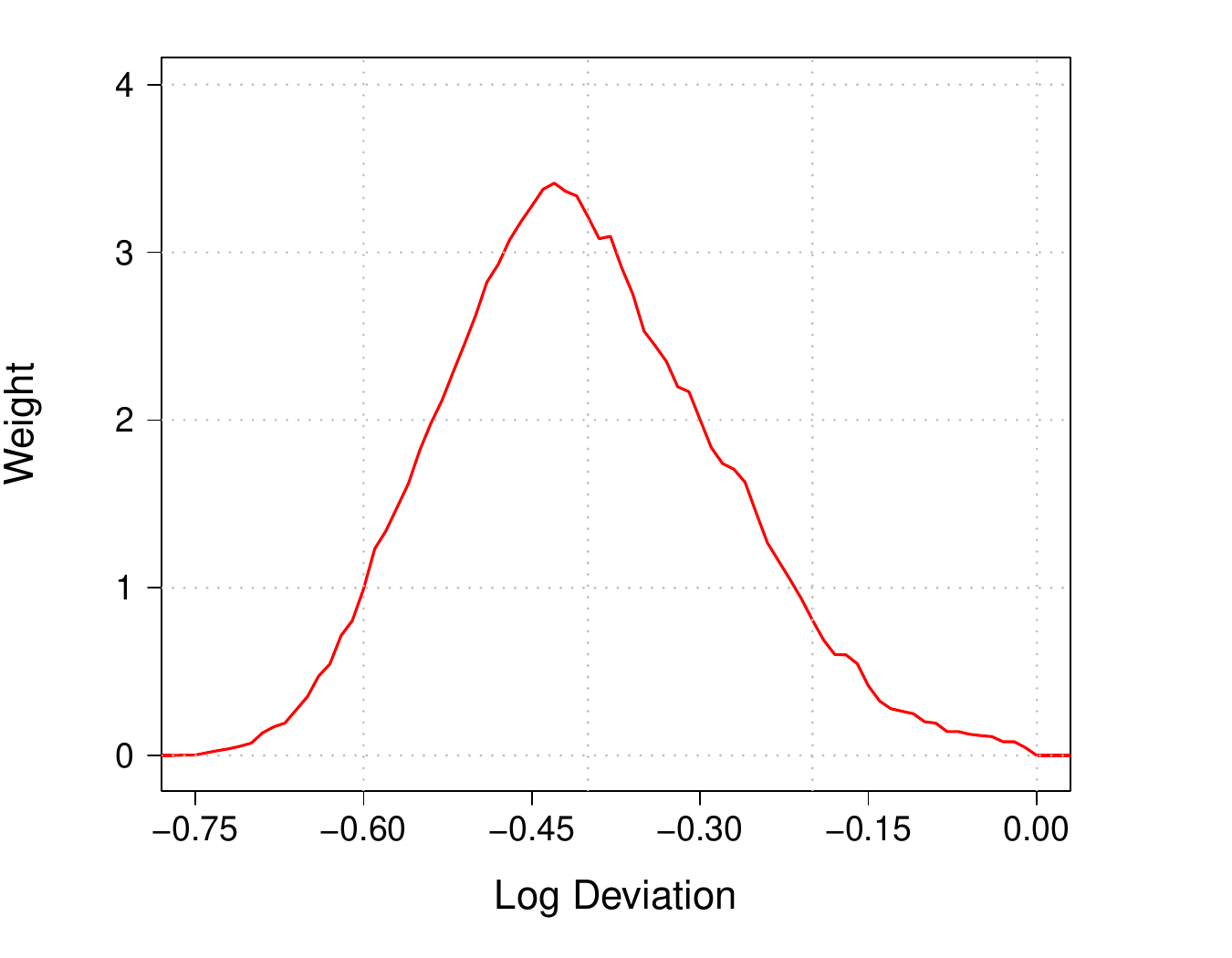}%
\end{minipage}

\vspace{0.5em}
\begin{tablenotes}
\small

\item Note: The weights on treatment levels $d\in(\underline{d},\overline{d})$
are given by $\sum_{t=1}^{T}E\left[\psi_{it}^{*}(d)\right]$, defined
in equation \eqref{eq:psi_star_def} and integrate to $1$. 

\end{tablenotes}
\end{threeparttable}
\end{figure}

Figure \ref{Fig:Causal_weight_levels} explores whether any specific
minimum wage levels receive negative weights in causal interpretation.
Two different treatment variables are considered. Panel (a) simply
uses the log of the minimum wage as the treatment. Panel (b) uses
the deviation of the log minimum wage from the predicted log 25th
percentile wage of prime-age (25--54) workers. This alternative treatment
variable is motivated by the potential concern raised earlier---if
the minimum wage effects differ systematically across states based
on how their minimum wages compare to labor productivity levels, then
causal interpretation of the TWFE coefficient may be compromised.
The predicted log 25th percentile wage is obtained by regressing the
observed raw state--year values on state and year indicators, ensuring
that this redefinition of the treatment variable does not alter the
TWFE coefficient.

Across both panels, the weights assigned to the different minimum
wage levels are predominantly positive. This suggests that even when
accounting for potential nonlinearity across minimum wage levels,
the TWFE coefficient can be interpreted as a meaningful weighted average
of the causal effects.

\end{document}